\documentclass[floatfix,showpacs,twocolumn,preprintnumbers,amsmath,amssymb]{revtex4}

\usepackage{graphicx}
\usepackage{dcolumn}
\usepackage{amsmath}
\usepackage{subfigure,hyperref}
\usepackage[normalem,normalbf]{ulem}

\newcommand{\be}{\begin{equation}}
\newcommand{\ee}{\end{equation}}         
\newcommand{\etal}{{\it et al.}}

\newcommand{\Min}{ {\text{\scriptsize Min } } }
\newcommand{\tr}{\text{Tr}}
\newcommand{\trg}{\text{\bf Tr}}
\newcommand{\gk}{\Gamma_k}

\newcommand\cg[1]{ \boldsymbol{ \sf #1}}
\newcommand\ie{{\it i.e. }}

\newcommand\Hi{H^{\text{int}} }
\def\nbOne{{\mathchoice {\rm 1\mskip-4mu l} {\rm 1\mskip-4mu l} {\rm 
1\mskip-4.5mu l} {\rm
1\mskip-5mu l}}}
\def\nbC{{\mathchoice {\setbox0=\hbox{$\displaystyle\rm C$}%
\hbox{\hbox to0pt{\kern0.4\wd0\vrule height0.9\ht0\hss}\box0}} 
{\setbox0=\hbox{$\textstyle\rm
C$}\hbox{\hbox to0pt{\kern0.4\wd0\vrule height0.9\ht0\hss}\box0}} 
{\setbox0=\hbox{$\scriptstyle\rm
C$}\hbox{\hbox to0pt{\kern0.4\wd0\vrule height0.9\ht0\hss}\box0}}
{\setbox0=\hbox{$\scriptscriptstyle\rm C$}\hbox{\hbox to0pt{\kern0.4\wd0\vrule
height0.9\ht0\hss}\box0}}}}
\def\nbQ{{\mathchoice {\setbox0=\hbox{$\displaystyle\rm 
Q$}\hbox{\raise 0.15\ht0\hbox
to0pt{\kern0.4\wd0\vrule height0.8\ht0\hss}\box0}} 
{\setbox0=\hbox{$\textstyle\rm Q$}\hbox{\raise
0.15\ht0\hbox to0pt{\kern0.4\wd0\vrule height0.8\ht0\hss}\box0}} 
{\setbox0=\hbox{$\scriptstyle\rm
Q$}\hbox{\raise 0.15\ht0\hbox to0pt{\kern0.4\wd0\vrule 
height0.7\ht0\hss}\box0}}
{\setbox0=\hbox{$\scriptscriptstyle\rm Q$}\hbox{\raise 0.15\ht0\hbox 
to0pt{\kern0.4\wd0\vrule
height0.7\ht0\hss}\box0}}}}
\def\nbT{{\mathchoice {\setbox0=\hbox{$\displaystyle\rm 
T$}\hbox{\hbox to0pt{\kern0.3\wd0\vrule
height0.9\ht0\hss}\box0}} {\setbox0=\hbox{$\textstyle\rm 
T$}\hbox{\hbox to0pt{\kern0.3\wd0\vrule
height0.9\ht0\hss}\box0}} {\setbox0=\hbox{$\scriptstyle\rm 
T$}\hbox{\hbox to0pt{\kern0.3\wd0\vrule
height0.9\ht0\hss}\box0}} {\setbox0=\hbox{$\scriptscriptstyle\rm T$}\hbox{\hbox
to0pt{\kern0.3\wd0\vrule height0.9\ht0\hss}\box0}}}}
\def\nbS{{\mathchoice {\setbox0=\hbox{$\displaystyle     \rm 
S$}\hbox{\raise0.5\ht0%
\hbox to0pt{\kern0.35\wd0\vrule height0.45\ht0\hss}\hbox 
to0pt{\kern0.55\wd0\vrule
height0.5\ht0\hss}\box0}} {\setbox0=\hbox{$\textstyle        \rm 
S$}\hbox{\raise0.5\ht0%
\hbox to0pt{\kern0.35\wd0\vrule height0.45\ht0\hss}\hbox 
to0pt{\kern0.55\wd0\vrule
height0.5\ht0\hss}\box0}} {\setbox0=\hbox{$\scriptstyle      \rm 
S$}\hbox{\raise0.5\ht0%
\hboxto0pt{\kern0.35\wd0\vrule height0.45\ht0\hss}\raise0.05\ht0%
\hbox to0pt{\kern0.5\wd0\vrule height0.45\ht0\hss}\box0}} 
{\setbox0=\hbox{$\scriptscriptstyle\rm
S$}\hbox{\raise0.5\ht0%
\hboxto0pt{\kern0.4\wd0\vrule height0.45\ht0\hss}\raise0.05\ht0%
\hbox to0pt{\kern0.55\wd0\vrule height0.45\ht0\hss}\box0}}}}
\def\nbZ{{\mathchoice {\hbox{$\sf\textstyle Z\kern-0.4em Z$}} 
{\hbox{$\sf\textstyle Z\kern-0.4em Z$}}
{\hbox{$\sf\scriptstyle Z\kern-0.3em Z$}} 
{\hbox{$\sf\scriptscriptstyle Z\kern-0.2em Z$}}}}

\makeatletter
\renewcommand{\@thesubfigure}{\alph{subfigure}) \space}
\renewcommand{\p@subfigure}{}
\makeatother
\begin{document}
\title{Nonperturbative renormalization group approach to frustrated magnets}

\author{B. Delamotte} \email{delamotte@lpthe.jussieu.fr}
\affiliation{Laboratoire de Physique Th\'eorique et Hautes
Energies, CNRS UMR 7589 Universit\'es Paris VI-Pierre et Marie Curie - Paris
VII-Denis Diderot, 2 Place Jussieu, 75251 Paris Cedex 05, France.  }%
\author{D. Mouhanna} \email{mouhanna@lpthe.jussieu.fr}
\affiliation{Laboratoire de Physique Th\'eorique et Hautes
Energies, CNRS UMR 7589 Universit\'es Paris VI-Pierre et Marie Curie - Paris
VII-Denis Diderot, 2 Place Jussieu, 75251 Paris Cedex 05, France.  }%
\author{M. Tissier} \email{tissier@lptl.jussieu.fr}
\affiliation{Laboratoire de Physique Th\'eorique des Liquides, CNRS UMR 7600
Universit\'e Paris VI-Pierre et Marie Curie, 4  Place Jussieu, 75252 Paris Cedex
05, France.}%
\date{\today}

\begin{abstract}
 This article is devoted to the study of the critical properties of
 classical XY and Heisenberg  frustrated magnets in three dimensions.
 We first  analyze the experimental and numerical situations.  We show
 that the unusual behaviors encountered in these systems, typically
 {\it nonuniversal scaling}, are  hardly compatible with the hypothesis 
of a second order phase transition. Moreover,  the fact  that the  
 scaling laws are significantly violated and that the anomalous dimension
 is negative in many cases provides  strong indications  that the transitions
 in frustrated magnets  are most probably of very weak first order. We then
review the various perturbative and early nonperturbative approaches
used to investigate these systems. We argue that none of them
provides a completely satisfactory description of the three-dimensional 
critical behavior.  We then recall  the principles of the  nonperturbative
approach --- the effective average action method --- that we have used to
 investigate the physics of frustrated magnets. First, we recall the treatment
 of the   unfrustrated --- $O(N)$  --- case with this method. This  allows 
 to introduce its technical aspects. Then, we show how this method unables  
 to clarify most of the problems encountered in the previous theoretical
 descriptions of frustrated magnets.  Firstly, we get an explanation of the
 long-standing mismatch between  different perturbative approaches which 
consists in a   nonperturbative mechanism  of annihilation of fixed points 
 between two and three  dimensions. Secondly, we get a coherent picture of the
 physics of frustrated magnets in qualitative and  (semi-) quantitative 
agreement with the numerical and experimental results.  The central feature 
that emerges from our approach is the existence of scaling behaviors 
{\it without } fixed or  pseudo-fixed  point and that  relies on a 
slowing-down of the renormalization group flow  in a {\it whole}  region 
in the coupling constants  space.  This phenomenon allows to explain the 
occurence of {\it generic}  weak first order behaviors and to understand
the absence of universality in the critical behavior of frustrated magnets.
\end{abstract}

\pacs{75.10.Hk,64.60.-i,11.10.Hi}

\maketitle

\begin{widetext}
\tableofcontents
\end{widetext}

\newpage

\section{Introduction}

Understanding phase transitions and, specifically, critical phenomena
has been one of the central issues of statistical mechanics during
these last decades \cite{wilson74} and the field theoretical
renormalization group (RG) approach to these phenomena has been one of the
great successes of theoretical physics. This is so true that it is
generally believed that, apart from specific problems --- disordered
and glassy systems for instance ---,  an almost complete understanding
of the physics occurring at a phase transition has been reached. This
is certainly due to the fact that it is indeed the case for all the
systems belonging to the so-called Wilson-Fisher  universality classes
of $d$-dimensional  systems whose symmetry breaking scheme is given by
$O(N)\to O(N-1)$. In fact, although they have
become the archetype of systems  displaying critical phenomena well
described by  perturbative field theoretical approaches, these $O(N)$ symmetric
systems turn out to be exceptions rather than the rule. For most
systems a quantitative and, even sometimes, a qualitative description of the critical
physics is either still lacking or very difficult to obtain by perturbative RG methods.
This is the case, for instance, in the Potts model
\cite{amit76,priest76}, in magnetic systems with disorder
\cite{holovatch02}, in superconductors \cite{lubensky74,dasgupta81},
in Josephson junction arrays \cite{teitel83}, in  He$_3$
\cite{jones76,bailin77}, in smectic liquid crystals \cite{halperin74},
in electroweak phase transitions \cite{lawrie83,marchrussel92} and  in
frustrated magnets like helimagnets or geometrically frustrated
magnets (triangular for instance) which are our main purpose  in this article (see \cite{diep94} 
for a review).

Actually, it should not be surprising that a qualitative difference
exists between the critical behaviors of systems belonging to the
$O(N)$ universality class and the others: among the systems where the
order parameter has $N$ real components, $O(N)$ corresponds to the
maximal symmetry and, thus, to the simplest situation.  Think, for
instance, at a unit-norm constraint imposed on the microscopic degrees
of freedom ($\vec{S}^2=1$): the maximal symmetry compatible with it is
indeed $O(N)$. From a perturbative point of view, this means that the
Ginzburg-Landau-Wilson (GLW) Hamiltonian of an $O(N)$ symmetric model
involves only one (marginal) interaction term --- $({\vec{\phi}}^2)^2$
--- and, thus, only one coupling constant.  As a consequence, the
perturbative RG flow of the critical theory takes place in a
one-dimensional space of coupling constant and is thus simple.  In
particular, only one nontrivial perturbative fixed point can exist
\cite{Itzykson89}.  On the contrary, the Hamiltonian of systems having
a $N$-dimensional order parameter and displaying a symmetry group $G$
smaller than $O(N)$ involves also terms that {\it explicitly} break
$O(N)$. It thus contains several interaction terms and, therefore,
several coupling constants.  The RG flow then takes place in a
multi-dimensional space and is thus far less simple: it can, in
particular, involve fluctuation-induced first order transitions ---
runaway in a region of instability --- and several fixed points with
different symmetries.  Universality itself is not guaranteed in these
systems since the basins of attraction of the fixed points can be highly nontrivial.

 Of particular interest for us, it is generically observed in these
systems that, by varying $N$ and/or $d$, the critical physics changes
qualitatively: low dimensions ($d\to 2$) and  large number of spin
components  ($N\to\infty$)
favor smooth fluctuations and, thus, second order phase transitions,
while larger dimensions ($d\to 4$) and small $N$ ($N\sim 1$) favor
larger fluctuations and thus first order transitions \footnote{This
behavior is observed for theories with continuous order
parameters. For the Potts model, the situation is different since
large numbers of states favor first order phase transitions.}.
Therefore, in many systems --- and notably in frustrated systems --- 
the critical behavior changes
qualitatively {\it i)} for the physically interesting values of $N$ 
---  $N\sim 1$ ---  
when the dimension varies between  $d=2$ and $d=4$  {\it ii)} at fixed dimension 
when the number  $N$ of components varies between  $N=\infty$ and $N\sim 1$.
Thus, the different perturbative approaches are in the worst possible position:
it is  quite difficult to obtain definite conclusions in $d=3$ and for $N\sim 1$  from
extrapolations of perturbative results even if they are valid in the domains where 
they have been established: $d=2$ for the
Nonlinear Sigma (NL$\sigma$) model,  $d=4$ for the GLW model and for large $N$ in a $1/N$ expansion.  This
is one of the reasons why, after more than twenty five years of
considerable efforts, the situation is still not clear for most three-dimensional systems that 
do not belong to the $ O(N)/O(N-1)$
universality classes.

Let us now discuss two concrete problems encountered in the
perturbative RG studies performed on the three-dimensional systems we
are interested in. First, the computational difficulties encountered
in perturbation theory are non negligible.  Within
the NL$\sigma$ model approach, the series are generally considered as
useless due to the lack of Borel-summability (see however
\cite{kleinert00}).  Within the GLW approach, the perturbative
computations almost always call for resummation procedures. In
general, these  procedures are not as easy as they are  in
the ${O(N)/O(N-1)}$ models.  The series are either not proven to be
Borel summable or are even suspected to be non Borel summable. This is
the root of a lot of difficulties encountered in this approach (see
\cite{bray87,mckane94,shalaev97,folk99,alvarez00,carmona00,loison00};
for a review in the case of the diluted Ising model, see \cite{folk00},
for the presence of non analyticities in perturbative series, see
\cite{sokal94,sokal95,bagnuls97,pelissetto98} and, for a general review, 
see \cite{pelissetto01c}). The second point is more conceptual:
although it is generically possible to perform a perturbative expansion of the
critical theory around $d=2$ --- within the NL$\sigma$ model approach
--- and around $d=4$ --- within the GLW approach --- it has not been
possible to relate these two expansions within the usual field
theoretical approach (except for large enough $N$ where the $1/N$
expansion  allows to recover, at leading orders, the
perturbative results obtained in the NL$\sigma$ and GLW approaches).
From this point of view, even high-order perturbative calculations
performed in the GLW model do not help since the perturbative
expansion cannot be extrapolated down to $d=2$ for $N\ge 2$. For instance, the
critical exponent $\nu$  diverges in $d=2$ when  it is calculated
as a power series in $\epsilon=4-d$. This  fact, which is not
crucial for systems whose critical behavior does not change
qualitatively between $d=2$ and $d=4$ (e.g.  the $ O(N)/O(N-1)$
models) forbids for the others to obtain a completely coherent picture
of the physics between $d=2$ and $d=4$. Most
of the time one of the perturbative approach --- usually the NL$\sigma$ one --- 
is dismissed without real justifications  and the other
is blindly trusted.  Since all the RG equations are smooth in $N$ and
$d$, it is not clear  if and also,  why,  this procedure is  legitimate.  It
would, of course, be much more satisfactory to have a unified approach
not linked to a particular value of $d$ or $N$ and that allows to
interpolate between both  approaches.

All these drawbacks of the usual perturbative RG methods call for a
nonperturbative approach.  Such an approach is, in fact, already  known and its
foundations go back to Kadanoff and Wilson with the idea of block spin
and effective, scale-dependent theory \cite{kadanoff66,wilson74}.  It
is sometimes called the exact renormalization group method but we
prefer to call it the nonperturbative renormalization group (NPRG)  method
(for contributions of different authors to the early attempts
to use NPRG, see \cite{wegner73,nicoll76,nicoll77,hasenfratz86},  for an exhaustive bibliography of the
 subject, see \cite{bagnuls01}).  This
idea has been turned into an efficient computational tool during the last
ten years, mainly by Ellwanger
\cite{ellwanger93a,ellwanger93b,ellwanger94a,ellwanger94b,ellwanger94c},
Morris \cite{morris94a,morris94b} and Wetterich
\cite{wetterich91,wetterich93c,tetradis94,berges02}.  It has allowed
to determine the critical exponents of the $O(N)$ models with high
precision without having recourse to resummation techniques
\cite{aoki98,morris98c,seide99,berges02,canet03a,canet03b}.  It has
also allowed, for the first time \cite{wetterich93}, to relate, for
any $N$, the results of the $O(N)/O(N-1)$ model obtained near $d=4$
and $d=2$, a fact of major importance for our purpose.  Also important
for the present purpose, it has allowed to tackle with  genuinely nonperturbative
situations. For instance, the Berezinskii-Kosterlitz-Thouless phase
transition \cite{berezinskii70,kosterlitz73} has been recovered directly from a study of the GLW model,
\ie {\it without} introducing explicitly the vortices
\cite{grater95,gersdorff01}.  To cite just a few other successes of
this method, let us mention low-energy Quantum Chromodynamics
\cite{berges02}, the abelian Higgs model relevant for
superconductivity \cite{bergerhoff96}, the study of the Gross-Neveu
model in three dimensions \cite{rosa01,hofling02}, phase transitions
in He$_3$ \cite{kindermann01}, the study of cubic anisotropy in all
dimensions as well as the randomly diluted Ising model
\cite{tissier01b}, the two-dimensional Ising multicritical points
\cite{morris95b},  etc.

In this article, we study by means of NPRG methods one of the most
famous systems exhibiting the changes of critical behavior  previously described:
it is the system consisting of XY or Heisenberg spins on the triangular lattice (stacked triangular
in $d=3$) with antiferromagnetic nearest neighbor interaction (Section
\ref{chapitre_model}).  This system is the archetype of frustrated
spin systems and is supposed to be in the same universality class as
another set of frustrated magnets: the helimagnets. Almost all these
systems have been intensively studied both numerically and
experimentally these last twenty five years (see Section
\ref{chapitre_experimental}).  However, their behavior remains unclear
and displays quite unconventional features.  For instance, almost all
experiments exhibit scaling laws around the transition temperature ---
which suggests a second order phase transition --- but with critical exponents
that  depend on the particular material studied, on microscopic details, etc,  which is incompatible
 with the standard phenomenology of a second order phase transition.  In some experiments or
numerical simulations, the scaling laws are sometimes significantly
violated while the anomalous dimension $\eta$ is found 
negative, a fact forbidden by first principles if the theory is
$\phi^4$ GLW-like (see the following).  The theoretical situation in
these systems is also not clear from the perturbative point of view
(Sections \ref{chapitre_chronological} and \ref{chapitre_perturbatif}):
first, independently of the experimental context, the results obtained
within the usual perturbative approaches --- in dimensions
$d=2+\epsilon$ and $d=4-\epsilon$ --- conflict. Second, neither the
low-temperature expansion around $d=2$ nor high-order weak-coupling
calculations performed around $d=4$ or directly in $d=3$ succeed in
reproducing satisfactorily the  phenomenology.  We
show, in this
 article, that
the NPRG approach (Section \ref{chapitre_effective}) to frustrated  systems
(Section \ref{chapitre_on_o2}) clarifies almost entirely the
situation. First, it allows to smoothly interpolate between $d=2$ and
$d=4$ and to clarify the mismatch between these approaches. In
particular, a mechanism of annihilation of fixed points, already
identified  for a long time around $4-\epsilon$ dimensions for $N\sim 21.8$  is shown to operate
  around two dimensions for $N\sim 3$ {\it  nonpertubatively}  with respect to  the low-temperature
approach of the NL$\sigma$ model \cite{tissier00b,tissier00}. This
 explains
the irrelevance in $d=3$ of the $O(4)$ fixed point obtained within a
low-temperature approach in $d=2+\epsilon$.  Second, our approach
provides a description of the physics in $d=3$, in terms of weakly
first order behaviors, compatible with the phenomenology (Sections
\ref{chapitre_test} and \ref{chapitre_d=3}). In this respect, an
important feature of our work is that it explains the occurence of
scaling in frustrated magnets  {\it without}  fixed   or
 pseudo-fixed \cite{zumbach93,zumbach94,zumbach94c}  point.   This phenomenon 
 relies on a slowing-down of the RG flow in  a whole  region in coupling constants space. 
This allows to explain one of the most puzzling aspect of the critical
physics of these systems, {\ie}the occurence of scaling {\it without}
universality. We discuss (Section \ref{chapitre_checkscenario})
possible experimental and numerical tests of our scenario. We then
comment  (Section \ref{chapitre_consequences}) the consequences of our
work for the perturbative approaches that have been used to
investigate the physics of frustrated magnets. Finally, we give our
conclusions (Section \ref{chapitre_conclusion}).

\section{The STA model and  generalization}
\label{chapitre_model}

\subsection{The lattice model, its continuum limit  and  symmetries}

 We now describe the archetype of frustrated spin systems, the Stacked
Triangular Antiferromagnets (STA). This system is composed of
two-dimensional triangular lattices which are piled-up in the third
direction. At each lattice site, there is a magnetic ion whose
spin  is described by a classical vector. 
The interaction between the spins  is given  by the usual
lattice Hamiltonian:
\begin{equation} 
H=\sum_{\langle ij\rangle}J_{ij}\, \vec{S}_i. \vec{S}_j
\label{hmicroscopique} 
\end{equation}
where, depending on the
anisotropies, the $\vec{S}_i$ are two or three-component vectors and the sum
runs on  all pairs of nearest neighbor spins. The coupling constants
$J_{ij}$ equals $J_\perp$ for a pair of sites inside a plane and
$J_\parallel$  between planes.

\begin{figure}[htbp] 
\begin{center}
\includegraphics[width=.8\linewidth,origin=tl]{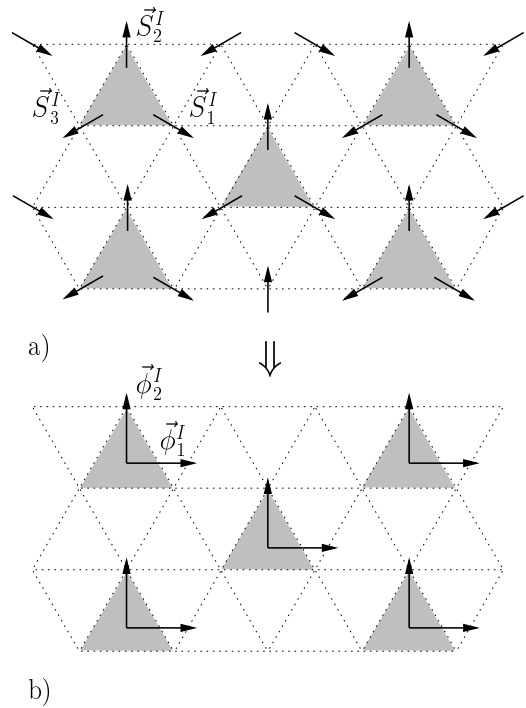}\hfill%
\end{center} 
\caption{The ground state configurations a) of the spins on the
triangular lattice and b) of the order parameter made of two
orthonormal vectors. The plaquettes, which constitute the magnetic
cell, are indexed by $I$ and are  shaded.}
\label{triangulaire} 
\end{figure}

The interactions between nearest neighbor spins
within a plane is antiferromagnetic, \ie $J_\perp>0$. This induces
frustration in the system and, in the ground state, gives rise to the
famous 120$^\circ$ structure of the spins, see
Fig.~\ref{triangulaire}a.
This nontrivial magnetic structure is invariant under translations of
length $\sqrt 3$ times the initial lattice spacing. The magnetic cell,
indexed by $I$,
which is replicated all over the system, is a plaquette of  three spins $\vec{S}_1^{I}$, $\vec{S}_2^{I}$
and $\vec{S}_3^{I}$, see Fig.~\ref{triangulaire}a.

Note that the  nearest-neighbor out-of-plane interaction  $J_\parallel$ is,
depending on the compounds, ferro- or antiferro-magnetic, but the two
cases can be treated simultaneously since no extra frustration appears
through this interaction. Finally, interactions between more distant spins
(next-to-nearest neighbors, etc) also exist but are neglected  in the
following since they are supposed to be irrelevant.

There have been numerous derivations of the long distance effective
field theory supposed to describe the critical physics of this system
\cite{garel76,yosefin85,dombre88,kawamura88}. We here sketch the
derivation which is the most appropriate for our purpose. The Hamiltonian
(\ref{hmicroscopique}) has the usual rotational symmetry acting on the
spin components: $O(2)$ or $O(3)$ for XY or Heisenberg spins,
respectively. To identify the order parameter, it is also necessary to
consider the symmetry of the magnetic cell. For the triangular
lattice, this is the $C_{3v}$ group that interchanges the spins inside
a plaquette \footnote{The symmetry of the full Hamiltonian
$H$ is the space group of the triangular lattice.  However, as 
 we are most of the time interested in
Hamiltonian densities --- for instance when the continuum limit has
been taken ---,  the relevant symmetry is $C_{3v}$.}.

The identification of the order parameter is close in spirit to what
is done in the nonfrustrated case, {\it e.g.} for the antiferromagnets
on a square lattice.  At zero temperature, the sum of the three spins
for a given plaquette~$I$:
\begin{equation}
\vec{\Sigma}^{I}=\vec{S}_1^{I}+\vec{S}_2^{I}+\vec{S}_3^{I} 
\end{equation}
is vanishing (Let us  note that $\vec{\Sigma}^I$ is analogous to the
local magnetization of  nonfrustrated  antiferromagnets --- 
$\vec{\Sigma}^I=\vec{S}_1^{I}+\vec{S}_2^{I}$ in this last case --- that also
 vanishes in the ground state). In average, this is also the
case at any finite temperature so that the thermal average of:
\begin{equation}
\vec{\Sigma}=\sum_I\vec{\Sigma}^{I}\ ,
\end{equation}
where the sum runs on all plaquettes, cannot be an order parameter: the associated modes are never critical.
We therefore replace $\vec{\Sigma}^{I}$ by its average value:
\begin{equation}
\vec{\Sigma}^{I}\to\langle \vec{\Sigma}^{I}\rangle=\vec{0}
\label{eq3}
\end{equation}
which is equivalent to freezing the fluctuations of the spins inside
each plaquette.  The constraint $\vec{\Sigma}^{I}=\vec 0$ is called
the ``local rigidity constraint''.  Having eliminated $\vec{\Sigma}$,
we keep only two vectors per plaquette
$(\vec{\phi}_1^{I},\vec{\phi}_2^{I})$ which represent the local order
parameter.  For $\vec{\phi}_2^{I}$, we choose one of the spins of the
plaquette, see Figs.~\ref{triangulaire}a and b. For the other,
$\vec{\phi}_1^{I}$, we choose the linear combination of the spins
which is orthogonal to $\vec{\phi}_2^{I}$ and of unit norm, see Fig.~\ref{triangulaire}b. 
The local order parameter thus obeys on each plaquette:
\begin{equation}
\vec{\phi}_i^{I}.\vec{\phi}_j^{I}=\delta_{ij}\  \   \  \  \   \  \
\hbox{with}\ \ \ \ \ \ i,j\in\{1,2\}\ .
\label{eq4}
\end{equation}
The dihedral  $(\vec{\phi}_1^{I},\vec{\phi}_2^{I})$  plays  a role analogous to 
 the  staggered magnetization in the nonfrustrated case.

As usual, once the model is reformulated in terms of its order
parameter, the effective interaction --- from pla\-quet\-te to
plaquette --- becomes ferromagnetic, see Fig.~\ref{triangulaire}b.
By taking the dihedral $(\vec{\phi}_1^{I},\vec{\phi}_2^{I})$ on the
center of the plaquette $I$, we indeed   find that it interacts
ferromagnetically with the dihedral $(\vec{\phi}_1^{J},\vec{\phi}_2^{J})$ defined
on the center of the plaquette $J$ --- the plaquettes $I$ and $J$
being nearest neighbours --- such that $\vec{\phi}_1^{I}$ interacts
only with $\vec{\phi}_1^{J}$ and $\vec{\phi}_2^{I}$ only with
$\vec{\phi}_2^{J}$. A more detailed analysis shows that the two
vectors $\vec{\phi}_1^{I}$ and $\vec{\phi}_2^{I}$ play symmetric roles
\cite{yosefin85}. As a consequence, the effective Hamiltonian reads:
\begin{equation}
 H=-J \sum_{\langle
I,J\rangle} \left(\vec{\phi}_1^{I}.\vec{\phi}_1^{J}
+\vec{\phi}_2^{I}.\vec{\phi}_2^{J} \right)
\label{stiefel}
\end{equation}
with the same coupling constant $J>0$ for the $\vec{\phi}_1^{I}$'s and for the
$\vec{\phi}_2^{I}$'s.  Moreover, since the anisotropies resulting from the
stacked structure of the lattice are supposed to be irrelevant, we
take the same coupling constant for the interactions inside a plane
and between the planes.  The continuum limit is now trivial and
proceeds as in the usual ferromagnetic case.  The effective
Hamiltonian in the continuum thus writes, up to constants:
\begin{equation}
H=-\int d^d \cg x  \left( (\partial \vec{\phi}_1(\cg x))^2 +
(\partial  \vec{\phi}_2(\cg x))^2  \right) 
\label{continu} 
\end{equation}
with the constraint that $\vec{\phi}_1$ and $\vec{\phi}_2$ are
orthonormal.  This model is called the Stiefel V$_{N,2}$ model with
$N=2$ in the XY case and $N=3$ in the Heisenberg case. In $V_{N,2}$
the index 2 means that we are considering {\it two} orthonormal
vectors $\vec \phi_1$ and $\vec \phi_2$.

It is convenient to gather the vectors $\vec{\phi}_1$ and
$\vec{\phi}_2$ into a rectangular matrix:
\begin{equation}
\Phi=(\vec{\phi}_1,\vec{\phi}_2)
\label{matriceparametreordre}
\end{equation}
and to rewrite $H$ as:
\begin{equation}
H=-\int  d^d\cg x \,   \hbox{Tr}\left(\partial^{\;t}\!\Phi(\cg x).
\partial\Phi(\cg x)\right)
\label{eq9}
\end{equation}
where $(^{t}\!\Phi)_{ij}=\Phi_{ji}$.

In the following two sections, we consider successively the case of
Heisenberg and XY spins.

\subsection{The Heisenberg case}
\label{Heisenbergcase}

In this case, $H$ is invariant under the usual left $O(3)$ rotation
and inversion group acting on the spins:
\begin{equation}
 \Phi'= R \Phi,\ \ \ \ \ \ \ R\in O(3) \ .  
\label{o3gauche}
\end{equation}
It is also invariant under a right $O(2)$: 
\begin{equation}
 \Phi'=\Phi U,\ \ \ \ \
\ \ U\in O(2) \ .
\label{o2droit} 
\end{equation}
This last symmetry encodes the fact that $\vec{\phi}_1$ and
$\vec{\phi}_2$ play the same role which, itself, is reminiscent of the
$C_{3v}$ symmetry of the triangular plaquette.  The system is thus
symmetric under $G=O(3)\times O(2)$. In the low-temperature phase, a
typical ground state configuration is given by (see Fig.~\ref{triangulaire}b):
\begin{equation} 
\Phi_0\propto
\begin{pmatrix}
1&0\\
0&1\\
0&0
\end{pmatrix}
\ .
\end{equation}
It is symmetric under the diagonal group --- $O(2)_{\hbox{\small
diag}}$ --- built from the right $O(2)$ and from  a particular left
$O(2)$ in $O(3)$:
\begin{equation}
\Phi_0=
\left(
\begin{array}{ccc}
\epsilon \cos\theta & -\sin\theta  & 0 \\
\epsilon \sin\theta & \cos\theta   & 0 \\
 0         &     0        &  1
\end{array}
\right) \Phi_0 \left(
\begin{array}{cc}
\epsilon \cos\theta  &  \epsilon \sin\theta  \\
-\sin\theta  &  \cos\theta
\end{array}
\right)
\end{equation}
where $\epsilon=\pm 1$ encodes the ${\nbZ}_2$ part of
$O(2)_{\hbox{\small diag}}$.  Apart from the previous ${\nbZ}_2$
contained in the $O(2)_{\hbox{\small diag}}$, another ${\nbZ}_2$ is
also left unbroken. It is the combination of a ${\nbZ}_2$
included into the right $O(2)$ of $G$, Eq.~(\ref{o2droit}), and of a
rotation of $\pi$ around the $x$-axis  contained in the rotation group
$SO(3)$ of $G$. Thus, $G$  is spontaneously broken down to
$H={\nbZ}_{2}\times O(2)_{\hbox{\small diag}}$. As
a consequence, the symmetry breaking scheme reads:
\begin{equation}
 G=O(3)\times O(2) \to H={\nbZ}_{2}\times O(2)_{\hbox{\small diag}}
\label{symbroken} 
\end{equation}
which is often referred to, once all the ${\nbZ}_{2}$  groups have been cancelled, 
as the $SO(3)\times SO(2)/SO(2)$ model.

 Here appears the main feature of frustrated magnets: the $SO(3)$ group
is {\it fully} broken  in the low-temperature phase whereas it is only broken down to $SO(2)$ in
 nonfrustrated magnets. This has two important consequences that  are  at the very
origin of the nontrivial critical behavior encountered in frustrated
magnets.

 First, there are three Goldstone modes in the broken phase instead of two in
the nonfrustrated case.  This implies a  physics of spin waves different from that of  the $O(3)/O(2)$ model. 
Second, the order parameter space $SO(3)$
having a nontrivial first homotopy group \cite{mermin79}:
\begin{equation}
\pi_1(SO(3))={\nbZ}_2
\label{homotopie}
\end{equation}
there exist stable nontrivial topological configurations called
vortices. Because of the $\nbZ_2$ homotopy group, only one kind of
vortex exists, contrarily to the well-known case of XY ferromagnets
where there are infinitely many different kinds of vortices, each one
being indexed by an integer, the winding number. 

 It has been established firstly by Kawamura and Miyashita 
\cite{kawamura84}
that the existence of  vortices is important  at finite temperature in two dimensions. This has been
largely confirmed  by subsequent works studying the temperature
 dependence of thermodynamical quantities such
as the correlation length, the spin-stiffness, etc
\cite{apel92,southern93,wintel94,southern95,stephan00,caffarel01}. Actually,
although this  has not been directly established, they
certainly also play an important role for the critical physics of the STA in
three dimensions. A simple argument allows  to argue to that end:
let us  go back on the lattice and introduce, on
each plaquette $I$, together with $\vec\phi_1^I$ and $\vec\phi_2^I$, a
third vector $\vec\phi_3^I$ defined by:  
\begin{equation}
\vec\phi_3^I=\vec\phi_1^I\wedge\vec\phi_2^I\ .
\label{thirdvector}
\end{equation}
Let us then  gather them  into a 3$\times$3 matrix:
\begin{equation}
\Phi^I=\left(\vec\phi_1^I,\vec\phi_2^I,\vec\phi_3^I\right)\ . 
\end{equation}
Since ($\vec\phi_1^I,\vec\phi_2^I,\vec\phi_3^I$) are three orthonormal
vectors, one has $^t\Phi^I\Phi^I=\nbOne$ and,  therefore,  $\Phi^I$ is a
$SO(3)$ matrix. This allows to rewrite  the Hamiltonian (\ref{stiefel})
 on the lattice as:
\begin{equation}
H=-\sum_{\langle I,J\rangle} \hbox{Tr}\left({\cal{P}}\ ^t\Phi^I.\Phi^J\right)
\label{hamilmatrix}
\end{equation}
where ${\cal P}$ is a diagonal matrix of coupling constants that
characterizes the interaction between the $\vec\phi_1^I$'s,  between the
$\vec\phi_2^I$'s and  between  the $\vec\phi_3^I$'s. One deduces  from the
microscopic derivation that  ${\cal P}=\hbox{diag}(J,J,0)$, {\ie that the
  interaction is the same between  the $\vec\phi_1^I$'s and beween  the
  $\vec\phi_2^I$'s and that there is no interaction between the
  $\vec\phi_3^I$'s. However, for the present purpose, we
  consider, without loss of generality,  the case where the interaction is nonvanishing
 and  identical  between all  vectors.  One thus has ${\cal  P}=J\nbOne$.   Now,  we use the decomposition
  of a rotation matrix $\Phi^I$ of $SO(3)$ in terms of a four-component {\it unit}
  vector $\widetilde{S^I}=(S_0^I, S_1^I, S_2^I, S_3^I)$: 
\begin{equation}
\Phi_{kl}^I=2\left(S_k^I S_l^I-{1\over 4}
\delta_{kl}\right)+2\epsilon_{klm}S_0^I S_m^I+2\left({S_0^I}^2-{1\over
  4}\right)\delta_{kl}\ . 
\end{equation}
In terms of the vector $\widetilde{S^I}$, the Hamiltonian
(\ref{hamilmatrix}}) writes: 
\begin{equation}
H= -4J{\displaystyle \sum_{\langle I,J\rangle}}
\left(\widetilde{S^I}.\widetilde{S^J}\right)^2\  
\label{nematic}
\end{equation}
which is the  Hamiltonian for {\it four}-component nonfrustrated
spins  with a particularity that  each vector
$\widetilde{S^I}$ appears quadratically.  Therefore, the Hamiltonian 
(\ref{nematic}) is  invariant under a global $O(4)$ group and under a 
{\it local} --- gauge ---  $\nbZ_2$ group that changes $\widetilde{S^I}$ to $-\widetilde{S^I}$. It
corresponds to  the
$RP^3=SO(4)/(SO(3)\times \nbZ_2$) model.  Note that, had we  kept  the
  microscopical coupling constants:  ${\cal P}=\hbox{diag}(J,J,0)$,  the
  Hamiltonian (\ref{nematic}) would be  supplemented
  by terms breaking the  $SO(4)$  global symmetry and  leaving untouched
  the $\nbZ_2$ local symmetry which is the important point for our
  purpose (see \cite{caffarel01} for details). For
three-component spins,  an  analogous  Hamiltonian --- the $RP^2$
model --- had been  introduced by Maier and Saupe \cite{maier59} and by
Lebwohl and Lasher \cite{lebwohl72} to investigate  the
isotropic-nematic transition in liquid crystals. An extensive study of
the  $RP^2$  model,  as well as  a detailed investigation of
the role of vortices in  this  transition, has been performed by
Lammert \etal\ \cite{lammert93,lammert95}.  These authors  have shown, in particular,
that these  nontrivial topological configurations favor the first order character
 of the  transition. In the
case of  four-component spins, no such  detailed analysis has been
performed. However,  the  $RP^N=SO(N)/(SO(N-1)\times \nbZ_2)$
models that  generalizes  Hamiltonian (\ref{nematic})  to  $N$-component  spins
  have been numerically studied in \cite{kohring87} for  $2\le N\le 4$. These systems
have been shown to  undergo a first order phase transition. Since the only difference between
the $RP^N$ and the  $O(N)/O(N-1)$ --- or, equivalently, $SO(N)/SO(N-1)$ --- models  lies
in their  topological properties,  one   is naturally led  to attribute the origin of the
first order character of the phase transition in the $RP^N$  models
to the $\nbZ_2$  vortices. Finally, since the Hamiltonian (\ref{hamilmatrix}), relevant to STA, can be mapped
 onto the Hamiltonian (\ref{nematic}) --- up to the $O(4)$-breaking terms ---
 one can expect that the  topological configurations  also  favor
first order  phase transitions in  frustrated magnets in three dimensions.

\subsection{The XY case}

 In the XY case, the Hamiltonian (\ref{eq9}) is still invariant
under a  right $O(2)$ group, see Eq.~(\ref{o2droit}), while the left
symmetry group becomes $O(2)$. In the low-temperature phase, the
rotational symmetry is broken and, since the spins are constrained to
be in a plane, the permutation symmetry between $\vec{\phi}_1$ and
$\vec{\phi}_2$ is also broken.  As a consequence,  the symmetry breaking
scheme is:
\begin{equation}
G=O(2)\times O(2) \to H=O(2)_{\hbox{\small diag}}.
\end{equation}
This symmetry-breaking scheme is usually referred to as $SO(2)\times
\nbZ_2\to \openone$.  The ${\nbZ}_2$ degrees of freedom are known as
chirality variables
\cite{villain77,kawamura84,miyashita85,kawamura88}.

In this case,  there also  exist topological defects since:
 \be
\pi_1(SO(2))=\nbZ\ .  
\ee 
These defects are identical to those of
the ferromagnetic XY model that drive the famous
Berezinskii-Kosterlitz-Thouless transition in two dimensions
\cite{berezinskii70,kosterlitz73}. However, in the frustrated case, they 
very likely   interact non trivially with the chirality degrees of freedom
which are critical in $d=3$ at the same temperature as the spin wave
degrees of freedom.  This is apparent from the fact that one
observes  a unique phase transition and not two distinct Ising-like and XY-like transitions
\cite{plakhty00}. As a consequence,  one can expect, in the  frustrated case,  a physics different
 and probably  more complicated than  in  the nonfrustrated $O(2)$ model that undergoes a 
standard second order phase transition in three dimensions.

\subsection{Generalization}

For reasons that will become clear,  we consider the
generalization of Hamiltonian (\ref{hmicroscopique}) to $N$-component
spins.  It is straightforward to extend the previous considerations to
this case.  One finds the symmetry breaking scheme:
\begin{equation}
G=O(N)\times  O(2) \to H=O(N-2)\times O(2)_{\hbox{\small diag}}\ .  
\end{equation}

In the following, we shall drop the ``diag'' index for simplicity.
Note that the previous  Heisenberg and XY  cases are recovered trivially provided
 that we identify  $O(0)$ with the trivial group $\openone$ and $O(1)$ with $\nbZ_2$.

We now give a review of the experimental and numerical results for
both the XY and Heisenberg systems. We will argue that a critical
analysis of these results is crucial to understand that, up to now,
the critical behavior of these systems is still unexplained.

\section{Experimental and numerical  situations}
\label{chapitre_experimental}
\subsection{Preliminaries}
\label{preliminaries}

 In this section, we  analyze the experimental and numerical
results relevant to the physics of frustrated magnets. Our aim is to
show that these data are hardly compatible with a second order phase
transition since, in particular, they  show that frustrated magnets display   scaling {\it without}
universality. Moreover, we show that  there are even some direct indications for weak
first order behaviors in these systems.  We recall that a
phase transition is said to be weakly of first order when, at the
transition, the  jump  of the order parameter is small and the
correlation length is large. Thus,  scaling behaviors can
be observed on a large range of temperatures so that these transitions
look like second order phase transitions except very close to the
critical temperature where scaling aborts. 

We emphasize that, by itself, the analysis of the experimental and
numerical results would not be sufficient to firmly conclude on the
first order nature of the transitions. It has to be seen as one of the
pieces of the argumentation that, together with a theoretical
analysis, will lead to a coherent picture of the critical physics of
frustrated systems.

To perform this analysis we need, in our discussion, to compare
experimental results among themselves, as well as with numerical and
theoretical calculations.  Let us explain how we extract average
values and error bars out of a set of experimental determinations of
critical exponents. In the experimental literature, only one error bar
is quoted, which merges the systematic and statistical errors. Our
first --- minimal --- hypothesis is that error bars have a purely
statistical origin (no systematic error).  Under this assumption, we
can trivially compute the (weighted) average values of the exponents
together with their error bars. This is the meaning of the numbers we
give in the following when we deal with average values of the critical
exponents. It is clear that this hypothesis is too simple to be
realistic since the experimental systematic errors cannot be
neglected. Thus, the values we compute, especially the error bars,
should be taken with caution. We however show in the course of this
article that our conclusions are  robust to a possible underestimation of the error bars
in our calculations, see Section \ref{chap_critical_remark}.

Let us also notice that a possible source of error in the estimation
of the critical exponents themselves could be the existence of
corrections to scaling that could bias all the results.  As we now
argue, we can however expect that these effects are not dramatic.  Let
us consider the well-documented case of the ferromagnetic Ising model
in $d=3$.  Most of the time corrections to scaling are not considered
in the determination of the critical exponents and the associated
error bars. When they are taken into account, they induce a tiny
change in the critical exponents, \ie at most of few percents (see for
instance \cite{defotis81} and \cite{blote95} for a review).  It is
therefore reasonable to think that neglecting corrections to scaling
induce an error of few percents on the critical exponents 
while this probably leads to largely underestimated error bars when
those are announced to be of the order of 1\% \footnote{Let us
emphasize that for other realizations of the Ising universality class,
such as demixion or liquid-vapor phase transitions, where the scaling
domains extend on three to four decades of reduced temperature, the
corrections to scaling are indeed necessary to fit the data on the
whole range of temperature.  The problem comes  from the
region of high reduced temperature.}.

In the case of frustrated magnets, if we make the assumption that the
corrections to scaling are comparable with those found in the
ferromagnetic Ising model  and bear in mind that the error bars
quoted in the literature are of the order of $5-10\%$ (see Tables
\ref{table_exp_crit_STA_XY_exp}, \ref{table_exp_crit_heli_XY_exp} and
\ref{table_exp_crit_Heis_exp}), we are led to the conclusion that
corrections to scaling are significant neither for the exponents nor
for the error bars.

\subsection{The XY systems}

Let us first discuss the XY case  since the experimental situation is
richer than in the Heisenberg case. Also, the symptoms of the
existence of a problem in the interpretation of the results are
clearer than in this latter case for reasons that shall be explained
in this article and particularly in Section \ref{chapitre_d=3}.

\subsubsection{The experimental situation}

Two classes of materials are supposed to be described by the Hamiltonian
(\ref{eq9}). The first one is made of ABX$_3$ hexagonal perovskites
--- where  A is an alkali metal, B a transition metal and X a halogen atom
--- which are physical realizations of XY STA.  The most studied ones
are CsMnBr$_3$, CsCuCl$_3$, CsNiCl$_3$ and CsMnI$_3$ (see
\cite{collins97} for a review and \cite{weber96} for RbMnBr$_3$. We have excluded this material since the measurement of its specific heat  presents a 
shoulderlike anomaly near $T_c$ which renders the determination of  $\alpha$ and $\beta$ doubtful). The second one is made of rare earth helimagnets:
Ho, Dy, Tb.  For most materials, the transitions are found continuous
but {\it not}  with the same critical exponents. For CsCuCl$_3$, the transition is
found to be weakly of first order, \ie with small discontinuities.
The results are summarized in Tables \ref{table_exp_crit_STA_XY_exp}
and \ref{table_exp_crit_heli_XY_exp}.
\begin{table}[htbp]
\centering
\begin{tabular}{|l|c|c|c|c|c|}
\hline
Compound&Ref.&$\alpha$&$\beta$&$\gamma$&$\nu$\\
\hline
\hline
CsMnBr$_3$&\cite{plakhty00}&&0.21(1)&&\\
&\cite{gaulin89}&&0.24(2)&&\\
&\cite{mason89}&&0.21(2)&1.01(8)&0.54(3)\\
&\cite{ajiro88}&&0.25(1)&&\\
&\cite{mason87}&&0.22(2)&&\\
&\cite{wang91}&0.39(9)&&&\\
&\cite{deutschmann92}&0.40(5)&&&\\
&\cite{deutschmann92}&0.44(5)&&&\\
&\cite{kadowaki88}&&&1.10(5)&0.57(3)\\
\hline
CsNiCl$_3$&\cite{weber95}&0.37(8)&&&\\
&\cite{weber95}&0.37(6)&&&\\
&\cite{enderle94}&0.342(5)&&&\\
&\cite{enderle97}&&0.243 (5)&&\\
\hline
CsMnI$_3$&\cite{weber95}&0.34(6)&&&\\
\hline
CsCuCl$_3$&\cite{schotte94}&&0.23-0.25(2)&&\\
&\cite{weber96}&0.35(5)&1$^{\hbox{st}}$ order&&\\
\hline
\end{tabular}
\caption{The critical exponents of the XY STA.}
\label{table_exp_crit_STA_XY_exp}
\end{table}
\begin{table}[ht]
\begin{tabular}{|c|c|c|c|c|c|}
\hline
Compound&Ref.&$\alpha$&$\beta$&$\gamma$&$\nu$\\
\hline
\hline
Tb&\cite{jayasuriya84}&0.20(3)&&&\\
&\cite{tang95}&&0.23(4)&&\\
&\cite{tang92}&&0.21(2)&&\\
&\cite{hirota91}&&&&0.53\\
\hline
Ho&\cite{tindall77}&1$^{\hbox{st}}$\hbox{order}& & & \\
&\cite{jayasuriya85b}&0.27(2)& & & \\
&\cite{wang91}&0.10-0.22& & & \\
&\cite{thurston94}&&0.30(10)&1.24(15)&0.54(4)\\
&\cite{thurston94}&&0.37(10)&&\\
&\cite{brits88}&&0.39(3)&&\\
&\cite{duplessis95}&&0.39(2)&&\\
&\cite{eckert76}&&0.39(4)&&\\
&\cite{helgesen94}&&0.39(4)&&\\
&\cite{helgesen94}&&0.41(4)&&\\
&\cite{gaulin88}&&&1.14(10)&0.57(4)\\
&\cite{plakhty01}&&0.38(1)&&\\
\hline
Dy&\cite{loh74}&&0.335(10)&&\\
&\cite{duplessis83}&&0.39$^{+0.04}_{-0.02}$&&\\
&\cite{duplessis95}&&0.38(2)&&\\
&\cite{brits88}&&0.39(1)&&\\
&\cite{gaulin88}&&&1.05(7)&0.57(5)\\
&\cite{jayasuriya85a}&0.24(2)&&&\\
\hline
\end{tabular}
\caption{The critical exponents of the XY helimagnets.}
\label{table_exp_crit_heli_XY_exp}
\end{table}

We highlight four striking characteristics \cite{tissier01} of these
data. Their consequences for the physics of frustrated magnets will be
discussed in more details in the following.

{\it i) There are two groups of incompatible exponents}. In the
following discussion, we mainly use the exponent $\beta$ to analyze
the results since, as seen in  Tables
\ref{table_exp_crit_STA_XY_exp} and \ref{table_exp_crit_heli_XY_exp},
it is by far the most precisely measured exponent. Clearly, there are
two groups of materials, each of which being characterized by a set of
exponents, $\beta$ in particular.

 In the first one --- that we call
group 1 --- made up of:
\be 
 \hbox{group 1}: \hbox{CsMnBr}_3, \hbox{CsNiCl}_3, \hbox{CsMnI}_3,  \hbox{Tb}
\label{materiauxgroup1}
\ee
one has:
\be 
\beta\sim 0.237(4)\ .
\label{betagroup1}
\ee
 Note that, as far as we know, there is no determination of the exponent $\beta$ for
 $\hbox{CsMnI}_3$ that, being given its composition,  has been   included in the
 group 1 of materials. Anyway, our conclusions are not affected by this fact.

In the second --- group 2  --- made up of
\be 
 \hbox{group 2}: \hbox{Ho}, \hbox{Dy}
\label{materiauxgroup2}
\ee
one has:
\be
 \beta\sim 0.389(7)\ .
\label{betagroup2}
\ee 
These exponents are clearly  incompatible.  Actually, we find for
the average exponents of CsMnBr$_3$ alone --- the most and best
studied material of group 1 ---:
\begin{equation}
\begin{split}
\beta=0.228(6),\nu=0.555(21),\alpha=0.416(33),\gamma=1.075(42)\ .
\label{expoCsMnBr3}
\end{split}
\end{equation}
If we consider all the materials of group 1 (except Tb for which the
results are not fully under control, but perhaps $\beta$) we find:
\begin{equation}
\begin{split}
&\beta=0.237(4),\nu=0.555(21),\alpha=0.344(5),\gamma=1.075(42)\ .
\label{expogroup1}
\end{split}
\end{equation}
For materials of group 2 (Ho and Dy) we find:
\begin{equation}
 \beta=0.389(7),\ \ \nu=0.558(25),\ \ \gamma=1.10(5)\ .
\label{expogroup2}
\end{equation}
We do not give a value for $\alpha$ which is poorly determined.  

Let us indicate that the exponents vary much from compound to compound
in group 1. Although less accurately determined than $\beta$, $\alpha$
is only marginally compatible between CsNiCl$_3$ and CsMnBr$_3$. Note
moreover that, even for the same material, the data are not fully
compatible among themselves: $\beta$ in CsMnBr$_3$ shows a somewhat
too large dispersion.

{\it ii) The anomalous dimension $\eta$ is negative for group 1 which is impossible.} If we 
assume that the transition is of second order for group 1, we can use
the scaling relations to compute $\eta$. In particular, the precise
determination of $\beta$ allows to use $\eta=2\beta/\nu-1$ to
determine rather accurately $\eta$. The exponent $\nu$ itself can be
obtained directly from the experiments or deduced using the scaling
relation:
\begin{equation}
 \nu=(2-\alpha)/3\ .
\label{nualpha} 
\end{equation}
The large number of experiments devoted to the determination of
$\alpha$ allows a precise determination of $\nu$. By using the
scaling relation Eq.~(\ref{nualpha}), we find $\nu=0.528(11)$ if we
consider the experimental results for CsMnBr$_3$ alone and
$\nu=0.552(2)$ if we consider CsMnBr$_3$, CsNiCl$_3$ and CsMnI$_3$.
 By  using the relation $\eta=2\beta/\nu-1$ together with Eq.~(\ref{nualpha}) or 
the relation $\eta=6\beta/(2-\alpha)-1$ and by considering the data of CsMnBr$_3$
alone  or the data of the materials of group 1 (except Tb for which it
is not sure that the data are reliable) we can obtain four
determinations of $\eta$. In the four cases, we find
$\eta$ negative by at least 4.1 standard deviations and the
probability to find it positive always less than $ 10^{-5}$. In fact,
the most precise determination is obtained by combining all the data
of group 1, Eq.~(\ref{expogroup1}), and by using
the relation $\eta=6\beta/(2-\alpha)-1$. In this case, we obtain $\eta=-0.141(14)$
and thus a (almost) vanishing probability to find it positive. Note also that, although $\beta$
 and $\nu$ are less accurately known
in Tb --- for which experiments are anyway delicate ---, $\eta$ is
also found negative.

{\it However},  we stress that $\eta$ {\it cannot} be negative in a true
second order phase transition. This is a general result, based on
first principles of field theory, that $\eta$ is always positive if
the theory describing the transition is a unitary GLW $\phi^4$-like
model \cite{zinn_eta_pos} as it is the case here (see Appendix
\ref{annexe_eta}).

{\it iii) For group 2, the scaling relation} $\gamma + 2 \beta -
3\nu=0$ {\it is violated}. From Eq.~(\ref{expogroup2}) it is possible
to check the scaling relations.  We find $\gamma + 2 \beta -
3\nu=0.202(92)$ and thus a violation by 2.2 standard deviations.

{\it iv) CsCuCl$_3$ undergoes a weak first order phase transition.} 
Until recently, CsCuCl$_3$ was believed to undergo a second order
phase transition with exponents compatible with those of group 1, see
Table \ref{table_exp_crit_STA_XY_exp}. It has been finally found to
display a weak first order phase transition \cite{weber96}.

\subsubsection{The numerical situation }

 Monte Carlo simulations   have been performed on 
five  different kinds  of  XY  systems. The first one is the STA itself
\cite{kawamura92,kawamura89,kawamura87,kawamura86,plumer94,boubcheur96,itakura03}.
The second model is the STAR (where R stands for rigidity) which
consists in a STA for which the local rigidity constraint ---
Eq.~(\ref{eq3}) --- has been imposed on each plaquette at all temperatures
\cite{loison98}. The third model is the Stiefel $V_{2,2}$ model whose Hamiltonian is
given by Eq.~(\ref{stiefel}) \cite{kunz93,loison98}.  This is a hard spin,
discretized version of  the NL$\sigma$  model relevant to
frustrated magnets.  Note that,  for this last model,  the triangular structure
is irrelevant since the interaction is  ferromagnetic;  a  cubic lattice can be
chosen. Also a soft spin, discretized version of the GLW model
has recently been studied by Itakura \cite{itakura03}  who also re-studied the
 STA model for large sizes. Finally, a  helimagnetic system defined on a body-centered-tetragonal (BCT)
lattice --- the ``BCT model'' ---  has  been investigated~\cite{diep89}. 

Here, we emphasize that the local rigidity constraint ---
Eq.~(\ref{eq3}) --- as well as the manipulations that lead to the STAR,
Stiefel  $V_{2,2}$, GLW and BCT  models only affect  the {\it massive} --- noncritical --- modes. Thus, all 
the STA, STAR, Stiefel $V_{2,2}$,  GLW and BCT
models have the same {\it critical} modes, the same symmetries and the
same order parameter. Therefore, one could expect a common critical
behavior for all these systems.

Let us comment the results of the simulations given in Table
\ref{table_exp_crit_XY_num}.  Note that, due to the its novel character, we shall 
 comment  the recent work of Itakura \cite{itakura03} separately.
\begin{table}[htbp]
\centering
\begin{tabular}{|l|l|l|l|l|l|l|}
\hline
System&Ref.&$\alpha$&$\beta$&$\gamma$&$\nu$&$\eta$\\
\hline
\hline
STA&\cite{kawamura86,kawamura92}&0.34(6)&0.253(10)&1.13(5)&0.54(2)&-0.09(8)\\
&\cite{plumer94}&0.46(10)&0.24(2)&1.03(4)&0.50(1)&-0.06(4)\\
&\cite{boubcheur96}&0.43(10)&&&0.48(2)&\\
\hline
STA&\cite{itakura03}&\multicolumn{5}{c|}{1$^{\hbox{st}}$ \hbox{order}}\\
\hline
STAR&\cite{loison98}&\multicolumn{5}{c|}{1$^{\hbox{st}}$ \hbox{order}}\\
\hline
V$_{2,2}$&\cite{loison98}&\multicolumn{5}{c|}{1$^{\hbox{st}}$ \hbox{order}}\\
\hline
BCT&\cite{diep89}&\multicolumn{5}{c|}{1$^{\hbox{st}}$ \hbox{order}}\\
\hline
GLW&\cite{itakura03}&\multicolumn{5}{c|}{1$^{\hbox{st}}$ \hbox{order}}\\
\hline
\end{tabular}
\caption{Monte Carlo critical exponents of  XY  systems. Note that the exponent $\eta$ 
is computed from $\gamma/\nu=2-\eta$.}
\label{table_exp_crit_XY_num}
\end{table}

{\it i) For STA,  scaling laws are found with exponents compatible with
those of group 1.} Let us however notice that, similarly to what
happens for the materials of group 1 there exists, in the numerical
simulations of STA, a rather large dispersion of the results.  For
instance, the two extreme values of $\nu$ differ by 2.1 standard
deviations.

Let us make two other remarks. First, the good agreement between the numerical
 results for STA and the experimental ones for materials of group 1 has been repeatedly
interpreted in the literature as a proof of the existence of a second
order transition and even as an evidence of the existence of the
chiral fixed point of the GLW model~\cite{kawamura88}. We
emphasize here that the fact that a  Monte Carlo simulation reproduces 
experimental results only means  that the Hamiltonian of the simulated system
is a good approximation of the microscopic  Hamiltonian describing the physics of
 real materials.   However, this  neither  explains nor proves 
anything else --- and  {\it certainly not}  the existence of a second order phase
 transition  --- since Monte Carlo simulations suffer from problems analogous to
those encountered in experiments: a weakly first order phase transition is very
difficult to identify and  to distinguish from a second order one.

Let us now come to our second remark. In a beautiful experiment, Plakhty \etal{}~\cite{plakhty00} have measured the so-called
chiral critical exponents $\beta_c$ and $\gamma_c$~\cite{kawamura92}
in CsMnBr$_3$. They have found values compatible with those found
numerically in STA~\cite{kawamura92}. Let us emphasize, again, that this
agreement simply means that the   parameters  characterizing the numerical simulations  are not too far from
 those associated to the experiments. By no means it implies --- or gives a new indication of the
existence of --- a second order transition. Let us notice that $\beta_c$
has also been measured in Ho~\cite{plakhty01}. The value found
completely disagrees with the result found in STA and in CsMnBr$_3$.

{\it ii) The anomalous dimension $\eta$ is negative for STA.} As shown
in \cite{loison98}, $\eta$ is found negative using the two scaling
relations $\eta=2\beta/\nu -1$ and $\eta=2-\gamma/\nu$ for the two
simulations where these calculations can be performed.

{\it iii) The simulations performed on STAR,  $V_{2,2}$  and BCT models give
 first order transitions.} Therefore, the  modifications in
the microscopic details  which change STA into STAR, $V_{2,2}$  and  BCT  affect
 drastically  the scaling behavior.

 {\it iv)  In a remarkable work, Itakura has recently performed  Monte Carlo and 
Monte Carlo  RG approaches  of the STA and its GLW model version that  has led to a
clear first  order behavior \cite{itakura03}}.  Itakura has performed standard Monte
Carlo
 simulations of the STA involving sizes up to  $126\times 144\times 126$ leading
to  clear first order transitions. In particular, for these  lattice sizes, 
the double-peak of the probability distribution of the energy at
 the transition is clearly  identified. Itakura has also used  an improved Monte
 Carlo RG simulation of the STA and its GLW model version. One advantage of this
approach compared with previous RG Monte Carlo studies is that it allows to  reach the
asymptotic critical behavior using systems of moderately large lattice sizes. Within this
approach, Itakura has   found several evidences for a first order  behavior with, notably,  a 
runaway behavior of the RG flow and the  absence of any nontrivial fixed point.

\subsubsection{Summary}

We now summarize the results of our analysis of both experiments and
numerical simulations for XY frustrated magnets.

1) Scaling laws are found in STA and helimagnetic materials on a
rather wide range of temperature. This is also the case within all ---
but an important one \cite{itakura03} --- numerical simulations of the STA.

2) There are two groups of systems that differ by their critical exponents. The
first one includes the group 1 of materials and the numerical STA
model. The second one corresponds to the group 2 of materials.  One also observes
variations of critical exponents inside a given group of exponents.

3) The anomalous dimension $\eta$ is negative for the materials of
group 1 and for the numerical STA model.  This is very significant
from the experimental results, less from the numerical ones.

4) For group 2, the scaling relations are violated by 2.2 standard
deviations.

5) CsCuCl$_3$ is found to undergo a weak first order transition.

6)  STAR, $V_{2,2}$ and  BCT  models  undergo strong first order transitions.

7) Recent Monte Carlo and Monte Carlo RG approaches  of STA and the soft spin
discretized version of the GLW model give  clear indications of
first order behaviors.
 
\subsubsection{Conclusion: five  possible scenarios} 
\label{scenarios}

Let us now propose five  possible scenarios to explain the
phenomenology of XY frustrated systems.

{\it Scenario I.} 

This scenario is --- together with the second one --- the most often
invoked: the critical behavior of frustrated magnets, when they
display scaling, is controlled by a {\it unique} fixed point of the RG
flow  which is associated to a {\it new universality class} \cite{garel76,yosefin85,kawamura85,kawamura86,kawamura87,kawamura88}. Although, from
point 1) above, XY frustrated magnets appear to display rather generic
scaling behaviors, the examination of the experimental and numerical
data provides clear indications against this first scenario. Indeed,
from point 2), there is a manifest lack of universality in the scaling
behavior of frustrated magnets. Also several points, from 3) to 7), strongly
militate in favor of first order behaviors.

{\it Scenario II.} 

In the second scenario, the two sets of exponents corresponding to
groups 1 and 2 are, in fact, associated with two true second order
phase transitions from which result two distinct universality
classes. This scenario is ruled out by the fact ---  see point 3) ---  that
the anomalous dimension $\eta$ is negative for group 1 and for the
numerical STA model. Thus, provided {\it i)} the quoted error bars in
the literature are reliable, {\it ii)} our hypothesis of a purely
statistical origin of the errors does not completely bias our analysis
and {\it iii)} corrections to scaling do not alter drastically all the
results, we are led to the conclusion that the behavior of the
materials of group 1 and of the numerical STA model {\it cannot} be
explained by the existence of a fixed point in the GLW model. In the
simplest hypothesis, these systems must undergo first order phase
transitions. This last hypothesis seems to be confirmed by several
other facts. Firstly, CsCuCl$_3$, whose exponents are close to those
of group 1 has been finally found to undergo first order phase
transitions, see point 5) of the summary. Secondly, point 6), numerical models very close
to STA --- STAR, $V_{2,2}$ and BCT --- also undergo first order phase
transitions.  Finally, the  hypothesis of a first order phase transition for  STA itself  is corroborated by
 the fact, point 7), that  recent Monte
Carlo  and Monte Carlo RG simulations of this system  predict a first order phase transition \cite{itakura03}.

{\it Scenario III.}

In the third scenario, materials of group 2 undergo a second order
phase transition --- $\eta$ is found positive there --- while those of
group 1 as well as the numerical STA model all undergo weakly first
order phase transitions. Within this scenario, the critical exponents
of materials of group 1 should be considered as {\it effective} or
{\it pseudo}-critical exponents, characterizing the {\it
pseudo}-scaling observed, valid for temperatures far enough from the
critical temperature. There is no direct and definitive argument
against this scenario.  Of course, violation of the scaling relations
for materials of group 2, point 4), makes doubtful a second order
behavior. However, this violation is too small to definitely reject
it. Actually, the drawback with this third scenario is its lack of
naturalness. Indeed, it implies a very specific fine-tuning of the
microscopical coupling constants --- \ie of the initial conditions of
the RG flow --- for materials of group 1.  Their representative points
in the coupling constant space must lie outside the basin of
attraction of the fixed point governing the critical behavior of
materials of group 2 but very close to its border so that the
transitions are weakly of first order.

{\it Scenario IV.}

In the fourth scenario, all frustrated magnets undergo first order
phase transitions that almost generically appear to be weak or very
weak and are characterized by pseudo-scaling and pseudo-critical
exponents. This fourth scenario, compared with the third one, could
thus seem even more unnatural. This is true, but only within the usual
explanation of weak first order phase transitions where the weakness
of the first order transition is obtained by fine-tuning of
parameters.  Actually, we shall provide arguments in favor of the present 
scenario and shall show that the genericity of pseudo-scaling
has, in fact, a natural explanation relying neither on the existence
of a fixed point nor on a fine-tuning of parameters.

{\it Scenario V.} 

Finally, one can imagine several variants of these scenarios. For instance,  we have 
adopted the standard  position that  consists in  associating  a unique set of critical exponents to a 
 fixed point. On the contrary, Calabrese \etal\  \cite{calabrese02,calabrese03b} have
suggested  that a unique fixed point could lead to a whole spectrum of critical
exponents. This scenario, which would  explain the occurence of a spreading of critical
exponents in the experimental and numerical contexts,  will be discussed
in details in the following.

We now review the experimental and numerical results obtained for the
Heisenberg systems.

\subsection{The Heisenberg systems}
\subsubsection{The experimental situation}
\label{chap_exp_heis}

Contrarily to the XY case, there is no Heisenberg helimagnets (see
however \cite{garel76}). Therefore there remain, {\it a priori}, only
the Heisenberg STA materials. In fact, the A/B phase transition of
He$_3$ can be described by the same GLW Hamiltonian as the Heisenberg
STA \cite{jones76,bailin77}. It is thus a candidate. Unfortunately,
the narrowness of the critical region of this transition does not
allow a reliable study of the critical behavior of this system and
there are no available data about it.

 Three classes of Heisenberg STA materials have been studied.  First,
systems like VCl$_2$, VBr$_2$,  Cu(HCOO)$_2$2CO(ND$_2$)$_2$2D$_2$O an
 Fe[S$_2$CN(C$_2$H$_5$)$_2$]$_2$Cl which are  generically  quasi-XY
except in  a particular range  of temperature where  their
anisotropies are irrelevant. Second, those which become
isotropic thanks to a  magnetic field  that  exactly counterbalances
the anisotropies. This is the case of CsNiCl$_3$ and CsMnI$_3$ at
their multicritical point. Finally, those which become isotropic
because they have been prepared in a fine-tuned st\oe chiometry such
that the Ising-like and XY-like anisotropies cancel each other to form
an isotropic material. This is the case of
CsMn(Br$_{0.19}$I$_{0.81}$)$_3$.

Let us comment the experimental results summarized in Table
\ref{table_exp_crit_Heis_exp}.
\begin{table}[htbp]
\centering
\begin{tabular}{|c|l|c|c|c|c|}
\hline
Compound  & Ref.            &  $\alpha$ & $\beta$  & $\gamma$ &  $\nu$ \\
\hline
\hline
 VCl$_2$   &\cite{kadowaki87}&           & 0.20(2)  & 1.05(3)  & 0.62(5) \\ 
\hline
 VBr$_2$   &\cite{wosnitza94}&  0.30(5)  &          &          &         \\
 \hline
 A         &\cite{koyama85}  &           & 0.22(2)  &          &         \\
 \hline
B&\cite{defotis78,defotis81,defotis86}&  & 0.24(1)  &  1.16(3) &         \\
 &\cite{defotis02}           &0.244(5)   &          &          &         \\
 \hline
CsNiCl$_3$&\cite{weber95,beckmann93}&0.25(8)&       &          &          \\
           &\cite{enderle94} &   0.23(4) &          &          &          \\
           &\cite{enderle97} &           & 0.28(3)  &          &          \\
 \hline
 CsMnI$_3$ &\cite{weber95}   &  0.28(6)  &          &          &          \\
 \hline 
C& \cite{bugel01}&0.23(7)&  &    &          \\
           &\cite{kakurai98}     &           & 0.29(1)  &[0.75(4)] &[0.42(3)] \\
           &\cite{ono99}        &           & 0.28(2)  &          &          \\
 \hline
\end{tabular}
\caption{The critical exponents of the Heisenberg STA. The abbreviations A, B and C stand
for Cu(HCOO)$_2$2CO(ND$_2$)$_2$2D$_2$O, Fe[S$_2$CN(C$_2$H$_5$)$_2$]$_2$Cl and CsMn(Br$_{0.19}$I$_{0.81}$)$_3$
respectively. The data in brackets are suspected to be incorrect. They
are given for completeness.}
\label{table_exp_crit_Heis_exp}
\end{table}

{\it i) As in the XY case, the Heisenberg materials fall into two
groups}.  The  group 1, made up of:
\begin{eqnarray}
\hbox{group 1}: \ &&\hbox{Cu(HCOO)}_2  2\hbox{CO(ND}_2)_2 2\hbox{D}_2\hbox{O},\nonumber\\
&&\hbox{Fe[S}_2 \hbox{CN(C}_2 \hbox{H}_5)_2]_2\hbox{Cl},\nonumber\\
&&\hbox{VCl}_2,  \hbox{VBr}_2
\label{materiauxgroup1Heisenberg}
\end{eqnarray}
is characterized by:
\be
\beta=0.230(8)
\ee
while for group 2,  made up of:
\be
\hbox{CsNiCl}_3,\hbox{CsMnI}_3, \hbox{CsMn(Br}_{0.19},\hbox{I}_{0.81})_3
\ee
one finds:
\be
\beta=0.287(8)\ .
\ee
Note that, strictly speaking, the values of $\beta$ for VBr$_2$ and for
CsMnI$_3$ are not known and, thus, our classification is somewhat
improper.  It seems however logical to suppose that VBr$_2$ is close
to VCl$_2$ and CsMnI$_3$ close to CsNiCl$_3$. Anyway, it will be clear
in the following that our analysis is almost insensitive to this
point.

For group 1, the  average values of the critical exponents are given by:
\begin{equation}
\begin{split}
\beta=0.230(8),\alpha=0.272(35),\nu=0.62(5),\gamma=1.105(21)\ .
\label{expoheisgroup1}
\end{split}
\end{equation}

A very severe difficulty in the study of the materials of group 1 is
their two-dimensional  character and Ising-like anisotropies.  The temperature
range where the systems  behave effectively as  three-dimensional  Heisenberg systems is narrow. This is
 the case of VCl$_2$ where this
range is less than two decades and where, closer to the critical
temperature, the system becomes Ising-like.  For this group of
materials the exponent $\beta$ is very small and the authors of
\cite{kadowaki87} have noticed that such small values have also been
found in materials where dimensional cross-over is suspected. Thus, it
is not clear whether the whole set of results really corresponds to a
three-dimensional Heisenberg STA.

For group 2, the experimental situation seems to be better under
control. The average values of the critical exponents are given by:
\begin{equation}
\begin{split}
\beta=0.287(9),\alpha=0.243(3),\nu=0.585(9),\gamma=1.181(33)
\label{expoheisgroup2}
\end{split}
\end{equation}
where the scaling relations have been used to compute $\nu$ and
$\gamma$. Note that the values of $\nu$ and $\gamma$ thus obtained
differ significantly from those of CsMn(Br$_{0.19}$I$_{0.81}$)$_3$
whose critical behavior has been claimed to be perturbed by disorder
(see however \footnote{The values of $\gamma$ and $\nu$ found in
CsMn(Br$_{0.19}$I$_{0.81}$)$_3$ are very small compared with all known
values in spin systems as well as what is expected from scaling
relations and from numerical simulations (see Table
\ref{table_exp_crit_heis_num}). The authors of \cite{kakurai98} argue
that the randomness of the magnetic anisotropy and of the exchange
interaction in the c-plane existing in their sample is responsible for
these small values.  Although it is very difficult up to now to have
definitive statements about disordered systems, it seems probable for
both randomly diluted systems and systems with random fields that
$\nu$ obeys the constraint $\nu>2/d$ where $d$ is the space dimension,
see \cite{chayes86,aharony98} and references therein. Thus, it is far from
 being straightforward that disorder is responsible for
the smallness of $\nu$.}).

{\it ii) For group 1, the anomalous dimension $\eta$ is significantly
negative.}  Using the two exponents that have been measured at least
twice in group 1 --- $\beta$ and $\gamma$ --- we can compute the
anomalous dimension from the scaling relation $\eta= (4\beta-\gamma)/(2\beta+\gamma)$.  We
find $\eta=-0.118(25)$ which   is thus negative by 4.8 standard
deviations.

{\it iii) For group 2, the anomalous dimension $\eta$ is marginally
negative.}  Using the critical exponents given in Eq.~(\ref{expoheisgroup2}),
one obtains, for the anomalous dimension: $\eta=-0.018(33)$. Thus $\eta$ is
found negative but not significantly, contrarily to what happens
in group 1.

{\it iv) For group 1, the scaling relations $\gamma + 2\beta - 2 +
\alpha=0= 2\beta+\gamma-3\nu$ are violated}. Indeed, $\gamma + 2\beta
- 2 + \alpha=-0.135(56)$ and $2\beta+\gamma-3\nu=-0.29(15)$.  Of
course, none of these violations is completely significant in itself
because of the lack of experimental data. However, since they are both
independently violated it remains only a very small probability that
the scaling relations are actually satisfied.

\subsubsection{The numerical situation}

 In the Heisenberg case, as in the XY case, five different kinds of systems:  STA, STAR,
 Stiefel ($V_{3,2}$ in this case), BCT  and GLW  models have been  studied.  The results of the
simulations are given in Table \ref{table_exp_crit_heis_num}.

\begin{table*}[htbp]
\centering
\begin{tabular}{|l|l|l|l|l|l|l|}
\hline
System&Ref.&$\alpha$&$\beta$&$\gamma$&$\nu$&$\eta$\\
\hline
\hline
STA&\cite{kawamura85,kawamura92}&0.240(80)&0.300(20)&1.170(70)&0.590(20)&\ 0.020(180)\\
&\cite{mailhot94}&0.242(24)&0.285(11)&1.185(3)&0.586(8)&-0.033(19)\\
&\cite{bhattacharya94}&0.245(27)&0.289(15)&1.176(26)&0.585(9)&-0.011(14)\\
&\cite{loison94}&0.230(30)&0.280(15)&&0.590(10)&\ 0.000(40)\\
&\cite{peles02}&&&&0.589(7)&\\
\hline
STA&\cite{itakura03}&\multicolumn{5}{c|}{1$^{\hbox{st}}$ \hbox{order}}\\
\hline
STAR&\cite{loison00b}&0.488(30)&0.221(9)&1.074(29)&0.504(10)&-0.131(13)\\
\hline
V$_{3,2}$&\cite{loison00b}&0.479(24)&0.193(4)&1.136(23)&0.507(8)&-0.240(10)\\
&\cite{kunz93}&0.460(30)&&1.100(100)&0.515(10)&-0.100(50)\\
\hline
V$_{3,2}$&\cite{itakura03}&\multicolumn{5}{c|}{1$^{\hbox{st}}$
  \hbox{order}}\\
\hline
BCT&\cite{loison99}&0.287(30)&0.247(10)&1.217(32)&0.571(10)&-0.131(18)\\
\hline
GLW&\cite{itakura03}&\multicolumn{5}{c|}{1$^{\hbox{st}}$ \hbox{order}}\\
\hline
\end{tabular}
\caption{Monte Carlo critical exponents of the Heisenberg systems.
$\eta$ is computed by $\gamma/\nu=2-\eta$ and, apart in
\cite{kawamura92} and \cite{kunz93}, $\alpha$ is computed by
$3\nu=2-\alpha$. }
\label{table_exp_crit_heis_num}
\end{table*}

Let us comment them. Again, we put aside the work of Itakura \cite{itakura03}. 

{\it i) For the STA, scaling laws are
found with an exponent $\beta$ close to that of group 2}. The average
values for the exponents of STA are:
\begin{equation}
\beta=0.288(6),\ \ \gamma=1.185(3),\ \ \nu=0.587(5) \ .
\end{equation}
$\beta$ is thus extremely close to the experimental value of group 2
while  $\nu$ and $\gamma$ are extremely close to the experimental values
deduced from the scaling relations, Eq.~(\ref{expoheisgroup2}).  The
scaling relation $\gamma+2 \beta-3\nu=0$ is very well verified since
$\gamma+2 \beta-3\nu=10^{-4}\pm 6. 10^{-2}$.

{\it ii) For the STA, $\eta$ is negative}. Using the values of
$\beta/\nu$ and $\gamma/\nu$ obtained directly in the simulations, one
can compute the average value of $\eta$: $ -0.0182(89)$. 
The probability of it to be positive is $0.02$ and
is thus small although  not vanishing. 

{\it iii) For the  STAR,  $V_{3,2}$ and BCT models, the values of
$\beta$ are all incompatible with that of STA (three standard
deviations at least) and are all incompatible among each others.}
This has been interpreted as an indication of very weak first order
phase transitions \cite{loison00b}. This is to be compared with the XY
case, where the transitions for STAR and the $V_{2,2}$ model are
strongly of first order.

{\it iv) For the BCT, STAR and $V_{3,2}$ models, $\eta$ is always
found significantly negative}, see Table
{\ref{table_exp_crit_heis_num} where $\eta$ has been calculated from
$\gamma/\nu$.

 {\it v) The  Monte Carlo and Monte Carlo RG approaches of the STA, $V_{3,2}$ and GLW model performed by
 Itakura has led to clear first order behaviors} \cite{itakura03}. For Heisenberg STA, contrarily to the XY
 case, even for the largest lattice sizes --- $84\times 96\times 84$ --- 
 the double-peak of the probability  distribution of the energy is not observed.  
However, the  $V_{3,2}$ model displays a clear double-peak. Moreover, for the STA and the $V_{3,2}$ model, 
 the RG flow clearly does not exhibit any fixed point.  Instead, a runaway of the RG
flow toward the region of instability is found which indicates first
order transitions.  The transitions are thus  ---  weakly --- of first order. The transition is also weaker
of first order for Heisenberg than for XY spins.

\subsubsection{Summary}

 We now summarize the experimental and numerical situations for frustrated magnets 
with  Heisenberg spins. Here,  the experimental situation is much
poorer than in the XY case and is still unclear on many aspects. On
the contrary, the numerical results are numerous and more precise than
in the XY case.

1) Scaling laws are  found in STA materials on a rather wide range of
temperatures as well as in  all Monte Carlo simulations  --- apart that
based on Monte Carlo RG.

2) There are two groups of materials that do not have the same
exponents. The exponent $\beta$ of the numerical STA model agrees very
well with that of group 2.

3) The anomalous dimension is manifestly negative for group 1 and 
marginally negative for group 2. For the numerical STA model, $\eta$ is found
negative although not completely significantly.  For STAR, $V_{3,2}$
and BCT, $\eta$ is found significantly negative.

4) For group 1, the scaling relations are violated.

5) STAR, $V_{3,2}$ and BCT exhibit scaling behaviors without
universality.  Also, the results are incompatible with that of the
numerical STA model.

6) A Monte Carlo RG approach of the STA, $V_{3,2}$ and GLW models has
led to clear first order behaviors.

\subsubsection{Conclusion}

Let us now draw some conclusions about the Heisenberg case. The
experimental and numerical data reveal  the
same problems as those encountered in the XY case: the different
materials split into two groups, the anomalous dimension is found
negative in many materials and in most numerical simulations, the
scaling relations are violated in some materials and there is no
universality in the exponents found in the simulations. The same kind
of conclusion as in the XY case follows (see Section \ref{scenarios}) : the first scenario, that of
an explanation based on the existence of a unique fixed point appears
unlikely. There are also signs of first order behaviors but less
significantly than in the XY case. Thus, at this stage, it is
impossible from the experimental and numerical data alone to
discriminate between the different scenarios II, III,  IV and V. It is
therefore important to gain insight from the theoretical side.

Before discussing this, let us mention another interesting numerical
result.

\subsection{The $N=6$ STA}

Let us quote a  simulation of the STA with six-component spins that 
has been performed by Loison \etal{}~\cite{loison00}. The results are given in Table
\ref{table_exp_crit_6_num}. Six-component spins were chosen since it
was expected that the transition was of second order.  Loison {\it et
al}. have  clearly identified scaling laws at the transition 
with a positive anomalous dimension. Let us emphasize that, even if the transition is actually of first order,
 as  suggested by the recent results of Calabrese \etal{}~\cite{calabrese03b}, it  should be extremely weakly
 of first order --- see the following. Thus, scaling laws
 should hold for all temperatures but those very  close to $T_c$. In this respect, 
 the exponents for $N=6$ are therefore  very trustable 
so that reproducing them is a challenge for the theoretical approaches. 
\begin{table}[htbp]
{\begin{tabular}{|l|l|l|l|l|}
\hline
System&$\alpha$&$\beta$&$\gamma$&$\nu$\\
\hline
\hline
 $N=6$ STA&-0.100(33)&0.359(14)&1.383(36)&0.700(11)\\
\hline
\end{tabular}}
\caption{Monte Carlo critical exponents for six-component spins in the 
STA system~\protect\cite{loison00}.}
Note that  using the results of  Loison \etal\  and the relation
$\eta=2-\gamma/\nu$,  one finds $\eta=0.025(20)$.

\label{table_exp_crit_6_num}
\end{table}

\section{A brief chronological survey of the theoretical approaches}
\label{chapitre_chronological}

Let us briefly review the most important theoretical developments
concerning this subject.

The first microscopic derivation and RG study --- at one- and
two-loop order  in $d=4-\epsilon$ --- of the  effective GLW model relevant for the STA --- see
below --- was performed for He$_3$ by Jones \etal{} in 1976
\cite{jones76} and by Bailin \etal{} in 1977 \cite{bailin77}.  The
model was re-derived and re-studied in the context of helimagnets (for
general $N$) by several groups including Bak \etal{} (1976)
\cite{bak76}, Garel and Pfeuty (1976) \cite{garel76} and Barak and
Walker (1982) \cite{barak82}.  It was established at that time that,
around $d=4$, the transitions for Heisenberg spin systems had to be of
first order.  More precisely, these authors found that there exists a
critical value $N_c(d)$ of the number $N$ of spin components above
which the transition is of second order and below which it is of first
order.  They found \cite{jones76,bailin77} : 
\be
\begin{aligned}
N_c(d=4-\epsilon)&=\displaystyle{4(3+\sqrt 6)-4\left(3+\frac{7}{\sqrt 6}\right)\,\epsilon+ O(\epsilon^2)}\\
&\sim 21.8 -23.4 \, \epsilon + O(\epsilon^2)
\label{eq24} 
\end{aligned}
\ee
with $\epsilon=4-d$.  A first large-$N$ expansion was also studied, in
particular for $d=3$, by Bailin \etal{} in 1977 \cite{bailin77}.

A group theoretical derivation of the GLW model relevant to the XY STA
was performed by Yosefin and Domany~\cite{yosefin85} in 1985.  They
found the same Hamiltonian as for helimagnets.  Between 1985 and 1988
Kawamura~\cite{kawamura85,kawamura86,kawamura87,kawamura88} have  performed this  analysis for $N$-component STA.
 He has shown that 
the Hamiltonian is the same as  for He$_3$ or
helimagnets, the RG analysis giving obviously  the same results.  This author has
also extrapolated the two-loop result for $N_c(d)$ of Eq.~(\ref{eq24})
in $d=3$ and found $N_c(d=3)<2$.  This led him to conjecture the
existence of a second order phase transition for frustrated magnets associated with
 a new universality class. However, as well-known, this direct extrapolation cannot  be reliable  
since it is notorious that the perturbative series must be resummed.

In 1988, Dombre and Read~\cite{dombre88} derived, in the quantum case, the Nonlinear
Sigma  (NL$\sigma$) model relevant  to frustrated magnets. In 1990, Azaria \etal{} studied 
the classical
version of this  NL$\sigma$ model around $d=2$. They found a fixed point
of the RG flow in a two-loop calculation for any $N\ge 3$
\cite{azaria90}.  For $N=3$, they found the phenomenon of enlarged
symmetry: at the fixed point the symmetry becomes $SO(3)\times
SO(3)\sim SO(4)$ instead of $SO(3)\times SO(2)$.  Thus, their
conclusion was that, if the transition is of second order, it is
characterized by $O(4)/O(3)$ critical exponents --- at least for
$\nu$. Another possibility proposed by these authors was that the
transition could be also mean-field tricritical or of first order.
However, {\it none} of the experimental or numerical results are
compatible with the $O(4)/O(3)$ or mean-field tricritical
exponents. Note finally that these authors supposed that, if
tricritical, the behavior at the transition should be mean-field
tricritical in $d=3$, something which is mandatory only for
$O(N)/O(N-1)$ models, but not for more complex models.

The first nonperturbative approach to frustrated magnets  was performed by
Zumbach in 1993 \cite{zumbach93,zumbach94,zumbach94c}.  He wrote down the
NPRG equations for the  GLW  models suited to the
description of these systems. He studied them within the Local
Potential Approximation (LPA) of the Wilson-Polchinski equation ---
analogous to the Wegner-Houghton approximation \cite{wegner73} --- and
found $N_c(d=3)\sim 4.7$.  Since he found no fixed point for $N=2$ and
$N=3$ he claimed that the transition is of first order in these cases.
In the case $N=3$, there is a minimum in the RG flow, a pseudo-fixed
point, that fakes a true fixed point (see below for details).  The
transition was thus conjectured to be {\it weakly} of first order with
pseudo-scaling characterized by pseudo-critical exponents.  Note that,
within the LPA, all derivative terms in the Hamiltonian are neglected
so that the anomalous dimension is vanishing. This has two important
consequences. First, the pseudo-critical exponents found by Zumbach
were not very reliable and thus difficult to compare with the
experimental and numerical results.  Second, this approach neglects
terms --- the so-called current-term (see below) --- that are
fundamental within the perturbative approach of the NL$\sigma$ model
performed around two dimensions. Thus, within Zumbach's approach, it was {\it not} possible to
match with these results.  Finally note
that, in the $N=2$ case, {\it no} minimum in the RG flow was found and,
thus, no pseudo-critical exponent was  obtained, in contradiction with the
scaling behaviors observed in the experimental and numerical contexts.

Then,  three-loop calculations have been  performed by Antonenko \etal{} in 1994
and 1995 on the GLW model.  In $d=3$ this has led, after Pad\'e-Borel
resummation, to $N_c(d=3)=3.91$ \cite{antonenko94}.  In 
$d=4-\epsilon$, they have determined the three-loop
contribution --- 7.1 $\epsilon^2$ --- to  $N_c(d=4-\epsilon)$, see
Eq.~(\ref{eq24}). This has led to $N_c(d=3)=3.39$
\cite{antonenko95}. These authors have mentioned that, contrarily to the
$O(N)$ models, their three-loop results  were not well
converged.

In 1996, Jolic\oe ur and David studied a generalization of the Stiefel
model that involves $N$ vectors with $N$ components
\cite{jolicoeur96}. They showed within a mean field approximation and
a one-loop calculation performed in $d=2+\epsilon$ that a first order
line should appear in a nontrivial dimension above two. It should
isolate the chiral fixed point in the metastability region in such a
way that this point should no longer play any role.  Above this
dimension, the transition should therefore be of first order.

Then, in 2000-2002, using the technique of the effective average
action, including derivative terms, the present authors performed a
nonperturbative study of frustrated magnets for any dimension between  two and
four  \cite{tissier00b,tissier00,tissier01}. They recovered {\it all}
known perturbative results at one-loop in two and four dimensions as
well as for $N\to \infty$. They determined $N_c(d)$ for all $d$ and
found $N_c(d=3)=5.1$. Accordingly, for $N=6$, they found a second
order phase transition. Their exponents were in very good agreement
with those found numerically. For $N=3$ \cite{tissier00}, they
recovered Zumbach's results  --- presence of a minimum in the RG flow
--- and improved his approach: they found pseudo-critical exponents in
good agreement with {\it some} experimental realizations of frustrated
magnets. {\it However}, regarding the spreading of the experimental and
numerical data, the recourse to a minimum, leading to a {\it unique}
set of pseudo-critical exponents, was clearly not the end of the story. During
the study of the $N=2$ case \cite{tissier01}, the present authors
realized that the property of pseudo-scaling and even more, generic
pseudo-scaling, does not rely on the concept of minimum of the flow.
Pseudo-scaling appears as a consequence of the existence of a {\it  whole
region} in the flow diagram in  which the flow is slow. This allowed
them to account for the {\it nonuniversal scaling} that occurs in XY
as well as in Heisenberg frustrated magnets. The present article
accounts for these last developments.

 In 2001, Pelissetto \etal{}~\cite{pelissetto01a} derived  the  six-loop series  for the  GLW model. They used
 sophisticated resummation methods  in order to find the
 fixed points and to  determine the critical exponents of the model.  For $N\gtrsim 7$, they found a fixed point
 of the same nature as  the one  obtained at large-$N$ and in the $4-\epsilon$ expansion. Thus, a second order
 phase transition is expected in this case. For  $5\lesssim  N \lesssim 7$,
they  considered that their resummed series were  not well converged, the number of 
fixed points depending  strongly of the number of loops considered. This  led them 
to interpret this result as an indication that $N_c(d=3)\sim 6$. Finally and surprisingly, for $N\lesssim 5$ and,
 in particular, for the physically relevant cases  $N=2$ and $N=3$, they  found stable fixed points. Thus,  a 
second order phase transition was also  predicted in
these cases. However, the critical exponents found were far from all experimental and numerical data (see the 
following). Moreover, regarding again the spreading of these data, an interpretation in terms of a unique set of
 exponents was clearly insufficient.

In another work~\cite{pelissetto01b}, assuming that $N_c(d=2)=2$, Pelissetto \etal{}
have reformulated the three-loop version of the series of
Eq.~(\ref{eq24}) --- see below --- to make it compatible with this last
guess.  The series seemed  to have better convergence properties --- see however below --- and
allowed Pelissetto \etal{} to compute  $N_c(d)$. They found  $N_c(d=3)=5.3$, in
good agreement with the value --- $N_c(d=3)=5.1$ --- obtained from the NPRG
approach~\cite{tissier00,delamotte03}. 

Recent re-investigations  of the five and six-loop perturbative
 series~\cite{calabrese02,calabrese03b}  have led Calabrese \etal{} (see also \cite{parruccini03})   to 
conjecture that the fixed point  found  by Pelissetto \etal\ --- that  corresponds  to a
 {\it focus} fixed point  ---  could explain the existence of  the spreading of critical
 exponents encountered in frustrated magnets. Indeed, they observed that, due to the 
specific structure of the fixed point, the critical exponents display strong variations
  along the RG  trajectories that could explain the lack of universality observed 
experimentally  and numerically. They have also given  estimates of the critical
 number of spin components for which there is a change of the order of the phase transition.
 They have found that there is a first order phase transition in the whole domain
 $5.7(3)<N<6.4(4)$ and a second order phase transition for the other values of $N$ and,
 in particular, for $N=2$ and $N=3$.}

 Finally, a very recent computation of the five-loop $\beta$ function of the GLW model 
in a $4-\epsilon$ expansion has lead to a novel estimation of  $N_c(d)$. Calabrese and
Parruccini~\cite{calabrese03c} have found  the  value $N_c(d=3)=6.1(6)$ which is compatible
with the value $N_c(d=3)=6.4(4)$  found within the six-loop computation performed in three
dimensions~\cite{calabrese02,calabrese03b}.

Since several aspects  of the recent perturbative and nonperturbative
approaches  differ,  in particular in their interpretations of the origin of the
nonuniversal  scaling found in frustrated magnets,   we postpone the detailed discussion
of these  last developments of both methods  to the following sections.

\section{The perturbative  situation}
\label{chapitre_perturbatif}

Let us discuss in more details the perturbative approaches that have been used to investigate 
frustrated magnets. 

There are essentially two different methods to analyze the critical
behavior of the system described by the Hamiltonian (\ref{eq9}). They
correspond to two different methods to deal with the constraints  obeyed
by the microscopic degrees of freedom, Eq.~(\ref{eq4}). They lead to
the NL$\sigma$ and GLW models that have been both perturbatively
analyzed around their respective critical dimension  as well as, for the GLW model,
directly in three dimensions.  Let us review the results of these approaches.

\subsection{The Nonlinear Sigma (NL$\sigma$)  model approach}
\label{chapitre_NLS_onXo2}

The idea underlying the construction of this model is to consider the
system in its low-temperature --- symmetry broken --- phase and to
take into account small fluctuations of the fields around the
direction of  the order parameter.
The corresponding treatment is thus, by construction, a
low-temperature expansion. Its actual validity is in fact less
stringent than that: it is sufficient that the system
is {\it locally} ordered and that the temperature is small. This explains
why this approach is valid even in two dimensions for systems obeying
the Mermin-Wagner theorem. Note that this approach applies ---   a priori
(see  Section \ref{Heisenbergcase} and  the discussion at the end of this section)  --- only for
$N\ge 3$. Indeed, in the $N=2$ case, the low-temperature expansion of
the NL$\sigma$ model leads to a trivial result, \ie the theory is
perturbatively free.  This result is however not reliable since there
exist topological as well as Ising-like degrees of freedom in the XY
frustrated case (see Section {\ref{chapitre_model}}). These degrees of
freedom, that are completely missed within the low-temperature
perturbative approach, drastically affect the physics at finite
temperature as in the famous Berezinskii-Kosterlitz-Thouless phase
transition \cite{berezinskii70,kosterlitz73}.

Within the NL$\sigma$ model approach, the partition function of the
$SO(3)\times SO(2)$-symmetric model follows from the Hamiltonian
Eq.~(\ref{continu}) together with the constraints of Eq.~(\ref{eq4})
\cite{dombre88}:
\begin{equation}
\begin{split}
\mathcal Z=&\int \mathcal D\vec{\phi}_1 \mathcal D\vec{\phi} _2\,\,
\prod_{i\le j}\,\delta(\vec{\phi}_i .  \vec{\phi}_j -\delta_{ij})\,\ .\\
&\ \ \ . \exp\left(\displaystyle -{\frac 1 {2T}} \int d^d\cg x \left(
(\partial \vec{\phi}_1)^2 +(\partial\vec{\phi}_2)^2\right)\right)\ .
\label{eq14}
\end{split}
\end{equation}
The delta-functionals allow the integration of the three massive modes
among the six degrees of freedom of $\vec{\phi}_1$ and $\vec{\phi} _2$. Therefore, 
 only the three ---  Goldstone  --- modes $\vec\pi$ remain, in terms of which 
the partition function  writes~\cite{friedan85,azaria90,azaria93}: 
\begin{equation}
Z=\int_{\vert \vec\pi\vert\le 1} D{\vec \pi} \
\exp \left( -{{1\over 2T}\int d^d\cg x \ g_{ij}(\pi)
    \partial\pi^i\partial\pi^j} \right)\ .
\label{nonlinearGoldstone}
\end{equation}
The Eq.~(\ref{nonlinearGoldstone}), where $g_{ij}(\pi)$ embodies the interaction, is the suitable expression for
 a low-temperature expansion of the $SO(3)\times SO(2)/SO(2)$ NL$\sigma$ model.

The low-temperature expansion of such NL$\sigma$ models has been
studied in general but rather abstract terms by Friedan
\cite{friedan85}. The specific study of the $SO(3)\times SO(2)/SO(2)$
model and its generalization to $N$-component spins --- the
$O(N)\times O(2)/(O(N-2)\times O(2))$ model --- has been performed by
Azaria {\it et al}. \cite{azaria90,azaria93} (see also
\cite{pelissetto01b}). The RG analysis requires to consider the most
general Hamiltonian invariant under $O(N)\times O(2)$ and
renormalizable around $d=2$.  This Hamiltonian involves not only the
usual kinetic terms for $\vec{\phi}_1$ and $\vec{\phi}_2$,
Eq.~(\ref{eq14}), but also a nontrivial derivative term, called the
``current-term'', which reads:
\begin{equation}
\int d^d\cg x\left( \vec{\phi}_1\; .\;\partial\vec{\phi}_2 - \vec{\phi}_2\; .\;
\partial\vec{\phi}_1 \right)^2.
\label{eq15}
\end{equation}
This term must be included in the model since it has the right
symmetry, is power-counting renormalizable around $d=2$ and is thus
generated during the RG flow. The correct NL$\sigma$ model  --- in the
sense of stability under RG transformations --- is given by (for any
$N\ge 3$) \cite{azaria90}:
\begin{equation}
\begin{split}
H=\int d^d\cg x& \,\Bigg( \frac{\eta_1}{ 2}\left(
 \left(\partial\vec{\phi}_1 \right)^2+ \left(\partial\vec{\phi}_2
 \right)^2 \right) +\\ &+ \left( \frac{\eta_2}{ 8}- \frac{\eta_1}{
 4}\right) \left(\vec{\phi}_1\;.\;\partial\vec{\phi}_2-\vec{\phi}_2\;.
 \;\partial\vec{\phi}_1 \right)^2 \Bigg)
\end{split}
\label{hamiltonien_nlsigma} 
\end{equation}
where we have chosen to reparametrize the coupling constants in a way
convenient for what follows.  Now, the Hamiltonian of the na\"{\i}ve
continuum limit Eq.~(\ref{eq14}) is just the initial condition of the
RG flow corresponding to $\eta_1=\eta_2/2=1/T$.  Note that we have 
included  the temperature in the coupling constants.

For the special case $N=3$, it is convenient to rewrite the model
differently. We define, as in Eq.~(\ref{thirdvector}), a third vector $\vec{\phi}_3$ by:
\begin{equation}
 \vec{\phi}_3=\vec{\phi}_1\wedge
\vec{\phi}_2  \ .
\end{equation}
With this expression, it is easy to verify that the current-term,
Eq.~(\ref{eq15}), is nothing but a linear combination of the kinetic
terms of $\vec{\phi}_1$, $\vec{\phi}_2$ and $\vec{\phi}_3$:
\begin{equation}
\begin{split}
\int d^d\cg x&\left(\vec{\phi}_1.\partial
\vec{\phi}_2-\vec{\phi}_2.\partial \vec{\phi}_1 \right)^2=\\& 2\int
d^d\cg x\left(\left(\partial \vec{\phi}_1\right)^2 + \left(\partial
\vec{\phi}_2\right)^2 - \left(\partial\vec{\phi}_3\right)^2 \right)\ .
\end{split}
\end{equation}
One can then gather the three vectors $\vec{\phi}_1$, $\vec{\phi}_2$
and $\vec{\phi}_3$ into a $3\times 3$ matrix:
\begin{equation}
\Phi=\left(\vec{\phi}_1,\vec{\phi}_2, \vec{\phi}_3\right).  
\end{equation}
Since $(\vec{\phi}_1,\vec{\phi}_2, \vec{\phi}_3)$ are three
orthonormal vectors, one has $^t\Phi\,\Phi=\nbOne$ and $\Phi$ is
therefore a $SO(3)$ matrix.

The partition function thus reads:
\begin{equation}
\mathcal Z=\int \mathcal D\Phi\  \delta(^t\Phi\,\Phi-\nbOne)
\,e^{\displaystyle -\int d^d\cg x\, \hbox{Tr}\left({\cal P}\ \partial\,^t\Phi\,\partial \Phi\right)}
\label{eq19}
\end{equation}
where ${\cal P}$ is a diagonal matrix of coupling constants: ${\cal
P}=\hbox{diag}(p_1=p_2=\eta_2/4,p_3=\eta_1/2-\eta_2/4)$. 

 It is easy to check  on Eq.~(\ref{eq19}) that the model is invariant under
the right transformation:
\begin{equation}
\Phi\to \Phi.V  
\end{equation}
with $V$ being the subset of $SO(3)$ matrices that commute with ${\cal
P}$.  When $p_3\ne p_1$, \ie $\eta_1\ne \eta_2$, $V$  is
isomorphic to $SO(2)$. When  ${\cal P}$ is proportional to the
identity, $V$  is isomorphic to  the whole $SO(3)$ group.  In this last case, the
high-temperature symmetry group is $G=SO(3)\times SO(3)\sim
SO(4)$. Note that this identity has to be understood at the level of
the Lie algebras since  $SO(3)\times SO(3)$ and $SO(4)$  are locally isomorphic but
differ globally and have different topological properties. This fact
will be important in the following.

The RG equations for the $O(N)\times O(2)/(O(N-2)\times O(2))$ model
have been computed at two-loop order in $d=2+\epsilon$
\cite{azaria90,azaria93}. We recall here the one-loop result that will
be useful in the following:
\begin{equation}
 \left\{
\begin{aligned} 
\beta_{\eta_1}&=-(d-2) \eta_1 +N-2-\frac{\eta_2}{2\eta_1}\\
\beta_{\eta_2}&=-(d-2) \eta_2
+\frac{N-2}{2}\left(\frac{\eta_2}{\eta_1}\right)^2Ê\ .
\end{aligned}
\right.
\label{recursionnls}
\end{equation}
A fixed point is found for any $N\ge3$. For $N=3$, it corresponds to
$p_1^\star=p_3^\star$, \ie $\eta_1^\star=\eta_2^\star$ and, thus, to an
enlarged symmetry $SO(3)\times SO(3)/SO(3)\sim SO(4)/SO(3)$. This
fixed point has only one direction of instability --- the direction of
the temperature --- and thus corresponds to a second order phase
transition. Surprisingly, the critical behavior is thus predicted to
be governed by the usual ferromagnetic Wilson-Fisher fixed point with
the subtlety that it corresponds to {\it four-component}
spins.  Note that this precisely corresponds to the particular case  considered in Section 
\ref{Heisenbergcase}.  Another subtlety is that since, here, the order parameter is
 a matrix instead of a vector --- it is a $SO(4)$ tensor --- the
anomalous dimension is different from the usual anomalous dimension of
the four-component vector model. Only the
exponent $\nu$ is independent of the nature of the order parameter and
is thus identical to the usual value of $\nu$ of the Wilson-Fisher
$N=4$ universality class \cite{azaria90,azaria93}.

In fact, it is easy to convince oneself that the fixed point found exists
to  all order of perturbation theory. Actually, the crucial fact is
that, in $d=2+\epsilon$, the {\it perturbative} $\beta$ functions of a
NL$\sigma$ model associated with the symmetry breaking scheme $G\to H$
only depend on the {\it local geometrical} structure of the manifold
$G/H$ which is itself determined by the Lie algebras of $G$ and $H$
\cite{azaria90,azaria93}.  Since the Lie algebras of $SO(3)\times
SO(3)$ and of $SO(4)$ are identical, the perturbative $\beta$ function
for the --- remaining --- coupling constant of the model with
$p_1=p_3$ is identical at all orders to the perturbative $\beta$
function of the usual $SO(4)/SO(3)$ NL$\sigma$ model.  The existence
of a fixed point for the $SO(3)\times SO(3)/SO(3)$ NL$\sigma$ model at
all order of perturbation theory follows from the fact that its
existence makes no doubt for the $SO(4)/SO(3)$ NL$\sigma$ model. 

 At the  time of the first investigation of the $O(N)\times O(2)/(O(N-2)\times O(2))$   NL$\sigma$
model,  the most natural position was to extend this equivalence
{\it beyond} perturbation theory and to assume that the $SO(3)\times
SO(3)/SO(3)$ fixed point exists everywhere between two and four
dimensions, as it is the case for the $SO(4)/SO(3)$ fixed point. This
was, in particular, the position advocated by Azaria \etal{}
\cite{azaria90,azaria93}. The outstanding fact is that although the $SO(4)$ behavior has indeed
been seen numerically in $d=2$ \cite{southern93,caffarel01}, it
actually does not exist far from two dimensions. This is clear since
{\it no} such fixed point is found in $d=4-\epsilon$ and since, as
already emphasized, the $SO(4)$ behavior is not seen in any numerical
or experimental data in $d=3$. It is thus extremely probable that
either the fixed point disappears in a nontrivial dimension smaller
than 3 or it survives in $d=3$ while being no longer the usual $N=4$
fixed point.  Note that, in the first case, its $SO(4)$ nature can also
change before it disappears. Anyway, this fixed point must disappear
below $d=4$.  The situation is thus more involved than in the
``usual'' $SO(4)/SO(3)$ model. There must exist nonperturbative
reasons explaining the disappearance of the fixed point and/or the
loss of its $SO(4)$ character.

Actually, it is clear that the perturbative low-temperature expansion
performed on the NL$\sigma$ model misses several nonanalytic terms in $T$ --- typically terms that
behave as exp$(-1/T)$ --- that  could be responsible for the disappearance of the fixed point
and/or  its change of nature.  There are, at least, two origins for such terms.

1)  The first one consists in the nontrivial topological configurations
---  see the discussion in Section \ref{Heisenbergcase} following  Eq.~(\ref{homotopie}) --- that are
completely neglected in the low-temperature expansion of the
NL$\sigma$ models. Indeed this expansion relies, by construction, on
the local geometrical properties of the manifold ${G/H}$ and is
insensitive to its global --- topological --- structure. Thus it
ignores vortex-like configurations that likely  play an important role
in three dimensions.

2)  The second origin of nonanalytic corrections to the
low-temperature $\beta$ function is more technical.   The
low-temperature expansion is performed in terms of the Goldstone ---
or pseudo-Goldstone in $d=2$ --- modes that are represented by fields
constrained to have a modulus less than one, see Eq.~(\ref{nonlinearGoldstone}).  This
 inequality cannot be taken into account in the
perturbative treatment~\cite{zinnjustin89} and is thus relaxed,
leading to  neglect terms of order exp$(-1/T)$.  All these terms
are negligible for the critical behavior when the critical temperature
is very small, which is the case near $d=2$. However, they become
important when $T_c\sim 1$ which is typically the case in $d=3$.

Only a nonperturbative treatment can take into account these nonanalytic
 terms and thus allows  to follow, when the dimension is
increased, the fate of the $O(4)$ fixed point.

\subsection{The Ginzburg-Landau-Wilson (GLW) model approach}

\label{chapitre_GLW}

The GLW model for the $O(N)\times O(2)/(O(N-2)\times O(2))$ model can be
deduced from a generalization of Eq.~(\ref{eq14}) to $N$-component
vectors, by replacing the functional delta-constraint by the most
general potential that favors the field configurations obeying the
initial constraint. For convenience, we choose to parametrize it by:
\begin{equation}
\prod_{i\leq j}\,\delta(\vec\phi_i.\vec\phi_j -\delta_{ij})\ \to\ e^{-U} 
\end{equation}
with:
\begin{equation}
\begin{split}
U=\int d^d\cg x\,\bigg(&
\frac{r}{2}\left(\vec{\phi}_1^{\,2}+\vec{\phi}_2^{\,2}\right)+
\frac{\lambda+\mu}{16}\left(\vec{\phi}_1^{\,2}+\vec{\phi}_2^{\,2}\right)^2 -\\&-
\frac{\mu}{4}\left(\vec{\phi}_1^{\,2}\vec{\phi}_2^{\,2}-(\vec{\phi}_1.\vec{\phi}_2)^{\,2}
\right) \bigg)\ 
\label{hamglw}
\end{split}
\end{equation}
 where, as usual,  $r$ is  proportional to the reduced temperature  while  $\lambda$ and $\mu$ are $\phi^4$-like
 coupling constants.

All field-dependent  terms in Eq.~(\ref{hamglw}) can be rewritten in terms
of the rectangular matrix $\Phi$ defined in
Eq.~(\ref{matriceparametreordre}).  The corresponding Hamiltonian then reads:
\begin{equation}
H=\int\, d^d\cg x
\left(\frac{1}{2} \hbox{Tr}\, \left(\partial\, ^t\Phi \partial\Phi\right)+
\frac{r}{2} \rho +\frac{\lambda}{16}\rho^2+
\frac{\mu}{4}\tau \right) 
\label{hamilton}
\end{equation}
with $\rho= \hbox{Tr}(\,^t\Phi \Phi)$ and $\tau=
\frac12\hbox{Tr}\left( ^t\Phi \Phi -\nbOne\  \rho/2\right)^2$ being  the only  $
O(N)\times O(2)$ independent invariants that can be built out the
fields, see Appendix \ref{annexe_invariants}. Note that minimizing the term in front of $\mu$ corresponds to imposing 
 $^t\Phi \Phi\propto \nbOne$, \ie to imposing  that $\vec{\phi}_1$ and $\vec{\phi}_2$ are orthogonal and of the same
 norm in agreement with the characteristics of the ground state of frustrated magnets, see Fig.~\ref{triangulaire}b.

\subsubsection{The RG flow}

The RG equations for the coupling constants entering in Hamiltonian
(\ref{hamilton}) have been computed in the $\epsilon=4-d$-expansion  up to
five-loop order \cite{calabrese03c} and  in a weak-coupling expansion in $d=3$ up to six-loop order
\cite{pelissetto01a}.   We recall here only the
one-loop result of  the $\epsilon$-expansion  to discuss qualitatively the flow
diagram:
\begin{equation}
\left\{
\begin{aligned}
\beta_\lambda &= -\epsilon \lambda +\frac{1}{16\pi^2}\left(4
\lambda\mu +4 \mu^2 +\lambda^2(N+4) \right)\\ \beta_\mu &= -\epsilon
\mu+\frac{1}{16\pi^2}\left(6 \lambda\mu + N \mu^2 \right).
\end{aligned}
\right.
\label{recursionglw}
\end{equation}

\begin{figure}[tbp]
\centering \makebox[\linewidth]{ \subfigure[$N>N_c(d)$]{
\label{point_fixe_n_grand}
\includegraphics[width=.45\linewidth,origin=tl]{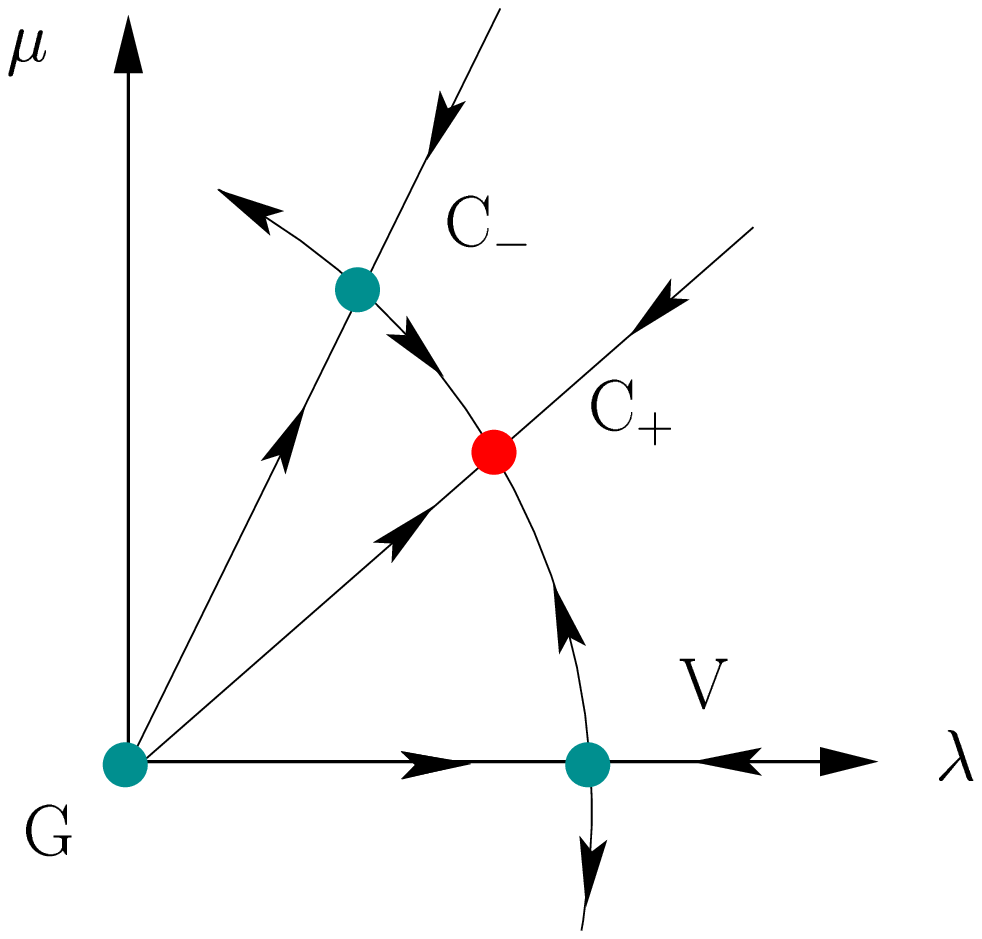}}\hfill%
\subfigure[$N_c'\le N\le N_c(d)$]{
\label{point_fixe_n_moyen}
\includegraphics[width=.45\linewidth,origin=tl]{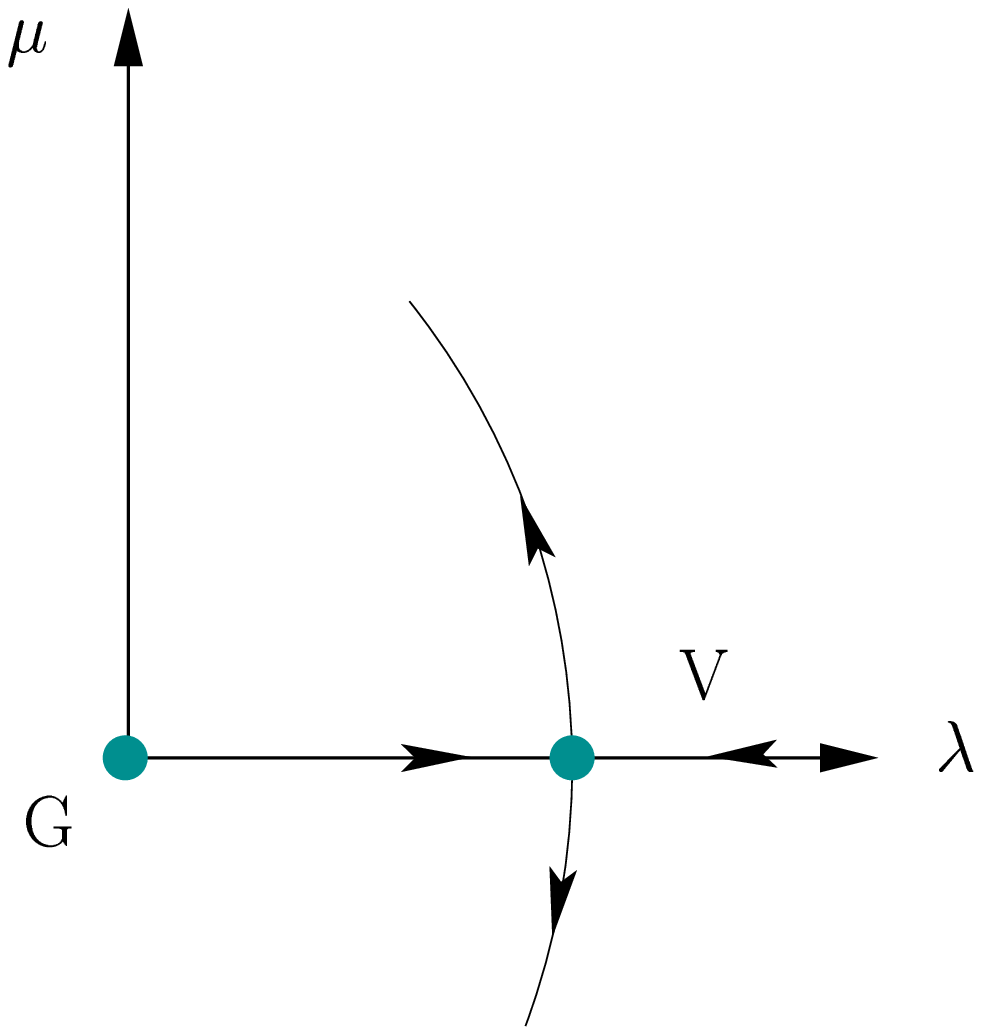}}\hfill%
}
\caption{Flow diagram for a) $N$ above $N_c(d)$ and b) $N$ below
$N_c(d)$. The fixed points $C_+$ and $C_-$ that exist above $N_c(d)$
coalesce at $N_c(d)$ and then disappear.  $G$ and $V$ are the Gaussian
and vector $O(2N)$ fixed points.  }
\label{ptfixe}
\end{figure}

 As well known, for any $N>N_c(d=4-\epsilon)=21.8+O(\epsilon)$  there exist four fixed points: 
the Gaussian --- $G$ --- the vector
$O(2N)$ ---   $V$ --- and two others, called the chiral ---  $C_+$ ---
and anti-chiral --- $C_-$ --- fixed points. Among these  fixed
points one, $C_+$, is stable and governs the critical properties of
the system and the others are unstable (see
Fig.~\ref{point_fixe_n_grand}). When, at a given dimension $d$ close to
four, $N$ is decreased, $C_+$ and $C_-$ move closer together, coalesce
at $N_c(d)$ and then disappear (see Fig.~\ref{point_fixe_n_moyen}). More precisely, for $N<N_c(d)$, the roots
of the $\beta$ functions acquire an imaginary part. Since no stable
fixed point exists below $N_c(d)$ and since the flow drives the system in
a region of instability, it is believed that the transition is of
first order. Note that for $N<N_c'(d=4-\epsilon)=2.2+O(\epsilon)$,  $C_+$ and $C_-$ reappear but not in
the  physically  relevant region to frustrated magnets. 

For completeness we give the exponent $\nu$ at one-loop:
\begin{equation}
\begin{split}
\label{nu_on_o2}
\nu=\displaystyle{1\over 2}+\epsilon &\left({(N-3)(N+4)\sqrt{48-24N+N^2}\over
  8(144-24N+4N^2+N^3)}+\right.\\
+&\left.\displaystyle{N(48+N+N^2)\over 8(144-24N+4N^2+N^3)}\right)\ 
\end{split}
\end{equation}
 and recall that the anomalous dimension vanishes at this order.
Note that the square root becomes complex for $2.2< N<21.8$, which is
reminiscent of the critical values $N_c(d)$ and $N_c'(d)$ of the number of
spin components, see above.

\subsubsection{The three and five-loop results in $d=4-\epsilon$}

 In $4-\epsilon$ dimensions, the critical value $N_c(d)$  has been computed at
three-loop order \cite{antonenko95} and, very recently, at five-loop order \cite{calabrese03c}:
\begin{equation}
\begin{array}{ll}
 N_c(d=4-\epsilon)= 21.80 & -  23.43 \epsilon +  7.09  \epsilon^2 \ - \\ & - 0.03 \epsilon^3 + 4.26\epsilon^4 + O(\epsilon^5) \ .
\label{ncritiquecinqboucles} 
\end{array}
\end{equation}
In fact, as it is often the case within this kind of expansion, the
series are not well behaved and it is  difficult
to obtain reliable results  even after resummation \cite{antonenko95,loison00,calabrese03c}.  We however
indicate the value found at three-loop order \cite{antonenko95}: $N_c(d=3)=3.39$ and at
five-loop order \cite{calabrese03c}: $N_c(d=3)=5.45$.

\subsubsection{The improved three and five-loop results}
\label{three-loopresults}

 It  has been conjectured by Pelissetto \etal{}~\cite{pelissetto01b} that $N_c(d=2)=2$,
a result which is however somewhat controversial \cite{calabrese03c}. It is possible to use
this nonperturbative information to reformulate the series obtained within the $4-\epsilon$
expansion.  Imposing the constraint $N_c(d=2)=2$ to the three-loop series, Pelissetto
\etal\  have obtained \cite{pelissetto01b}:
\begin{equation}
N_c(d=4-\epsilon)= 2 + (2-\epsilon)(9.90 - 6.77 \epsilon + 0.16 \epsilon^2 ) +
O(\epsilon^3)\ .
\label{nc(d)troisboucles}  
\end{equation}
Reformulated in this way, the coefficients of the
series decrease rapidly. It is thus reasonable to use this expression
to estimate $N_c(d=3)$.  Pelissetto \etal \ have thus obtained \cite{pelissetto01b}:
$N_c(d=3)=5.3(2)$ where the error bar indicates how $N_c(d=3)$ varies
from two to three loops.  However, Calabrese and Parruccini have
shown that, extended to five loops, the same series behaves
badly \cite{calabrese03c}:
\begin{equation}
\begin{array}{ll}
N_c(d=4-\epsilon)= 2 + &(2-\epsilon)(9.90 - 6.77 \epsilon + 0.16 \epsilon^2 + \\
& +\  0.06\epsilon^3 + 2.16 \epsilon^4) + O(\epsilon^5)\ .
\label{nc(d)cinqboucles}  
\end{array}
\end{equation}
Using  different kinds of tricks, notably the inverse of the series, they have obtained, from the
five-loop series, the value  $N_c(d=3)=6.1(6)$.

From this approach one is strongly tempted to conclude that, in the
physical cases $N=2$ and $N=3$, the transitions are of first order,
even if it is impossible to conclude about the strong or weak
character of this transition.

\subsubsection{The three-loop  results in $d=3$}

A weak-coupling analysis has been performed directly in $d=3$ at
three-loop order in \cite{antonenko94}. This leads to
$N_c(d=3)=3.91$. However, as already emphasized, this result is not
well converged.

\subsubsection{The large-$N$ results}

The large-$N$ expansion was first performed by Bailin \etal\ \cite{bailin77}. It was
then re-examined by Kawamura \cite{kawamura88} and Pelissetto \etal{}
\cite{pelissetto01b}. A fixed point is found within this expansion in
all dimensions between 2 and 4.  The exponents $\nu$ and $\eta$ have
been computed up to order $1/N^2$ in $d=3$ \cite{pelissetto01b}:
\begin{equation}
\left\{
\begin{aligned}
\nu&=1-\frac{16}{\pi^2}\frac{1}{N}-\left(\frac{56}{\pi^2}-\frac{640}{3\pi^4}\right)
\frac{1}{N^2} +O(1/N^3)\\ \eta&=\frac{4}{\pi^2}\frac{1}{N}-
\frac{64}{3\pi^4}\frac{1}{N^2} +O(1/N^3)\ .
\label{exposent1surN}
\end{aligned}
\right.
\end{equation}

Around $d=4$ and $d=2$ the perturbative results of, respectively, the
GLW and NL$\sigma$ models are recovered once the limit $N\to \infty$
has been performed. This suggests that, at least for sufficiently
large $N$, the two models belong indeed to the same universality class
in all dimensions. However, within this approach, no $N_c(d)$ line is
found (see however \cite{pelissetto01b}). It is thus impossible to
extrapolate to finite $N$ the results obtained in this approach.

\subsubsection{The six-loop results in $d=3$}
\label{chap_resommation}

 In three dimensions, a six-loop computation has been performed  by Pelissetto
 \etal{}~\cite{pelissetto01a} and re-examined by Calabrese
\etal{}~\cite{calabrese02,calabrese03b}, see below. The results are the following:

1) For $N$ sufficiently large --- $N>6.4(4)$  --- there exist 
four fixed points, one stable and three unstable, in agreement with
the usual picture given above, see Fig.~\ref{ptfixe}a. The transition is thus of
 second order.

2) For $5.7(3)<N<6.4(4)$, there is no nontrivial fixed point and the transition is expected to be of first order. 

3) For $N<5.7(3)$ and, in particular, for $N=2$ and $N=3$ a 
{\it stable fixed point} is found and  a second order phase transition is expected.

 According to Pelissetto \etal{}~\cite{pelissetto01a,pelissetto01b}, the fixed points  found for $N=2$
and $N=3$ should be non analytically connected with those found in the
$1/N$ and $4-\epsilon$ approaches. Therefore, it should be impossible
to obtain them by following smoothly those obtained at large-$N$ or
close to $d=4$.

The critical exponents obtained by Pelissetto \etal{} are given in
Table \ref{table_expsixloop}. Note that the error bars are about ten
times larger here than in the ferromagnetic $O(N)$ models
\cite{zinnjustin89,pelissetto01c} computed by the same method.  This
is an indication that the resummed perturbative series are
converging much slower than in the vectorial case.

Let us now discuss these results.

\begin{table}[htbp]
\begin{tabular}{|l|l|l|l|l|l|}
\hline
System&Ref.&$\alpha$&$\beta$&$\gamma$&$\nu$\\
\hline
\hline
XY&\cite{pelissetto01a}&0.29(9)&0.31(2)&1.10(4)&0.57(3)\\
\hline 
Heis.  &\cite{pelissetto01a}&0.35(9)&0.30(2)&1.06(5)&0.55(3)\\
\hline
\end{tabular}
\caption{The six-loop perturbative results in $d=3$.} 
\label{table_expsixloop}
\end{table}

{\it The XY case}. First, one should indicate that the exponents
$\gamma$ and $\nu$ computed from the six-loop approach compare
reasonably well with the data of group 1.  However, as already
mentioned, the value of $\eta$ found by the scaling relations must be
positive when there exists a fixed point. One finds, with the data of
Table \ref{table_expsixloop}, $\eta\sim 0.08$ which is significantly
positive. Let us recall that this is {\it not} the case for the
experiments performed on the materials of group 1 and for the
numerical simulations performed on STA.  Note, moreover, that the
value of $\beta$ found within the six-loop calculation, is very far
--- around four theoretical error bars --- from the average
experimental ones which are $\beta=0.228(6)$ for CsMnBr$_3$ alone,
$\beta=0.237(4)$ for the whole group 1 and far from the numerical
values obtained for STA $\beta=0.24-0.25$.  Thus, contrarily to what
is asserted in \cite{pelissetto01a}, it seems extremely improbable
that the exponents found at six loops could fit with those of group 1
and those of the numerical STA model. Actually, this is also the case
for the materials of group 2 for which the average $\beta$ equals 
$0.389(7)$.

{\it The Heisenberg case.} First, one notes that the agreement between
the $\gamma$ and $\nu$ exponents obtained from the six-loop approach
and from the experimental or numerical data is not as good as it is in
the XY case.  Concerning $\eta$, one finds, with the data of Table
\ref{table_expsixloop}, $\eta\sim 0.08$. This has to be compared with
the value of $\eta$ obtained {\it i)} for the materials of group 1,
which is significantly negative --- $\eta=-0.118(25)$ --- {\it ii)}
for materials of group 2, which is marginally negative ---
$\eta=-0.018(33)$ --- and {\it iii)} in the simulations of the STA
which is also negative although not completely significantly:
$\eta=-0.0182(89)$. The negativity of $\eta$ is an  indication of
a mismatch between the six-loop results and the data for the Heisenberg
systems even if it cannot be used as a definitive argument against a
second order phase transition. The exponent $\gamma$ obtained from the
numerical simulations of the Heisenberg STA model provides a further
information. The average value of this exponent --- $\gamma=1.185(3)$
--- is rather far --- 2.5 theoretical error bars --- from the six-loop
results \footnote{This result would be almost unchanged if the
numerical error bar --- and especially the one quoted in
\cite{mailhot94} --- was largely underestimated since it would surely
be much less than the theoretical one.}.

From the previous analysis one can conclude that,  as such,  the fixed point
obtained within the six-loop approach  turns out to be  {\it not} directly relevant to the
phenomenology of XY materials or simulated systems. This seems to  exclude the scenarios
 I, II and III that all assume that, at least, a certain  number of compounds or systems are well
described by a fixed point.

\subsubsection{Critical remarks}
\label{chap_critical_remark}

 As we mentioned at the beginning of our analysis of the experimental
results, see Section \ref{preliminaries}, we have made an assumption
on the nature of the experimental errors which is not realistic: the
systematic errors cannot be neglected.  We now come back on this point
and show that the conclusions we have drawn from our analysis persist
without this assumption.

Let us consider the XY case, where the symptoms of a mismatch between
the theoretical and experimental results are the clearest. We
concentrate on the materials of group 1 and on the exponent $\beta$
which is the best measured, see Table
\ref{table_exp_crit_STA_XY_exp}. With our assumption, we have found
$\beta=0.228(6)$. Let us suppose that, contrarily to our assumption,
the systematic error is large and dominates the total error. Let us
take:
\begin{equation}
\beta=0.23(2)
\end{equation}
so that all experimental and numerical results lie in the interval of
values, see Tables \ref{table_exp_crit_STA_XY_exp} and  \ref{table_exp_crit_XY_num}.  This estimate has to be
 compared with
the six-loop result:
\begin{equation}
\beta=0.31(2)
\label{betasixloop}
\end{equation}
where, in this case, the authors indicate that they have been very
conservative in the estimation of the error bar
\cite{pelissetto01a}. Although it is difficult to get fully
unambiguous conclusions out of these numbers, it is clear that the
agreement between them is not satisfactory. The
same considerations on group 2 of XY materials lead to suppose:
\begin{equation}
\beta=0.39(2)
\end{equation}
which, again, is far from being in agreement with the six-loop result
Eq.~(\ref{betasixloop}).

It is also possible to test the negativity of the anomalous dimension
$\eta$ with our new assumption. In the same spirit, one estimates
$\nu=0.555(30)$. We find, in this case:
\begin{equation}
-0.28\le\eta\le -0.048\ .
\end{equation}
Thus, $\eta$ is again found negative even in the most extreme
hypothesis.

We thus conclude that, although it is --- up to now --- impossible to
estimate rigourously the confidence level of our analysis of the
experimental data since only one error bar is given in the literature,
it appears to be very difficult to reconcile the experimental and
numerical data with the six-loop results.

\subsubsection{The  six-loop results in $d=3$ re-examined}

 In order to cope with the discrepancy between the six-loop results obtained by 
Pelissetto \etal{} and the experimental and numerical data, Calabrese \etal{} have
 reconsidered the resummed six-loop series~\cite{calabrese02,calabrese03b,parruccini03}. They claim  that they 
can account for the unusual properties of the
critical exponents for XY and Heisenberg frustrated systems in $d=3$
--- negative anomalous dimension and weak universality --- by the fact
that the RG trajectories around the stable --- focus --- fixed point found by Pelissetto \etal\ are spiral-like. 
 By integrating  the resummed $\beta$
 functions for the two coupling constants of the GLW model  and computing  the 
effective exponents $\eta$ and $\nu$ along the RG trajectories, they have found that
these exponents display large variations  in a transient regime. These authors  
argue that  the scaling properties of the system  are governed, over 
several decades of temperatures, by the preasymptotic regime so 
 that the effective exponents observed experimentally can differ significantly
 from their asymptotic values, {\ie} those defined at the fixed point. 

 Let us  underline  here several drawbacks of  the scenario of Calabrese \etal\ .

  First, it is based on the existence of stable fixed points  that  are  not related
to any  already known fixed point. In particular, the fixed points  found for $N=2$ and
$N=3$ within this computation in $d=3$ are, according to  Pelissetto \etal\ and
Calabrese \etal\   {\it non analytically} related to those found
 in the large-$N$ as well as in the  $4-\epsilon$  expansions. This means that there is
no way to  check their existence using these  perturbative methods.  This is specifically
problematic in the context of frustrated magnets where the properties of the fixed
points  appear to be very  unusual: {\it i)} the existence of  the stable  fixed points
strongly depends on the order  of perturbation ---  they are not present at three-loop
order and only show up, as far as we know, at  five-loop  order {\it ii)} the location
of the fixed points, as
$N$ and $d$ are varied, seems to have, in the
$(N,d)$ plane, a very particular structure since, in three dimensions,  they only exist 
when $N$ is below  {\it another} critical value of $N$ --- which is found to be equal to
 $5.7(3)$.

Second, it is very difficult, in the computation of Calabrese \etal\   , to relate --- 
even in principle --- the initial conditions   of the RG flow to the microscopic characteristics 
of real  systems. This would require to handle the infinity of coupling constants entering into 
the microscopic Hamiltonian obtained from the Hubbard-Stratonovitch transformation. This is
impossible, at least within the usual perturbation theory.  

Third, it is very
difficult to account, in this framework, for the first order behavior
deduced from  several  numerical simulations of XY and  Heisenberg 
systems~\cite{loison98,itakura03}. We have also already noticed that XY-systems have a stronger
tendency to undergo  first order transitions than Heisenberg systems. However, there is no
natural explanation for this phenomenon in the  scenario of Calabrese \etal\ .

Fourth, in this scenario, it is also very difficult to explain why there is no physical
 system characterized by  the asymptotic  critical exponents, {\ie} those  corresponding to
the fixed point found by Pelissetto \etal{}. This seems to require  very  unnatural 
experimental circumstances such that the initial conditions of the flow corresponding to
 the physical realizations of frustrated magnets are  such that  their  long distance
 properties are {\it never} controlled by the nontrivial 
fixed point. 

Finally, there  is no possible explanation of the breakdown of the NL$\sigma$ model 
predictions.

\subsection{Conclusion }

 XY and Heisenberg frustrated systems exhibit the kind of problems we
have described in the introduction: the perturbative results obtained
within a low-temperature expansion around two dimensions, within a
weak-coupling expansion around four dimensions or within a large-$N$
expansion fail to describe their critical physics in three
dimensions. Moreover, these different perturbative predictions are in
contradiction with each other. Contrarily to the $O(N)$ nonfrustrated
case, there is no possible smooth interpolation of these results
between two and four dimensions and,  at fixed dimension,  between $N=\infty$ and $N=2,3$.
More surprisingly and, again in contradiction with what happens in the
$O(N)$ nonfrustrated case, high-order calculations performed directly
in $d=3$ also fail to reproduce the phenomenology, at least  when they are interpreted in the usual way.
This situation reveals the difficulties of the conventional approaches  to
tackle with the physics of frustrated magnets. Only new interpretations or
methods can  allow to shed light  on the problems encountered here. We have
 presented  the solution proposed by  Calabrese \etal{} on the basis of  a high-order pertubative
 calculation and underlined its difficulties. We now present the nonperturbative method
 we have  used to explain the unusual behavior of frustrated magnets. This is 
the subject of the next sections. We start by a methodological introduction to this method and then apply it to the
 frustrated
systems.

\section{The Effective Average Action Method}
\label{chapitre_effective}

We now present the NPRG method we use: the effective average action
method \cite{wetterich91,wetterich93,wetterich93b,wetterich93c,tetradis94}.
The content of this section is neither original nor exhaustive.  There
exist several well-documented reviews on the subject
\cite{jungnickel99,bagnuls01,berges02}. Our aim here is only to
provide some of the physical ideas underlying this method --- notably
the block spin concept and its formulation in the continuum --- as
well as its technical implementation on the simple example of the
$O(N)$ model.

\subsection{Block spin in the continuum}

The effective average action method, as well as many other NPRG
techniques, is based on the well known concept of ``block spin''
\cite{kadanoff66,kadanoff67}: when dealing with any strongly
correlated system, it is fruitful to integrate out the fluctuations
step by step and, more precisely, scale by scale. In practice, one
first gathers the initial --- microscopic --- degrees of freedom into
small ``blocks''. It is then possible, at least formally, to integrate
out, in the partition function of the system, the internal
fluctuations of the blocks. This ``decimation'' is followed by a
rescaling of length-scales,  coupling constants and fields. In this way, starting
from a ``bare'' GLW Hamiltonian, one gets an effective Hamiltonian for
the block degrees of freedom, \ie for the low-energy modes. By
iterating this procedure, one generates a sequence of ---
scale-dependent --- effective Hamiltonians, parametrized by a running
scale $k$, that all share the same long-distance physics.   This
sequence defines a RG flow.  At a fixed point of this flow the system displays scale invariance. 
 This  allows  to obtain  the critical quantities
through an analysis of  the neighborhood of the fixed point in the flow
of effective Hamiltonians \cite{wilson74}.

To illustrate how this concept of block spin is implemented concretely
in the continuum, we consider the case of an Ising-like system,
initially defined on a lattice which, in the continuum, is described
by a scalar field $\zeta(\cg x)$.  If the lattice spacing is given by
$a$, the corresponding continuum field theory is characterized by an
overall momentum cut-off $\Lambda$ of order $a^{-1}$. The partition
function writes: 
\begin{equation}
{\cal Z}=\int {\cal D}\zeta\ e^ { -\frac{\scriptstyle 1}
{\scriptstyle 2}\displaystyle \,\zeta.C^{-1}_{\Lambda}.\zeta\ -\ \Hi_{\Lambda}[\zeta]}
\label{partition}
\end{equation}
where $\Hi_{\Lambda}[\zeta]$ stands for the interacting part of the
GLW Hamiltonian and:
\begin{equation}
\zeta.C^{-1}_{\Lambda}.\zeta=\int \frac{d^d\cg{q}}{ (2\pi)^d}\
\zeta(\cg{q}) C^{-1}_{\Lambda}(\cg{q})\zeta(-\cg{q})
\label{scalar}
\end{equation}
corresponds to the cut-off kinetic part.  In Eq.~(\ref{partition}) and
(\ref{scalar}), $C_{\Lambda}(\cg{q})$ is an ultra-violet \ ($UV$)
cut-off propagator that prevents the propagation of unphysical modes
with momentum higher than $\Lambda$. One writes it:
\begin{equation}
C_{\Lambda}(\cg{q})=\frac{F_{k=\Lambda}\left(\cg{q}^2\right)}{ \cg{q}^2}
\label{regulator}
\end{equation}
where $F_k(\cg{q}^2)$ is a function of the ratio $z=\cg{q}^2/k^2$ that
rapidly decreases when $z\to \infty$. One also imposes to
$F_k(\cg{q}^2)$ to be unity at the origin: $F_k(\cg{q}^2=0)=1$.  A
typical example of function $F$ is: $F_k(\cg{q}^2)=e^{-(\cg{q}/k)^2}$
but other forms can obviously be considered.
 
In Fourier space, the idea of block spin is specified by
separating the low- and high-momentum modes of the spin-field $\zeta$:
\begin{equation}
\zeta(\cg{q})= \zeta_{\scriptscriptstyle>}(\cg{q}) + \zeta_{\scriptscriptstyle<}(\cg{q})\ .
\label{separationchamp}
\end{equation}
The fields $\zeta_{\scriptscriptstyle>}(\cg{q})$ and
$\zeta_{\scriptscriptstyle<}(\cg{q})$ being unconstrained, the
separation between high- and low-momentum modes is actually realized
through their respective propagator. We thus write:
\begin{equation}
C_{\Lambda}(\cg{q})= C_k(\cg{q})+(C_{\Lambda}(\cg{q})-C_k(\cg{q}))\
\hat =\ C_{\scriptscriptstyle<}(\cg{q})+C_{\scriptscriptstyle>}(\cg{q})
\label{separationprop}
\end{equation} 
where $k$ is the typical scale that separates the high- and
low-momenta. In Eq.~(\ref{separationprop}),
$C_{\scriptscriptstyle>}(\cg{q})$ (resp.
$C_{\scriptscriptstyle<}(\cg{q})$) propagates
$\zeta_{\scriptscriptstyle>}$ (resp. $\zeta_{\scriptscriptstyle<}$),
the high- (resp. low-) momentum degrees of freedom of the field
$\zeta$.  This comes from a property of the Gaussian integral that can
be easily seen on a one-dimensional integral:
\begin{equation}
\begin{array}{ll}
\displaystyle\int dz &\exp\left(\displaystyle {-{z^2\over 2 
(\alpha+\beta)}+f(z)}\right)\propto\\
& \displaystyle\int \ dx \ dy\ \exp\left({\displaystyle -\frac{ x^2}{ 2 \alpha}-
\frac{y^2}{ 2 \beta}+f(x+y)}\right)\ .
\label{exemple}
\end{array}
\end{equation}
This result is easily obtained by changing, in the right hand side of
Eq.~(\ref{exemple}), the integration variables $x,y,$ into $z=x+y$ and
$t=x-y$ and by integrating on $t$.

Thus from Eqs.~(\ref{partition}), (\ref{separationchamp}),
(\ref{separationprop}) and (\ref{exemple}) one gets:
\begin{equation}
\begin{split}
{\cal Z}=\int {\cal{D}}\zeta_{\scriptscriptstyle<}\ 
{\cal{D}}\zeta_{\scriptscriptstyle>}\ &\exp\left( -\, \frac{1}{2}\,\zeta_{\scriptscriptstyle<}.C_{\scriptscriptstyle<}^{-1}.\zeta_{\scriptscriptstyle<}
  -\, \frac{1}{2}\,\zeta_{\scriptscriptstyle>}.C_{\scriptscriptstyle>}^{-1}.\zeta_{\scriptscriptstyle>}\right. \\ 
&\ \ \ \ \ \left. {\phantom{\frac{1}{2}}} -
\Hi _{\Lambda}[\zeta{\scriptscriptstyle<}+
\zeta_{\scriptscriptstyle>}]\right)\ .
\end{split}
\label{chisupchiinf}
\end{equation}  

The effective Hamiltonian $\Hi _k[\zeta_{\scriptscriptstyle<}]$ for
the low-momentum degrees of freedom $\zeta_{\scriptscriptstyle<}$ is
defined through the integration over the high-momentum degrees of
freedom in Eq.~(\ref{chisupchiinf}):
\begin{equation}
\begin{array}{l}
e^{\displaystyle-\Hi_k[\zeta_{\scriptscriptstyle<}]}\hat = 
{\displaystyle\int} {\cal{D}}\zeta_{\scriptscriptstyle>}
\exp\displaystyle{\left( -\frac{1}{2}\,\zeta_{\scriptscriptstyle>}. 
C_{\scriptscriptstyle>}^{-1}.\zeta_{\scriptscriptstyle>} \right.}\\
\hskip 4cm\left.{\phantom{\displaystyle\frac{ 1}{2}}}- 
\Hi _{\Lambda}[\zeta_{\scriptscriptstyle<}+
\zeta_{\scriptscriptstyle>}]\right) .
\end{array}
\label{effective}
\end{equation}
Integrating out the internal degrees of freedom of a block spin
between the scales $a$ and $a'>a$ corresponds, in this language, to
the integration of the modes $\zeta_{\scriptscriptstyle>}$ with
momenta between $k=a^{-1}$ and $k'={a'}^{-1}$.  The Equation
(\ref{effective}) implements the block spin procedure in the continuum
which is the starting point of any NPRG approach.

\subsection{The Polchinski equation}

The effective Hamiltonian $\Hi _k[\zeta_{\scriptscriptstyle<}]$
follows an exact equation describing its infinitesimal evolution when
the running scale $k$ is lowered. To establish this equation we
rewrite Eq.~(\ref{effective}) as:
\begin{align}
&e^{\displaystyle-\Hi _k[\zeta_{\scriptscriptstyle<}]}=\nonumber\\
&\ \ \ =\int {\cal{D}}\zeta\, \exp\left(
- \frac{1}{2} (\zeta-\zeta_{\scriptscriptstyle<})
\cdot C_{\scriptscriptstyle>}^{-1}\cdot(\zeta-
\zeta_{\scriptscriptstyle<})
-\Hi _{\Lambda}[\zeta]\right)\nonumber \\
&\hskip 0.33cm = \exp\left( \frac{1}{2}
 \frac{\delta}{\delta\zeta_{\scriptscriptstyle<}}\cdot 
C_{\scriptscriptstyle>}\cdot \frac{\delta}{
\delta\zeta_{\scriptscriptstyle<}}\right)\ 
e^{\displaystyle- \Hi _{\Lambda}[\zeta_{\scriptscriptstyle<}]}\ .
\label{effective2}
\end{align}

\begin{figure}[ht]
\parbox{2cm}{\vspace{-1.3cm}$${\partial_k \Hi _k=\frac 12}$$}
\includegraphics[width=.2\linewidth,origin=tl]{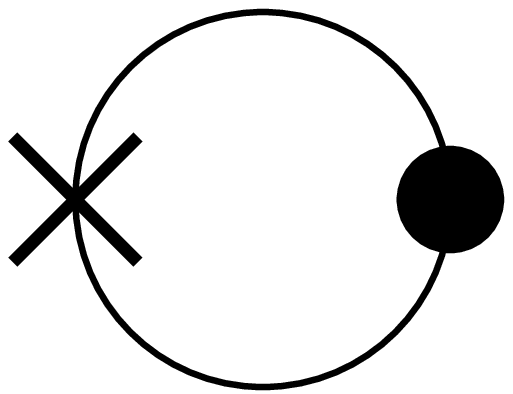}
\parbox{1cm}{\vspace{-1.3cm}$${-\frac 12}$$}
\includegraphics[width=.2\linewidth,origin=tl]{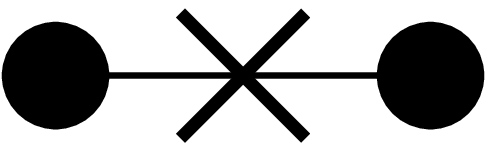}
\caption{{A graphical representation of the Polchinski equation. The
crosses represent the cut-off factor $\partial_k
C_{\scriptscriptstyle>}({\cg q})$. The black circles with $n$-external
legs correspond to the $n$-th functional derivative of $\Hi _k$ with
respect to the field.} }
\label{polchinskigraphic}
\end{figure}

This last  functional relation can be inferred from the one-dimensional identity:
\begin{equation}
\begin{array}{ll}
\displaystyle \int dx \ \exp\left(\displaystyle{-\frac{(x-y)^2}{ 2
\gamma}- f(x)}\right) \propto 
\displaystyle e^{\displaystyle{{\frac{\gamma}{2}}\frac{\partial^2}{\partial y^2}}} e^{\displaystyle - f(y)}\ .
\end{array}
\end{equation}
By differentiating each side of Eq.~(\ref{effective2}) with respect to
$k$ we obtain: 
\be
\begin{array}{ll}
\displaystyle{-(\partial_k \Hi _k){\phantom\int}\hskip-0.2cm e^{\displaystyle  -\Hi _k}}&=\\
&=\displaystyle{{1\over 2}
\left({\delta \over \delta \zeta_{\scriptscriptstyle<}}\cdot\partial_k C_{\scriptscriptstyle>}\cdot{\delta\over\delta \zeta_{\scriptscriptstyle<}}\right) \ e^{-\displaystyle  \Hi _k}}\\ 
\\&\hspace{-3cm}\displaystyle{={1\over 2}
\left({\delta \Hi _k\over \delta \zeta_{\scriptscriptstyle<}} \cdot
\partial_k C_{\scriptscriptstyle>} \cdot {\delta \Hi _k\over \delta
\zeta_{\scriptscriptstyle<}}-{\delta^2 \Hi _k\over \delta \zeta_{\scriptscriptstyle<} \delta
\zeta_{\scriptscriptstyle<}}\cdot \partial_k C_{\scriptscriptstyle>}\right) e^{-\displaystyle  \Hi _k}}.
\end{array}
\label{differentiate}
\ee 
Finally, the {\it exact} evolution equation for $\Hi _k$, known as
the Polchinski equation \cite{polchinski84} (see also \cite{ball94}),
writes explicitly:
\begin{equation}
\begin{split}
\partial_k &\Hi _k[\zeta_{\scriptscriptstyle<}]= {1\over 2} \int {d^d\cg{q}\over
(2\pi)^d} \ \partial_k C_{\scriptscriptstyle>}(\cg{q}).\\&. \left({\delta^2
\Hi _k\over \delta \zeta_{\scriptscriptstyle<}(\cg{q}) \delta \zeta_{\scriptscriptstyle<}(-\cg{q})}-{\delta
\Hi _k\over \delta \zeta_{\scriptscriptstyle<}(\cg{q})} {\delta \Hi _k\over \delta
\zeta_{\scriptscriptstyle<}(-\cg{q})}\right) \
\label{polchinski1}
\end{split}
\end{equation}
or, in real space:
\begin{equation}
\begin{split}
\partial_k &\Hi _k[\zeta_{\scriptscriptstyle<}]= {1\over 2} \int d^d\cg{x}\  d^d\cg{y}\
\partial_k C_{\scriptscriptstyle>}(\cg{x}-\cg{y}) .  \\&.\left({\delta^2
\Hi _k\over \delta \zeta_{\scriptscriptstyle<}(\cg{x}) \delta \zeta_{\scriptscriptstyle<}(\cg{y})}\ -{\delta
\Hi _k\over \delta \zeta_{\scriptscriptstyle<}(\cg{x})} {\delta \Hi _k\over \delta
\zeta_{\scriptscriptstyle<}(\cg{y})}\right) \
\label{polchinski2}
\end{split}
\end{equation} 
with $\partial_k C_{\scriptscriptstyle>}({\cg x}-{\cg y})=\int
{d^d{\cg q}\over (2\pi)^d}\ \partial_k C_{\scriptscriptstyle>}({\cg
q})\ e^{i{\cg q}({\cg x}-{\cg y})}$. Note
that in the preceding equations we have improperly used the same
notation for the field $\zeta_{\scriptscriptstyle<}$ and regulator
$C_{\scriptscriptstyle>}$ and for their Fourier transforms.  Note also
that, in the following, we shall use the notation $\zeta$ for
$\zeta_{\scriptscriptstyle<}$.  A graphical representation of the
Polchinski equation is given in Fig.~\ref{polchinskigraphic}.

Let us first make some remarks about Eq.~(\ref{polchinski1}). A first
feature we would like to emphasize is that this equation involves the
quantity $\Hi _k$ which is the effective Hamiltonian for the degrees
of freedom that have {\it not} yet been integrated out, namely
$\zeta_{\scriptscriptstyle<}$. The drawback with $\Hi _k$ is that it
is an abstract object that has no direct physical meaning since it is
a function of a field $\zeta_{\scriptscriptstyle<}$ that eventually
fully disappears in the physical limit $k\to 0$, \ie when {\it all}
fluctuations have been integrated out. In particular, one should
realize that $\zeta_{\scriptscriptstyle<}$ is {\it not} a precursor of
the order parameter, \ie {\it not} a local magnetization at scale
$k$. Indeed this magnetization should come from a {\it thermodynamical
average} at scale $k$ while $\zeta_{\scriptscriptstyle<}$ is just a
stochastic variable that represents the low-momentum part of the
original spin field and is thus, roughly speaking, a {\it spatial
average} of this field. Consequently, the effective Hamiltonians by
themselves do not contain all the information on the integration of
the high-momentum degrees of freedom
$\zeta_{\scriptscriptstyle>}$. For instance, the computation of
correlation functions for the high-energy field
$\zeta_{\scriptscriptstyle>}$ would require to first couple the system
to a source $J$ --- a magnetic field --- by adding in $\cal Z$ a term
$\exp (J.\zeta)$ and to follow the flow of this term in order to
obtain the full $J$-dependence of $\cal Z$, a rather difficult
task. Thus, Eq.~(\ref{polchinski1}) provides at best a flow of the
running coupling constants that parametrize the effective Hamiltonian
$\Hi _k$ at scale $k$.

As shown mainly by Wilson, equations like Eq.~(\ref{polchinski1}) are,
in principle, sufficient to compute the critical exponents once a
fixed point Hamiltonian ${\Hi _k}^*$ has been found. Actually, even
for the evaluation of the RG flow, Eq.~(\ref{polchinski1}) suffers
from an important difficulty: although this equation looks simple ---
its only nonlinearity is a term  quadratic in $\Hi _k$ --- it is nevertheless
a functional-partial-integro-differential equation that has no known
solution in general.  Therefore, in order to render it manageable, one
has to truncate the Hamiltonian $\Hi _k$.

\subsubsection{ Derivative expansion}

A natural truncation consists in an expansion of the effective
Hamiltonian in powers of the derivatives of the field
\cite{golner86,tetradis94,morris94b}. For instance, for a
one-component scalar field theory one has:
\begin{equation}
\Hi _k[\zeta]= \int d^d{\cg x}\left( \ U_k(\zeta)+{1\over 2} 
Z_k(\zeta) (\partial \zeta)^2+O(\partial^4)\right)
\label{truncatedaction}
\end{equation}
where $U_k(\zeta)$ stands for the potential --- \ie the
derivative-independent part --- of the effective Hamiltonian and
$Z_k(\zeta)$ is the quadratic --- {\it field-dependent} --- field
renormalization.  With such a truncation, one neglects higher-order
derivative terms. This is justified {\it i)} when one is interested in
the long-distance, low-energy physics, since these higher-order derivative
terms should correspond to less important operators and {\it ii)} when
there is no qualitative change of nature between the microscopic and
macroscopic degrees of freedoms --- such as the appearance of bound
states at a finite scale $k$ --- that could induce non localities
\cite{gies02}. A practical guide to evaluate the validity of the
derivative expansion is the value of the anomalous dimension
$\eta$. If this quantity is small, one can expect that the inclusion
of higher-order derivative terms provides small corrections to the
results.

At first order in the derivative expansion one sets $Z_k(\zeta)=0$ in
Eq.~(\ref{truncatedaction}) and derives an RG equation for
$U_k(\zeta)$ from Eq.~(\ref{polchinski1}). This corresponds to the
so-called {\it Local Potential Approximation} (LPA) which has been
intensively explored in the past
\cite{nicoll74,hasenfratz86,felder87,zumbach94b}. In particular, this
kind of approach has been used by Zumbach to analyze the physics of
frustrated magnets in three dimensions
\cite{zumbach93,zumbach94,zumbach94c}.  The problem with the LPA is that,
since by definition it neglects the field renormalization, it leads to
a trivial --- vanishing --- anomalous dimension. Consequently: {\it i)}
this prohibits to compare the results obtained within this approach to
that of a standard perturbative approach when this last one involves a
nontrivial anomalous dimension (this is, for instance, the case of the
NL$\sigma$ model around $d=2$ already at one-loop), {\it ii)} this
prevents an accurate evaluation of critical exponents for systems for
which the anomalous dimension is not expected to be small.  In the
context of frustrated magnets, these drawbacks are serious since we
are precisely interested in relating the different perturbative
approaches and, to some extent, by a satisfactory determination of the
critical exponents. We thus need to compute the field renormalization.

This kind of computation however encounters several
difficulties. First, whereas the RG equation for $U_k(\zeta)$ in the
LPA of the Polchinski equation is universal --- cut-off independent
---, the RG equation derived for $U_k(\zeta)$ and $Z_k(\zeta)$ at
second order in the derivative expansion explicitly depends on the
regulator $C_{\scriptscriptstyle>}({\cg q})$ chosen to separate the
high- and low-energy degrees of freedom in Eq.~(\ref{separationprop})
\cite{golner86,ball95}. Another related problem is that of
reparametrization invariance.  The partition
function (\ref{partition}) and, thus, the physical quantites like
critical exponents, are invariant under a general change of field of
the kind $\zeta\to \zeta+G(\zeta)$ where $G$ is an arbitrary function
starting at order $\zeta^2$. As a consequence of this invariance, the
normalization of the field $Z_k(\zeta=0)$ in the Hamiltonian is {\it a
priori} an arbitrary parameter.  Unfortunately, the reparametrization
invariance is broken as soon as one performs a truncation of the
Hamiltonian.  As a result the critical exponents and, in particular
$\eta$, depend on the normalization $Z_k(\zeta=0)$. It follows from
these considerations that, in any practical computation, one
encounters the problem that physical quantities depend on nonuniversal
parameters such as cut-off functions and normalizations. Different
techniques, such as the Principle of Minimum Sensitivity (PMS), have
been used to decrease the dependence of the critical quantities on the
cut-off function \cite{ball95,comellas98}. Also, some criterions have been
proposed to find the best normalization, \ie to find a value
$Z_k(\zeta=0)$ such that the derivative expansion converges the most
rapidly \cite{comellas98}. These considerations, having for aim to exploit the
Polchinski equation at the next to leading order in derivative
expansion, have led to the determination of rather satisfactory
critical exponents.

At the same time, there has been a great activity devoted to the
search of other formulations of the RG ideas that could avoid some of
the troubles encountered in the use of the Polchinski equation. The
effective average action method is the result of this search.

\subsection{The effective average action method}

The basic --- and physically very appealing --- idea of this new
formulation is to consider as the fundamental object, not the abstract
effective Hamiltonian $\Hi_k[\zeta]$, functional of the stochastic
low-energy field $\zeta_{\scriptscriptstyle<}$ but, rather, the Gibbs
free energy $\Gamma$ --- called effective action in field theory ---
functional of the order parameter $\phi=\langle \zeta\rangle$. To
implement this idea in the RG context, it is necessary to build a
running Gibbs free energy $\Gamma_k$ for the high-energy modes that
have already been integrated out at this scale $k$. The argument of
$\Gamma_k$ is, therefore, the order parameter at this scale that
eventually becomes, when $k\to 0$, the true order parameter.
 
These requirements imply several constraints on the definition of
$\Gamma_k$. First, at the scale of the lattice spacing,
$k=\Lambda=a^{-1}$, $\Gamma_{k}$ should correspond to the
microscopical Hamiltonian $H$ since no fluctuations have been taken
into account.  Second, when the running scale $k$ is lowered to 0,
$\Gamma_{k}$, which then includes {\it all} fluctuations, must
identify with the standard effective action $\Gamma$ from which all
thermodynamical quantities like magnetization, correlation length,
etc, are computed. To summarize, $\Gamma_k$ must respect the
constraints:
\begin{equation}
\left\{
\begin{array}{l}
\Gamma_{k=\Lambda}=H\ \\ 
\\
\Gamma_{k=0}=\Gamma
\label{limitesgamma}
\end{array}
\right.
\end{equation}
and has to interpolate smoothly between these two limits.

\subsubsection{Construction}

Let us again consider, for simplicity, the case of a system described
by a scalar field $\zeta(\cg x)$. The construction of the effective
average action proceeds in two steps. First, one should decouple the
low-energy modes --- with momenta $\cg q^2>k^2$ --- in the partition
function in order to get a theory involving only the high-energy ones
that will be summed over. Second, in this modified theory, one builds
the Gibbs free energy, as usual, by a Legendre transform. This gives
$\Gamma_{k}$. Let us now study how this is implemented in practice.

The first step is conveniently implemented by changing the partition
function $\cal Z$ into ${\cal Z}_k$ for which a $k$-dependent  term,
quadratic in the fields and thus analogous to a mass-term is added to
the microscopic Hamiltonian \cite{tetradis94,wetterich93c}.  With this
``mass-term'', the partition function in presence of a source $J$
writes:
\begin{equation}
{\cal Z}_k[J]=\int \mathcal D\zeta\ \exp\Big( -H[\zeta]-\Delta H_k[\zeta]\, +\,
J.\zeta\Big)
\label{partition2}
\end{equation}
with $J.\zeta=\int d^d \cg q\ J(\cg q) \zeta(- \cg q)$ and
\begin{align}
\Delta H_k[\zeta]&={1\over 2} \int {d^d\cg q d^d\cg {q'}\over (2\pi)^{2d}} \
{\mathcal R}_k(\cg q,\cg {q'}) \zeta(\cg q) \zeta(\cg {q'})\\
&={1\over 2} \int {d^d\cg q \over (2\pi)^{d}} \
{ R}_k(\cg q^2) \zeta(\cg q) \zeta(-\cg {q})
\label{regularisation_R_droit}
\end{align}
with ${\mathcal R}_k(\cg q,\cg {q'})=(2\pi)^d\ \delta(\cg
q+\cg{q'})R_k(\cg q^2)$. In Eq.~(\ref{regularisation_R_droit}),  $R_k(\cg q^2)$ is the cut-off function that
controls the separation between the low- and high-energy modes.  To
decouple the low-energy modes, it must act as a large-mass term for
small $\cg q$ whereas it must vanish for large $\cg q$ to keep
unchanged the high-energy sector of the theory. Thus:
\begin{equation}
R_k(\cg q^2)\sim k^2 \hspace{0.5cm} \hbox{for} \hspace{0.5cm} \cg q^2\ll k^2
\label{IR}
\end{equation} 
and
\begin{equation}
R_k(\cg q^2)\to 0 \hspace{0.5cm} \hbox{when}
\hspace{0.2cm} {\cg q}^2\gg k^2 \ .
\label{UlVi}
\end{equation} 
The first constraint means that, for momenta lower than $k$, $R_k(\cg
q^2)$ essentially acts as a mass --- \ie an IR cut-off --- which
prevents the propagation of the low-energy modes. The second ensures
that the high-energy modes are fully taken into account in ${\cal
Z}_k[J]$ and thus in the effective average action.  Moreover, since we
want to recover the original theory when $k\to 0$, \ie when all
fluctuations have been integrated out, $R_k(\cg q^2)$ must vanish in
this limit. Thus we require:
\begin{equation}
R_k(\cg q^2)\to 0\ \    \hbox{identically when} \hspace{0.3cm} k\to 0
\end{equation}
which ensures that ${\cal Z}_{k=0}[J]={\cal Z}[J]$. On the other hand,
when $k\rightarrow \Lambda$, \ie when no fluctuation has been
integrated out, $\Gamma_{k}$ should coincide with the microscopic
Hamiltonian. This is achieved by requiring (see below for the proof):
\begin{equation}
R_k(\cg q^2)\to \infty\ \  \hbox{identically when}  \hspace{0.3cm} k\to
\Lambda  \ .
\end{equation}
 Note that, since  we shall not be interested in the precise relation  between the
 microscopic characteristics --- defined at scale $\Lambda$ --- of a given system and its
  critical or pseudo-critical properties,  we set $\Lambda=\infty$ in
the following.

A widely used cut-off function is provided by \cite{berges02}:
\begin{equation}
R_k(\cg q^2)={Z {\cg q}^2\over e^{{\cg q}^2/k^2}-1}\ 
\label{cutoffexp}
\end{equation}
where $Z$ is the field renormalization. Including it in $R_k$ allows
to suppress the explicit $Z$ dependence in the final RG equations ---
see below. The cut-off function $R_k(\cg q^2)$ corresponding to
Eq.~({\ref{cutoffexp}) is plotted on Fig.~\ref{separationmodes}.
\begin{figure}[htbp]
\includegraphics[width=.55\linewidth,origin=tl]{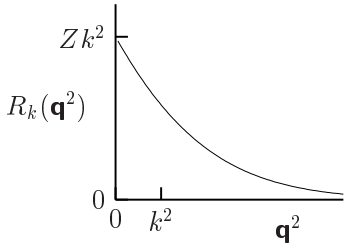}
\caption{{A typical realization of the separation of high- and
low-momentum modes provided by the cut-off
function $R_k(\cg q^2)$. At low momentum, $R_k(\cg q^2)$ acts as an effective mass of
order $Z k^2$ while the high-momentum behavior is not modified.}}
\label{separationmodes}
\end{figure} 
Another useful cut-off function,  called theta cut-off,  has been   proposed by Litim
\cite{litim02}. It writes:
\begin{equation}
R_k(\cg q^2)=Z \left(k^2- {\cg q}^2\right)\,\Theta\left(k^2-{\cg q}^2\right)
\label{cutoffstep}
\end{equation}
where $\Theta$ is the usual step function.

The second step consists in  defining  the effective average action. The free energy at scale
$k$ is given --- up to a factor $-k_B T$ --- by:
\begin{equation}
W_k[J]=\ln {\cal Z}_k[J]\ .
\label{freeenergy}
\end{equation}
From Eq.~(\ref{freeenergy}), one defines the order parameter $\phi_k(\cg
q)$ at scale $k$ as the average value of the microscopic field
$\zeta(\cg q)$ in the modified theory:
\begin{equation}
\phi_k(\cg q)=\langle \zeta(\cg q) \rangle={\delta W_k[J]\over \delta
J(- \cg q)}\ .
\label{orderparameter}
\end{equation}
Thanks to the properties of $R_k(\cg q^2)$, the contribution to the
average value in Eq.~(\ref{orderparameter}) coming from modes with
momenta $\cg q^2\ll k^2$ is strongly suppressed.  Also $\phi_{k}(\cg
q)$ identifies with the true order parameter in the limit $k\to
0$. Note that, for simplicity, we omit, in the following, the index
$k$ to $\phi_k$.

The effective average action is defined by \cite{wetterich93c}:
\begin{equation}
\Gamma_k[\phi]=-W_k[J]+ J.\phi -\Delta H_k[\phi]
\label{defgamma}
\end{equation} 
where $J=J[\phi]$, see Eq.~(\ref{orderparameter}). Thus
$\Gamma_k[\phi]$ essentially corresponds to a Legendre transform of
$W_k[J]$ for the macroscopic field $\phi$ --- up to the mass-like term
$\Delta H_k$. The relation (\ref{defgamma}) implies several
unconventional relations. First, taking its derivative with respect to
$\phi(\cg q)$ provides the relation between the source and
$\Gamma_k[\phi]$:
\begin{equation}
J(-\cg q)={\delta\Gamma_k\over \delta\phi(\cg q)}+\int \frac {d^d\cg
{q'}}{(2\pi)^{2d}}\mathcal R_k(\cg q,\cg {q'})\phi(\cg {q'})\ .
\label{J}
\end{equation} 
Taking the derivative of this relation with respect to $\phi(\cg q')$
implies a second important relation~\footnote{ From the relation $\int
d^d\cg y A(\cg x,\cg y)A^{-1}(\cg y,\cg z)=\delta(\cg x-\cg z)$, we
deduce the relation in the reciprocal space: $\int d^d\cg {p'} A(\cg
p,\cg {p'})A^{-1}(-\cg {p'},\cg {p''})=(2\pi)^{2d}\delta(\cg p+\cg
{p''})$. With this definition, if $A(\cg x,\cg y)=\tilde A (\cg x-\cg
y)$, then $A(\cg p,\cg q)=(2\pi)^d \delta(\cg p+\cg q)\tilde A(\cg
q)$, where $\tilde A(\cg q)$ is the Fourier transform of $\tilde A(\cg
x)$. Moreover, $A^{-1}(\cg p,\cg q)=(2\pi)^d \delta(\cg p+\cg
q)/\tilde A(\cg q)$ }:
\begin{align} 
{\Gamma^{(2)}_k}(\cg q,\cg q')+&\frac{\mathcal R_k(\cg q,\cg
{q'})}{(2\pi)^{2d}}={\delta J (-\cg q)\over \delta \phi(\cg q')}\\
&=(2\pi)^{-2d} \left(\delta^2 W_k\over \delta J(\cg q)\delta J(\cg
q')\right)^{-1}
\label{inverse}
\end{align}
where ${\Gamma^{(2)}_k}(\cg q,\cg q')={\delta^2 \Gamma_k/
\delta\phi(\cg q) \delta\phi(\cg q')}$.

Let us now show that the definition of $\Gamma_k$,
Eq.~(\ref{defgamma}), ensures that it satisfies the requirements given
in Eq.~(\ref{limitesgamma}), \ie that it  interpolates between the microscopic
Hamiltonian $H$ for $k=\infty$ and the (true) effective action
$\Gamma$ for $k\to 0$.  This last property follows directly from
Eq.~(\ref{defgamma}) and the fact that for $k=0$ the IR cut-off $R_k(\cg
q^2)$ identically vanishes.  The fact that $\Gamma_k$ identifies with
$H$ when $k\to\infty$ can be shown in the following way. One deduces  from
Eqs.~(\ref{partition2}), (\ref{freeenergy}), (\ref{defgamma}) and (\ref{J}) the functional
identity:
\begin{equation}
\begin{split}
e^{-\displaystyle \Gamma_k[\phi]}=&\int \mathcal D\zeta\ \exp\Big(
-H[\zeta]+\\
&+{\delta \Gamma_k[\phi]\over \delta \phi}.(\zeta-\phi)-\Delta
H_k[\zeta-\phi]\Big)\ .
\label{fonctionnal2}
\end{split}
\end{equation}
In the limit $k\to\infty$, $R_k(\cg q^2)$ goes to infinity. In this
limit, the mass-term $\exp(-\Delta H_k[\zeta-\phi])$ acts as a hard
constraint on the functional integration --- $\exp(-\Delta
H_k[\zeta-\phi])\simeq \delta[\zeta-\phi]$ --- so that
$\Gamma_{k=\infty}[\phi]=H[\phi]$.  With these properties,
$\Gamma_k[\phi]$ has the meaning of a coarse-grained Gibbs free energy
at scale $k^{-1}$: lowering $k$ corresponds to including more and more
fluctuations.

\subsubsection{The equation}

Let us now derive the exact RG equation for $\Gamma_k$. We start from
the expression:
\begin{equation}
e^{\displaystyle{W_k[J]}}=\int \mathcal D\zeta\ \exp\Big( -H[\zeta]-\Delta H_k[\zeta]\, +\,J.\zeta\Big)
\end{equation}
which results from Eqs.~(\ref{partition2}) and (\ref{freeenergy}). One first writes the variation of
$\exp\displaystyle(W_k[J])$ with respect to the scale $k$
\begin{equation}
\begin{array}{l}
\partial_k e^{\displaystyle{W_k[J]}}=\\
\\
\displaystyle{\ \ =- \int \mathcal D\zeta \Big(\partial_k\Delta
H_k[\zeta]\Big) \exp\Big( -H[\zeta]-\Delta H_k[\zeta]\, +\, J.\zeta\Big)}\\
\\
\displaystyle{\ \ =-\left(\partial_k\Delta H_k \left[{\delta\over \delta
J}\right]\right)\ e^{\displaystyle  W_k[J]}}\\
\\
\displaystyle{\ \ = -{1\over 2} \int \frac{d^d\cg
q}{(2\pi)^d}\left({\delta\over \delta J(\cg q)}\ . \ \partial_k R_k(\cg q^2)
\ . \ {\delta\over \delta J(-\cg q)}\right) e^{\displaystyle W_k[J]} }
\end{array}
\end{equation}
from which  follows:
\begin{equation}
\begin{split}
\partial_k W_k[J]=&-{1\over 2}\int\frac{d^d\cg q}{(2\pi)^d}
\partial_k R_k(\cg q^2)\ .\\ \ .&\Bigg({\delta W_k[J]\over \delta J(\cg q)} 
{\delta W_k[J]\over \delta J(-\cg q)} +
{\delta^2 W_k[J]\over \delta J(\cg q)\delta J(-\cg q)}\Bigg)\
\label{freeenergyevolution}
\end{split}
\end{equation}
which looks like the Polchinski equation (\ref{polchinski1}).

Let us now differentiate the expression (\ref{defgamma}) with respect
to $k$, at fixed $\phi$:
\begin{equation}
\begin{split}
\partial_k \Gamma_k{[\phi]}=-\partial_k W_k{[J]}{{\big |}_J}&-{\delta
W_k[J]\over \delta J}.\partial_k J +\\&+ (\partial_k J).\phi -\partial_k
\Delta H_k{[\phi]}\\
\\
 =-\partial_k W_k{[J]}&-\partial_k\Delta
H_k{[\phi]}
\end{split}
\end{equation}
since $\displaystyle \phi={\delta W_k\over \delta J}$. Together with
Eq.~(\ref{freeenergyevolution}) this gives:
\begin{equation}
\partial_k \Gamma_k{[\phi]}={1\over 2}\int \frac{d^d\cg q}{(2\pi)^d}
\partial_k R_k(\cg q^2){\delta^2 W_k[J]\over \delta J(\cg q)\delta
J(-\cg q)}\ .
\label{wetttilde} 
\end{equation}
Using Eq.~(\ref{inverse}) one obtains an equation involving only
$\Gamma_k$ and its second functional derivative $\Gamma_k^{(2)}$
\cite{wetterich93c,ellwanger93b,tetradis94,morris94a}:
\begin{align}
\partial_t\Gamma_k{[\phi]}=&{1\over 2} \int \frac{d^d\cg
q}{(2\pi)^d}\dot R_k(\cg q^2)\ .\nonumber\\
&\ . \hbox{Tr}\left((2\pi)^{2d}\Gamma_k^{(2)}[\phi]+\mathcal R_k\right)^{-1}(\cg
q,-\cg q)
\label{Wetterichfinal} 
\end{align}
with $t=\ln k$ and $\dot{R_k}=\partial_t R_k$.  In
Eq.~(\ref{Wetterichfinal}), $\hbox{Tr}$ must be understood  as a
trace on internal indices --- vectorial or tensorial --- if $\zeta$
spans a nontrivial representation of a group.

Let us finally give a form of Eq.~(\ref{Wetterichfinal}) more
convenient for practical use \footnote{Note that our factors $(2\pi)^d$ in 
Eqs.~(\ref{Wetterichfinal}) and (\ref{Wetterichfinalln}) are unusual. They come from our definition of the Fourier
 transform: $\displaystyle A(x)=\int {d^dq\over (2\pi)^d} \tilde A(q) \ e^{i qx}$ and from our definition of 
$\Gamma^{(2)}_k(\cg q,\cg q')$ as the second derivative of $\Gamma_k$ with respect to $\phi(\cg q)$ and not as
 the Fourier transform of $\Gamma^{(2)}(x,y)$.
These factors $(2\pi)^d$  always cancell in the $\beta$ functions  of the coupling constants with the  $(2\pi)^d$  
coming from their definitions, see for instance Eqs.~(\ref{defU2}) or (\ref{defZ0}).} :
\begin{equation}
\partial_t\Gamma_k{[\phi]}={1\over 2 (2\pi)^d} \tilde\partial_t
\trg\left(\ln \left((2\pi)^{2d}\Gamma_k^{(2)}[\phi]+\mathcal
R_k\right)\right)
\label{Wetterichfinalln} 
\end{equation}
where the ``time derivative'' $\tilde\partial_t$ only acts on $R_k$,
\ie $\tilde\partial_t\hat=\dot{R_k} {\partial/\partial R_k}$ and
where the trace $\trg$ now also means a momentum-integral $\int d^d\cg
qd^d\cg {q'} (2\pi)^{-d} \delta(\cg q+\cg
{q'})$. Eq.~(\ref{Wetterichfinal}) (or Eq.~(\ref{Wetterichfinalln}))
controls the evolution of $\Gamma_k$ with the running scale $k$. According to the preceding
discussion, it describes, when $k$ is lowered, how the running
effective average action is modified when more and more (low-energy)
fluctuations are integrated out.

\subsubsection{Properties}

We now give some important properties of
Eq.~(\ref{Wetterichfinal}). The reader interested in more details can
consult Ref. \cite{berges02}.

1) Eq.~(\ref{Wetterichfinal}) is {\it exact}.  It thus contains all
perturbative \cite{bonini93,morris99} and nonperturbative features of
the underlying theory: weak- or strong-coupling behaviors, tunneling
between different minima \cite{aoki02}, bound states
\cite{ellwanger98b,gies02}, topological excitations
\cite{gersdorff01}, etc.

2) While it has been derived here in the case of a one-component
scalar field theory, Eq.~(\ref{Wetterichfinal}) obviously holds for
any number of components and, more generally, for any kind of order
parameter. The extension to fermions is also trivial (see
\cite{berges02} for instance).

3) With a cut-off function $R_k(\cg q^2)$ which meets the condition
(\ref{IR}) or, more generally, with a finite limit when $\cg q^2\to 0$, 
the integral in Eq.~(\ref{Wetterichfinal}) is infrared (IR) {\it
finite} for any $k>0$. This IR finiteness is ensured by the presence
of the mass-term $R_k$ which makes the quantity $\displaystyle
{\Gamma_k^{(2)}[\phi]+R_k}$ positive for $k>0$ even {\it at} the
critical temperature. This allows to explore the low-temperature phase
even in presence of massless --- Goldstone --- modes. From the UV
side, the finiteness of the integral in Eq.~(\ref{Wetterichfinal}) is
ensured by a requirement of fast decaying behavior of $\dot R_k(\cg
q^2)$.

4) One can give a graphical representation of
Eq.~(\ref{Wetterichfinal}), see Fig.~\ref{wetterichgraphic}.  It
displays a {\it one-loop} structure.  Obviously, this one-loop
structure must not be mistaken for that of a weak-coupling
expansion. Actually, the loop involves here the {\it full} --- \ie
field-dependent --- inverse propagator $\Gamma_k^{(2)}[\phi]$ so that
the graphical representation of Fig.~\ref{wetterichgraphic} implicitly
contains {\it all} powers of the coupling constants entering in the
model. Note also that this one-loop structure automatically ensures
that all integrals over internal momenta involved in this formalism
have a one-loop structure and are thus one-dimensional. Thus they can
be easily evaluated numerically and, when some particular cut-off are
used, analytically. This radically differs from a weak-coupling
expansion which leads to multiple loop diagrams and thus, multiple
integrals. Another important feature of Eq.~(\ref{Wetterichfinal}) is
that very simple {\it ans{\"a}tze} on $\Gamma_k$ allow to
recover in a unique framework the one-loop perturbative results obtained
by standard perturbative calculations around two and four dimensions as well as 
in a large-$N$ expansion.

\begin{figure}[ht]
\parbox{2cm}{$${\partial_t \Gamma_k}=\frac12$$}
\parbox{1.5cm}{ \includegraphics[width=1.5cm]{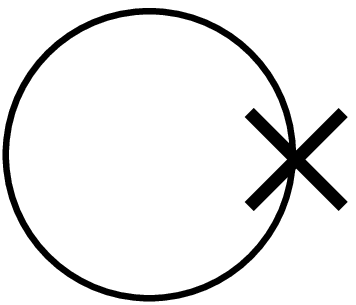}}
\caption{{A graphical representation of
Eq.~(\ref{Wetterichfinal}). The cross represents the function $\dot
R_k$ and the line the propagator
$\displaystyle{({\Gamma_k^{(2)}(\phi)+R_k}})^{-1}$.}}
\label{wetterichgraphic}
\end{figure}

Let us make a final remark.  The one-loop structure of
Eq.~(\ref{Wetterichfinal}) contrasts with the Polchinski equation
(\ref{polchinski1}) which, in addition to the loop term, involves a
tree part (see Fig.~\ref{polchinskigraphic}).  An interesting
consequence of the structure of this ``Legendre version'' of the NPRG
equation is that reparametrization invariance is preserved by the
derivative expansion when a power-law cut-off is used. This means that
with such a cut-off function, the anomalous dimension is no longer
ambiguously defined \cite{morris98b}. This would apparently select the
power-law cut-off as the best one. The situation is, in fact, more
involved since the power-law cut-off is afflicted with bad convergence
properties when used within the derivative expansion. It has appeared
that for instance, the exponential cut-off --- Eq.~(\ref{cutoffexp})
--- or the theta cut-off --- Eq.~(\ref{cutoffstep}) --- that do not
respect the reparametrization invariance of the RG equation, lead to
better results when optimization critera are used. We do not enter
into more details in these problems of reparametrization invariance 
\cite{morris94b,morris98b,morris98c,comellas98} and optimization of the results \cite{litim00,litim01,litim01b, litim01c,litim02,litim02b,canet03a,canet03b} and refer to the
literature. The main reason for this is that, as  we shall see in the following,  we
shall only deal with  pseudo-critical exponents that, being given their lack of
universality,
\ie their strong dependence with respect to the
 microscopic parameters, makes  superfluous  an optimization of the  computations.

\subsubsection{Truncations}

As it is the case for the Polchinski equation,
Eq.~(\ref{Wetterichfinal}) is too complicated to be solved
exactly. Its nonlinearities are even worse than in the Polchinski
case since it involves all powers of $\Gamma_k^{(2)}$. As a
consequence, the functional $\Gamma_k$ has to be truncated.  Different
kinds of expansions have been considered \cite{tetradis94}:

1) Field expansion where $\Gamma_k$ is expanded in powers of the order
parameter $\phi$. For a  scalar field theory, one has:
\begin{equation}
\Gamma_k[\phi]=\sum_{n=0}^{\infty} {1\over n!} \int \prod_{i=0}^{n}
d^d \cg{x_i}\ \phi(\cg{x_1})\dots\phi(\cg{x_n}) \ \Gamma_k^{(n)}(\cg{x_1},...,\cg{x_n})
\end{equation}
where $\Gamma_k^{(n)}(\cg{x_1},...,\cg{x_n})$ denotes the $n$-th
functional derivative of $\Gamma_k$.

2) Derivative expansion where $\Gamma_k$ is expanded in powers of the
derivatives of the order parameter. For  a  scalar field theory, one has:
\begin{equation}
\Gamma_k[\phi]= \int d^d{\cg x} \left( U_k(\phi)+{1\over 2} Z_k(\phi)
(\partial \phi)^2 +O(\partial^4)\right) 
\label{truncatedgammaderiv}
\end{equation}
$U_k(\phi)$ being the potential --- \ie derivative-independent --- part of $\Gamma_k$ while $Z_k(\phi)$ corresponds
 to the kinetic term.

3)  Combined derivative and field expansions  where one further expands
in Eq.~(\ref{truncatedgammaderiv}) the functions $U_k(\phi)$ and
$Z_k(\phi)$ in powers of $\phi$ around a given field configuration
$\phi_0$. Technically, this  kind of approximation allows
 to transform the functional equation (\ref{Wetterichfinal}) into a set of ordinary coupled differential equations
 for the coefficients of the expansion. In practice, it is 
 interesting to consider an
expansion around (one of) the field configuration $\phi_{0}$  that
minimizes the potential $U_k$.  For the simplest  ---  Ising --- model, this
 expansion writes: 
\begin{equation}
\begin{split}
&\Gamma_k[\phi]=\int d^d\cg x\;\Big(\\ &{1\over 2}
U_k''(\tilde\rho_0)(\rho-\tilde\rho_0)^2+{1\over 3!}
U_k^{'''}(\tilde\rho_0)(\rho-\tilde\rho_0)^3 +\cdots\\+&
\frac{1}{2}Z_k(\tilde\rho_0)(\partial\phi)^2+{1\over 2}
Z_k'(\tilde\rho_0)(\rho-\tilde\rho_0)(\partial\phi)^2+\cdots\Big)
\label{truncationmixteising}
\end{split}
\end{equation}
where $\tilde{\rho}={1\over 2}{\phi}^2$ and  $\tilde{\rho}_0={1\over 2}{\phi_0}^2$, 
 $\phi_{0}$ being the  magnetization at scale $k$.   The rational
behind this choice is that the minimum of $U_k$ is physically the
location that we want to describe the best since thermodynamical
quantities at vanishing external field are determined from the minimum
of $\Gamma_k$ at $k=0$. The relevance of such a  parametrization is confirmed by
many works showing that the convergence of the critical quantities,
when more and more powers of the field $\phi$ are added in the
truncation, is improved when compared with the same calculation
performed with an expansion of $U_k(\phi)$ and  $Z_k(\phi)$  around the $\phi=0$ configuration \cite{morris94c,aoki96}.

The choice of a good truncation is a complex problem. One has to
choose a manageable truncation that however encodes the relevant
physics. In practice, it appears that, surprisingly, even at
low-orders in powers of derivatives and fields,
Eq.~(\ref{Wetterichfinal}) provides correct qualitative features of
the RG flow. However, the precise determination of the critical
quantities requires to push the expansion to rather large orders in
the field and involves a heavy algebra
\cite{canet03a,canet03b}.

To illustrate how the technique works we now consider the simplest
case, \ie the vectorial $O(N)$ model.  The $O(N)\times O(2)$ model
relevant to frustrated magnets is technically more involved but the
procedure to derive the RG equations follows the same steps. Details
of the technicalities in this latter case will be given in
Section \ref{chapitre_on_o2}.

\subsection{The $O(N)$ model}
\label{chap_deriv_eq_on}

We present here the effective average action approach of the $O(N)$
model \cite{tetradis94,morris98d}.  We essentially follow the
presentation given, for instance, in \cite{tetradis94} with some
differences, ensuring a self-contained presentation. We use
a  truncation similar to the one  we use to deal with
frustrated magnets, \ie where $\Gamma_k$ is expanded both in
derivatives and fields.  Let us first consider the derivative
expansion of the effective average action for the $O(N)$ model at
order $\partial^2$:
\begin{equation}
\begin{split}
\Gamma_k[\vec\phi]=\int d^d \cg x \;\bigg(U_k(\rho)&+\frac {1}{2}Z_k(\rho)\:
(\partial \vec\phi)^2+\\& + {1\over 4}Y_k(\rho)\:(\partial \rho)^2
+ O(\partial^4)\bigg)
\label{tronqderive}
\end{split}
\end{equation}
where $\vec{\phi}$ is a $N$-component vector field and $\rho=
{\vec\phi}\,^2/2$.  In Eq.~(\ref{tronqderive}), $U_k(\rho)$ is the
potential --- \ie derivative-independent --- part of $\Gamma_k$ while
$Z_k(\rho)$ and $Y_k(\rho)$ correspond to kinetic terms.   These two
last terms embody the renormalization for the Goldstone and massive
fields, respectively.  Note that the term proportional to $(\partial
  \rho)^2$ is always  absent from the GLW  action used for a {\it perturbative}
    analysis in coupling constant. The reason for this is that, in
    this context,  it is
    power-counting irrelevant. On the contrary, in 
 the context of the effective
average action method, there is no perturbative expansion and, thus, no
power-counting argument works. One, however, expects  that the terms of lowest degrees
in the field (for $d>2$) and in the derivative are the most important for the long distance physics.

The case  $Z_k(\rho)=1$ and $Y_k(\rho)=0$ in Eq.~(\ref{tronqderive})
corresponds to the LPA. A nontrivial anomalous dimension is obtained
by going beyond this simplest truncation.  As said above, we use here
a truncation that mixes the derivative and field expansions. We thus
consider:
\begin{equation}
\begin{split}
&\Gamma_k[\vec\phi]=\int d^d\cg x\;\Bigg(\\ &{1\over 2}
U_k''(\tilde\rho_0)(\rho-\tilde\rho_0)^2+{1\over 3!}
U_k^{'''}(\tilde\rho_0)(\rho-\tilde\rho_0)^3 +\cdots\\+&
\frac{1}{2}Z_k(\tilde\rho_0)(\partial\vec\phi)^2+{1\over 2}
Z_k'(\tilde\rho_0)(\rho-\tilde\rho_0)(\partial\vec\phi)^2+\cdots\\
+&\frac{1}{4}Y_k(\tilde\rho_0)\:(\partial \rho)^2 +
\frac{1}{4}Y_k'(\tilde\rho_0)(\rho-\tilde\rho_0)\:(\partial
\rho)^2+\cdots\Bigg)
\label{truncationmixte}
\end{split}
\end{equation}
where $\tilde{\rho}_0={1\over 2}{\vec\phi_0}^2$ parametrizes the
$k$-dependent field configuration that minimizes $U_k$. Since our aim
here is only pedagogical and not devoted to the calculation of precise
critical quantities, we consider the following {\it ansatz} which is
limited to the smallest expression providing an nonvanishing anomalous dimension:
\begin{equation}
\Gamma_k[\vec\phi]=\int
d^d\cg x\;\Bigg(\frac12Z(\partial\vec{\phi})^2+\frac12
\tilde{u}_2(\rho-\tilde{\rho}_0)^2 \Bigg)
\label{developpement_general_action_on}
\end{equation}
where $Z\hat=Z_k(\rho_0)$ and
$\tilde{u}_2\hat=U_k''(\tilde{\rho}_0)$. This approximation looks very
much like the GLW Hamiltonian used to study perturbatively the $O(N)$
model, up to a trivial reparametrization.  There is, however, a major
difference. Here the {\it ansatz}
(\ref{developpement_general_action_on}) is {\it not studied
perturbatively} in the ${\vec\phi}\,^4$ coupling constant
$\tilde{u}_2$. It is to be inserted in the {\it exact} RG equation
(\ref{Wetterichfinal}).

Let us now establish the RG equations for the coupling constants
entering in Eq.~(\ref{developpement_general_action_on}). The
calculation proceeds in four steps:
 
{\it i)} We first define the running coupling constants
$\tilde{\rho}_0$, $\tilde{u}_2$ and $Z$ from functional derivatives of
the {\it ansatz} of $\Gamma_k$,
Eq.~(\ref{developpement_general_action_on}).  This is analogous to
imposing renormalization prescriptions for the renormalized coupling
constants in usual perturbative calculations.  As in this case, the
coupling constants are defined as (combinations of) functional
derivatives of $\Gamma_k$ --- the ``vertex functions'' --- taken in a
specific field-configuration of the model.  However, contrarily to the
perturbative approach which is generally performed in the
high-temperature phase and thus, around a zero field configuration, we
perform this expansion around a nontrivial  running field configuration
$\vec\phi_0$.

{\it ii)} We then apply the operator $\partial_t$ on these
definitions. This is implemented by the use of the evolution equation (\ref{Wetterichfinal})
or (\ref{Wetterichfinalln}). The flow equations for the coupling
constants are then expressed as traces of products  of vertex functions
that are evaluated from the {\it ansatz}
Eq.~(\ref{developpement_general_action_on}).

{\it iii)} The flow equations involve integrals over the internal
momentum.  It is convenient to express these integrals in terms of
dimensionless functions, known as threshold functions
\cite{tetradis94}. The properties of these functions are such that
they make explicit the phenomenon of decoupling of massive modes, see
below.

{\it iv)} Also, as usual, one introduces dimensionless renormalized
quantities to study the scale invariant solutions of the RG equations.

\subsubsection{Definition of the coupling constants}

Let us first choose one of the uniform field configurations that
minimize the effective average action $\Gamma_k$:
\begin{equation}
\vec{\phi}^{\,\Min}(\cg x)=
\begin{pmatrix}
\phi_0\\ 0\\ \vdots\\ 0\\
\end{pmatrix},
\label{defmin}
\end{equation}
or equivalently:
\begin{equation}
{\phi}^{\Min}_i(\cg q)=(2\pi)^d \phi_0\ \delta_{i1}\delta(\cg{q})\
\label{defmin2}
\end{equation}
where $\phi_0=(2\tilde{\rho}_0)^{1/2}$ is a $k$-dependent
quantity. Due to the $O(N)$ symmetry of $\Gamma_k$, which is preserved
at any $t$ by the RG flow Eq.~(\ref{Wetterichfinal}), the choice of a
particular direction for $\vec{\phi}^{\,\Min}$ is irrelevant and thus
does not affect the RG equations.

Let us now define the coupling constants. To do this we introduce the
notation:
\begin{equation}
\Gamma_{k\;\{\alpha_1,\cg{p_1}\},\cdots,\{\alpha_n,\cg{p_n}\}}^{(n)}=
{\delta^{n} \Gamma_k[\phi]\over \delta \phi_{\alpha_1}(\cg{p_1})
...\delta \phi_{\alpha_n}(\cg{p_n})}\ .
\label{defderivgamma}
\end{equation}

As said above, $\tilde{\rho}_0$ specifies the position of the ---
running --- minimum of $\Gamma_k$. It is implicitly defined by:
\begin{equation}
\Gamma_{k\;\{\alpha,\cg{0}\}}^{(1)} {\bigg |}_{{\Min}}=0\; 
\label{defrho0}
\end{equation}
where the notation ``Min'' refers to the configuration given in
Eq.~(\ref{defmin}). Because of our particular choice of
$\vec\phi^{\Min}$ the previous equality is trivially satisfied for
$\alpha=2,\dots,N$ and we shall consider only the case $\alpha=1$ in
the following.

The coupling constant $\tilde{u}_2$ is defined along the same line as:
\begin{equation}
\tilde{u}_2=\frac{(2\pi)^d}{2\tilde{\rho}_0 \delta(\cg{0})} \;
\Gamma_{k\;\{1,\cg{0}\},\{1,\cg{0}\}}^{(2)}\bigg |_{\Min}\;.
\label{defU2}
\end{equation}
Finally, the $k-$dependent field renormalization $Z$ is obtained by
considering the  term quadratic in momentum of a momentum dependent
configuration:
\begin{align}
Z&=\frac{(2\pi)^d}{\delta(\cg{0})}\lim_{\cg{p^2}\to
0}\frac{d\;}{d\cg{p}^2}\left(
\Gamma_{k\;\{2,\cg{p}\},\{2,-\cg{p}\}}^{(2)}\bigg |_{\Min}\right)\;.
\label{defZ0}  
\end{align}

In this last equation, the index 2 specifies a direction orthogonal to
that defined by the minimum (see Eq.~(\ref{defmin})). Note that one
could have considered any of the $N-1$ directions orthogonal to that
defined by the minimum.  Note finally that the ${\delta(\cg{0})}$ term
appearing in Eqs.~(\ref{defU2}) and (\ref{defZ0}) is the volume of the
system and is present here since $\Gamma_k$ is an extensive quantity
while the coupling constants are not.

\subsubsection{The $t$-derivation}
\label{chapter_t_derivative}

The flow equations for the coupling constants $\tilde{\rho}_0$,
$\tilde{u}_2$ and $Z$ are obtained by derivating, with respect to $t$,
the previous definitions Eqs.~(\ref{defrho0}), (\ref{defU2}) and
(\ref{defZ0}).

Let us start by $\tilde{\rho}_0$. One has to take care of the
$t$-dependence of both $\Gamma_{k\;\{1,\cg{0}\}}^{(1)}$ and its
argument, the configuration $\vec{\phi}^{\,\Min}$ ---  Eq.~(\ref{defmin}) ---  which has a nontrivial 
$t$-dependence through that of $\phi_0$:
\begin{equation}
\begin{split}
&\partial_t \left(\Gamma_{k\;\{\alpha,\cg{0}\}}^{(1)} {\bigg
|_{\Min}}\right)=\\&=\partial_t \Gamma_{k\;\{\alpha,\cg{0}\}}^{(1)}{\bigg|_{\Min}}+
\Gamma_ {k\;\{\alpha,\cg{0}\},\{1,\cg{0}\}}^{(2)}
{\bigg|_{\Min}}\frac{(2\pi)^d
\partial_t{\tilde{\rho}_0}}{\sqrt{2\tilde{\rho}_0}}\\&=0\ .
\end{split}
\end{equation}

The RG flow for $\tilde{\rho}_0$ follows from this equation, taken for
$\alpha=1$:
\begin{equation}
\partial_t{\tilde{\rho}_0}=-\frac{\sqrt{2\tilde{\rho}_0}}{(2\pi)^d\
\Gamma_ {k\;\{1,\cg{0}\},\{1,\cg{0}\}}^{(2)}}\,\partial_t
\Gamma_{k\;\{1,\cg{0}\}}^{(1)} {\bigg |_{\Min}}\ .
\label{flot_rho_interm}
\end{equation}

In the same way, one obtains:
\begin{equation}
\begin{split}
\partial_t{\tilde{u}_2}&=
\frac{(2\pi)^d}{2\tilde{\rho}_0\delta(\cg{0})}\,\partial_t
\Gamma_{k\;\{1,\cg{0}\},\{1,\cg{0}\}}^{(2)}\bigg
|_{\Min}+\\
&+\frac{(2\pi)^d
\partial_t{\tilde{\rho}_0}}{2\tilde{\rho}_0\delta(\cg{0})}
\bigg(-{1\over  {\tilde{\rho}_0}} \Gamma_{k\;\{1,\cg{0}\},\{1,\cg{0}\}
}^{(2)}\bigg |_{\Min}+\\
&+\frac{(2\pi)^d}{\sqrt{2
\tilde{\rho}_0}}\,{\Gamma_{k\;
\{1,\cg{0}\},\{1,\cg{0}\},\{1,\cg{0}\} }^{(3)}}\bigg |_{\Min} \bigg)\;
\label{evol_U2_interm}
\end{split}
\end{equation}
and:
\begin{equation}
\begin{split}
\partial_t{Z}=&\frac{(2\pi)^d}{\delta(\cg{0})}\lim_{\cg{p^2}\to
0}\frac{d\;}{d\cg{p}^2} \bigg(\partial_t
\Gamma_{k\;\{2,\cg{p}\},\{2,\cg{-p}\}}^{(2)}\bigg
|_{\Min}+\\
&+\frac{(2\pi)^d\partial_t{\tilde{\rho}_0}}{\sqrt{2\tilde{\rho}_0}}\,
\Gamma_{k\;\{2,\cg{p}\},\{2,-\cg{p}\},\{1,\cg{0}\}}^{(3)}\bigg
|_{\Min} \bigg)\;.
\label{evol_Z_interm}
\end{split}
\end{equation}

The RG flows for the coupling constants $\tilde{\rho}_0$, $\tilde{u}_2$
and $Z$ involve successive functional derivatives of $\partial_t
\Gamma_k$ with respect to different $\phi_{i}(\cg{q_j})$. These
quantities are easily obtained from Eq.~(\ref{Wetterichfinalln}). Let
us take its derivative with respect to $\phi_{i_1}(\cg{q_1})$. Using:
\begin{equation}
\begin{split}
\frac{\delta\;}{\delta\phi_{i_1}(\cg{q_1})}
\ln\Big((2\pi)^{2d}\Gamma_k^{(2)}&+ \mathcal
R_k\Big)=\\
&=(2\pi)^{2d}\Gamma^{(3)}_{k\;\{i_1,\cg{q_1}\}}.  \mathcal P_r,
\label{evol_G1_eq}
\end{split}
\end{equation}
where we have introduced the notation:
\begin{equation}
\mathcal P_r=\left((2\pi)^{2d}
\Gamma_k^{(2)} +\mathcal R_k\right )^{-1}
\end{equation}
one obtains:
\begin{equation}
\partial_t\Gamma_{k\;\{i_1,\cg{q_1}\}}^{(1)}= \frac{(2\pi)^d}2\, \tilde\partial_t
\tr \left\{\Gamma^{(3)} _{k\;\{i_1,\cg{q_1}\}}\,.\, \mathcal P_r\right\}
\label{evol_G1}
\end{equation}
for the one-point vertex function. Note that, in the right hand side
of the preceding expression, we have only specified the external
indices $\{i_1,\cg{q_1}\}$ and omitted the integrated (or summed over)
variables. The dot is here to remind that these integrations and
summations have to be performed.  The equation (\ref{evol_G1}) admits
a graphical representation:
\begin{equation}
\parbox{3.5cm}{$$\partial_t\Gamma_{k\;\{i_1,\cg{q_1}\}}^{(1)}=\frac {(2\pi)^d}2\,
\tilde\partial_t$$}
\parbox{2cm}{
\includegraphics[width=1.8cm]{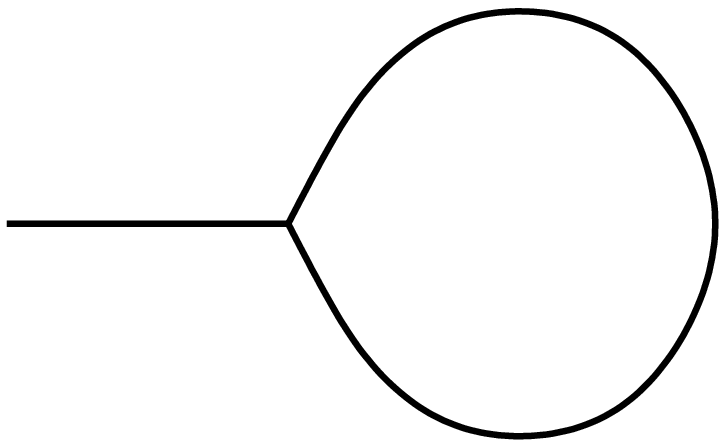}}\ .
\end{equation}

In this representation, the external leg implicitly carries an index
of internal symmetry $i_1$ and a momentum $\cg{q_1}$. Now taking the
derivative of Eq.~(\ref{evol_G1}) with respect to $\phi_{i_2}(\cg{q_2})$
and using:
\begin{equation}
\frac{\delta\;}{\delta\phi_i(\cg{q})}
\mathcal P_r=-\mathcal P_r\,.\,
\Gamma^{(3)}_{k\;\{i,\cg{q}\}}\,.\,
\mathcal P_r,
\end{equation}
one obtains:
\begin{equation}
\begin{split}
\partial_t &\Gamma_{k\;\{i_1,\cg{q_1}\},\{i_2,\cg{q_2}\}}^{(2)}=
\frac{(2\pi)^d}2\, \tilde\partial_t \tr \Big\{
\Gamma^{(4)}_{k\;\{i_1,\cg{q_1}\},\{i_1,\cg{q_2}\}}\, .
\, \mathcal P_r-\\&-\Gamma^{(3)}_{k\;\{i_1,\cg{q_1}\}}\, 
.\,  \mathcal P_r\,  .\, \Gamma^{(3)}_{k\;\{i_2,\cg{q_2}\}}\, 
.\,  \mathcal P_r\Big\}
\end{split}
\label{evol_G2}
\end{equation}
which can be  graphically represented by: 
\begin{equation}
\parbox{4.3cm}{$$\partial_t
\Gamma_{k\;\{i_1,\cg{q_1}\},\{i_2,\cg{q_2}\}}^{(2)}=\frac {(2\pi)^d}2\,
\tilde\partial_t\bigg($$}
\parbox{1.3cm}{\rule[-6mm]{1mm}{0mm}
\includegraphics[width=1.cm]{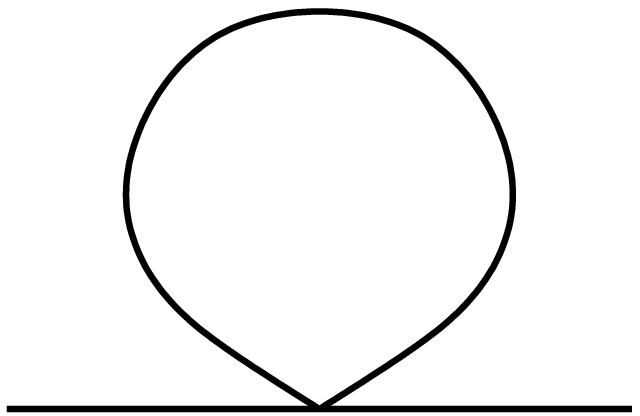} }
\parbox{5mm}{$\;-\;$}
\parbox{1.4cm}{ \includegraphics[width=1.6cm]{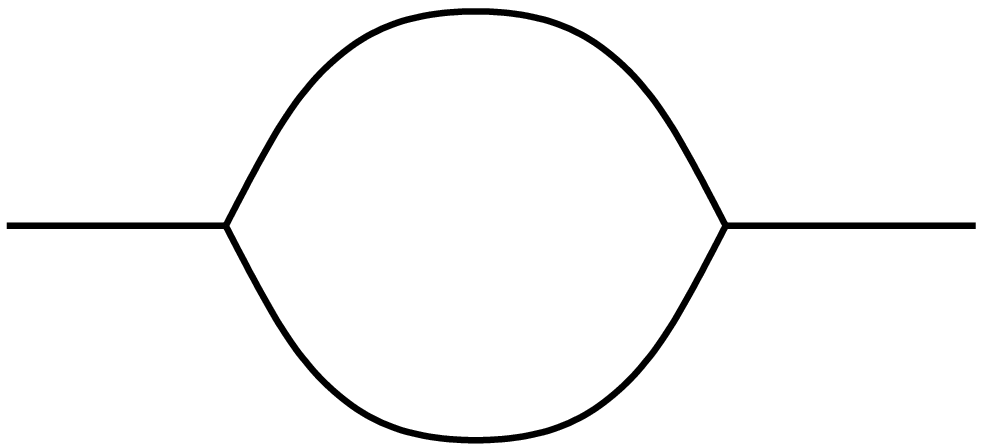}}
\parbox{5mm}{$$\bigg)$$}
\end{equation}

\subsubsection{The renormalization group  flow}
\label{chapitre_renormalisation_on}

We now explicitly write the flow equations for the coupling
constants. This requires to know the vertex functions taken at the
minimum $\Gamma_{k\;\{\alpha_1,\cg{p_1}\},\{\alpha_2,\cg{p_2}\},...,
\{\alpha_n,\cg{p_n}\}}^{(n)}\Big |_{\Min}$ appearing in
Eqs.~(\ref{flot_rho_interm}), (\ref{evol_U2_interm}) and
(\ref{evol_Z_interm}). To evaluate them, one uses the truncation
Eq.~(\ref{developpement_general_action_on}). One obtains:
\begin{widetext}
\begin{equation}
\left\{
\begin{aligned}
&\Gamma^{(1)}_{k\;\{i_1,\cg{q_1}\}}\Big|_{\Min}=0\\
&\Gamma^{(2)}_{k\;\{i_1,\cg{q_1}\},\{i_2,\cg{q_2}\}}\Big|_{\Min}
=\left( Z \cg{q_1}^2 \delta_{i_1i_2}+2 \tilde\rho_0 \tilde u_2 \delta_{i_11}
\delta_{i_2 1}\right)\frac{\delta(\cg{q_1}+\cg{q_2})}{(2\pi)^d}\\
&\Gamma^{(3)}_{k\;\{i_1,\cg{q_1}\},\{i_2,\cg{q_2}\},
\{i_3,\cg{q_3}\}}\Big|_{\Min}=
\sqrt{2 \tilde\rho_0} \tilde u_2 ( \delta_{i_1i_2} \delta_{i_31} +\delta_{i_2i_3}
\delta_{i_11} +\delta_{i_3i_1} \delta_{i_21}
)\frac {\delta(\cg{q_1}+\cg{q_2}+\cg{q_3})}{(2\pi)^{2d}}\\
&\Gamma^{(4)}_{k\;\{i_1,\cg{q_1}\},\{i_2,\cg{q_2}\}, \{i_3,\cg{q_3}\},
\{i_4,\cg{q_4}\}}\Big|_{\Min}=
\tilde u_2 ( \delta_{i_1i_2} \delta_{i_3i_4} +\delta_{i_1i_3}
\delta_{i_2i_4}+\delta_{i_1i_4} \delta_{i_2i_3}
)\frac {\delta(\cg{q_1}+\cg{q_2}+\cg{q_3}+\cg{q_4})}{(2\pi)^{3d}}\ .
 \end{aligned}
\right.
\label{vertex_phi_4}
\end{equation}
\end{widetext}

In this last set of equations,
$\Gamma^{(2)}_{k\;\{i_1,\cg{q_1}\},\{i_2,\cg{q_2}\}}\Big|_{\Min}$ is
of particular interest since its inverse provides the propagator
$\mathcal P_r$ at scale $k$ and, thus, the spectrum of excitations of the
theory, {\it at this scale}.  We easily get from
Eq.~(\ref{vertex_phi_4}):
\begin{equation}
\begin{split}
&\mathcal P_{r\{i_1,\cg{q_1}\} ,\{i_2,\cg{q_2}\}}\Big|_\Min =(2\pi)^d
\delta(\cg q_1+\cg q_2)\ . \\ &\ .
\begin{cases} 
\displaystyle
\frac 1 {Z \cg{q_1}^2+R_k(\cg{q_1}^2)}& \text{if }i_1=i_2=1 \\
\\
\displaystyle \frac 1 {Z \cg{q_1}^2+R_k(\cg{q_1}^2)+2 \tilde\rho_0 \tilde u_2}&
\text{if }i_1=i_2\neq 1\\
\\
0&\text{if }i_1\ne i_2\ 
\end{cases}
\end{split}
\label{massmatrix}
\end{equation}
where $R_k(\cg{q_1}^2)$ is the contribution of the regulating term
(\ref{regularisation_R_droit}).

It is clear on the expression (\ref{massmatrix}) that the $N\times N$
matrix $\mathcal P_{r\{i_1,\cg{q_1}\} ,\{i_2,\cg{q_2}\}}\Big|_\Min$ is
diagonal. This holds independently of the kind of truncation used.
The spectrum of excitations around the minimum, at scale $k$, is thus
directly red on Eq.~(\ref{massmatrix}).  We find --- up to the $R_k$ term
--- one massive mode of squared mass $2\tilde\rho_0 \tilde u_2$ in the
longitudinal direction and $N-1$ massless modes in the directions
orthogonal to the magnetization $\vec{\phi}_0$. The deformations of the vector $\vec{\phi}$ associated  to
 these modes are
represented in Fig.~\ref{spectrumON}.}
\begin{figure}[tbp]
\centering \makebox[\linewidth]{ \subfigure[massive excitation]{
\label{mode_massif_on}
\includegraphics[width=.45\linewidth,origin=tl]{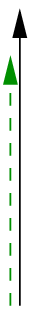}}\hfill%
\subfigure[$N-1$ ``Goldstone'' excitations]{
\label{mode_goldstone_on}
\includegraphics[width=.45\linewidth,origin=tl]{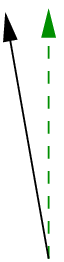}}\hfill%
}
\caption{Schematic description of the deformations  of the vector $\vec{\phi}$
associated with the proper excitations of the $O(N)$ model: dotted
arrows display the configuration chosen at the minimum of $\Gamma_k$
and plain arrows display the relevant deformations: a) massive
singlet, b) ``Goldstone'' $N-1$-uplet.}
\label{spectrumON}
\end{figure}

It is important to keep in mind that this spectrum corresponds to
effective masses --- at scale $k$ --- for which only high-momentum
fluctuations --- higher than $k$ --- have been considered. It is only
in the limit $k\to 0$ that one retrieves the physical spectrum.  In
particular, we stress that a qualitative change in the spectrum can
occur when $k$ is varied. For instance, the following situation can
happen: for large $k$, the minimum of $U_k$ is nonvanishing so that,
at this scale, the system behaves as if it was in its broken
phase. However, when $k$ is decreased, the minimum moves toward zero
and eventually vanishes for some $k>0$. Thus, while the system, for
$k=\Lambda$, looks as if it was in its broken phase, it is actually,
\ie for $k=0$, in the high-temperature phase.  This is what happens
when the temperature lies between the true critical and the mean-field
critical temperature. Another subtlety is that, in order to analyze
the critical behavior, we have to consider the dimensionless
renormalized quantities. Again, one has to be careful about the
conclusions deduced from the behavior of these quantities. For
instance, the dimensionless renormalized position of the minimum at
$k=0$ can be nonvanishing even at the critical temperature whereas the
``true'' magnetization is, of course, vanishing at $T_c$. This is
possible since the dimensionful quantities are the products of their
dimensionless counterparts and of positive powers of $k$.

Using Eqs.~(\ref{flot_rho_interm}), (\ref{evol_U2_interm}), (\ref{evol_Z_interm}), (\ref{evol_G1}), (\ref{evol_G2}) 
and (\ref{vertex_phi_4}), one
obtains the flow for $\tilde{\rho}_0$:
\begin{equation}
\begin{split}
\partial_t{\tilde\rho_0}=-{1\over 2}\,
\tilde\partial_t\int &\frac{d^d\cg{q}}{(2\pi)^d}\left(\frac{N-1}{Z
{\cg q}^2+R_k(\cg{q}^2)}+\right.\\
+&\left.\frac{3}{Z
\cg{q}^2+R_k(\cg{q}^2)+2\tilde\rho_0\tilde u_2}\right)\ ,
\end{split}
\end{equation}
for $\tilde{u}_2$:
\begin{equation}
\begin{split}
\partial_t{\tilde u_2}=-{\tilde u_2^2\over 2}\,
\tilde\partial_t\int &\frac{d^d\cg{q}}{(2\pi)^d}\left(\frac{N-1}{Z
{(\cg q}^2+R_k(\cg{q}^2))^2}+\right.\\
+&\left.\frac{9}{(Z
\cg{q}^2+R_k(\cg{q}^2)+2\tilde\rho_0\tilde u_2)^2}\right)\ ,
\end{split}
\end{equation}
and for the field renormalization $Z$:
\begin{equation}
\begin{split}
&\partial_t Z=-2\tilde{\rho}_0 \tilde{u}_2^2\,\lim_{\cg{p^2}\to
0}\frac{d\;}{d\cg{p}^2}
\Bigg(\tilde\partial_t\int \frac{d^d\cg{q}}{(2\pi)^d}\frac1{Z
\cg{q}^2+R_k(\cg{q}^2)}\ .\  \\&\ .\ \frac1{Z
(\cg{p}+\cg{q})^2+R_k((\cg{p}+\cg{q})^2)+2\tilde{\rho}_0
\tilde{u}_2}\Bigg)\;.
\end{split}
\end{equation}
The search for fixed point requires to introduce dimensionless
renormalized coupling constants. We define:
\begin{equation}
\left\{
\begin{aligned}
\rho_0&=Z k^{2-d} \tilde{\rho}_0\\
u_2&=Z^{-2}k^{d-4}\tilde{u}_2\;.
\end{aligned}
\right.
\label{kappa}
\end{equation}
These changes of variables are deviced so that $k$ and $Z$ disappear
from the flow equations of the renormalized dimensionless quantities.

The corresponding flow equations thus write:
\begin{equation}
\left\{
\begin{aligned}
\partial_t{\rho_0}&=-(d-2+\eta)\rho_0+\\
&+2v_d(N-1)l_1^d(0)+6v_d \; l_1^d(2u_2\rho_0)\\
\partial_t{u_2}&=(d-4+2\eta)u_2+\\
&+2v_d(N-1) u_2^2\; l_2^d(0)+18v_d
\; u_2^2\; l_2^d(2 u_2\rho_0)\ 
\end{aligned}
\right.
\label{flotu2rho}
\end{equation}
that depend  on $Z$ only through $\eta$, the  running ``anomalous dimension'':
\begin{equation}
\eta=-\partial_t \log Z\ .
\end{equation}
In our truncation, it is given by:
\begin{equation}
\eta=\frac{16v_d}{d} u_2^2 \rho_0\; m_{2,2}^d(2u_2\rho_0)\ .
\label{flot_eta_phi4}
\end{equation}
The usual anomalous dimension is given by the fixed point value of
Eq.~(\ref{flot_eta_phi4}). In Eqs.~(\ref{flotu2rho}) and
(\ref{flot_eta_phi4}), we have introduced the --- dimensionless ---
threshold functions $l_n^d$ and $m_{2,2}^d$:
\begin{equation}
\left\{
\begin{aligned}
l_n^d(w)&=-\frac{{Z}^n\, k^{-d+2n}}{4v_d}\ \tilde\partial_t\int
\frac{d^d\cg{q}}{(2\pi)^d}\\
. &\frac{1}{(Z \cg{q}^2+R_k(\cg{q}^2)+Z
k^2w)^n}
\\
m_{2,2}^d(w)&=-\frac{d\,{Z}^2\, k^{6-d}}{8v_d}\ \lim_{\cg{p^2}\to
0}\frac{d\;}{d\cg{p}^2}\Bigg( \tilde\partial_t \int
\frac{d^d\cg{q}}{(2\pi)^d}\ .\ \\ &\ .\ \frac{1}{Z
\cg{q}^2+R_k(\cg{q}^2)}\ .\ \\ &\ .\ \frac{1}{Z
(\cg{p}+\cg{q})^2+R_k((\cg{p+q})^2)+Z k^2w}\Bigg)
\end{aligned}
\right.
\end{equation}
with $v_d^{-1}=2^{d+1} \pi^{d/2} \Gamma(d/2)$. Some properties of
these threshold functions are provided in the Appendix
\ref{annexe_threshold}.  We concentrate here on the main physical
aspects of the threshold functions:

1) Note first that the arguments of the threshold functions entering
in Eqs.~(\ref{flotu2rho}) and (\ref{flot_eta_phi4}) are either 0 or
$2u_2\rho_0$ that are --- up to $R_k(\cg q^2)$ --- the dimensionless
renormalized square masses associated with the excitations around the
minimum.

2) The threshold functions $l_n^d(w)$ and $m_{2,2}^d(w)$ decrease as
power-laws when their arguments increase:
\begin{equation}
\left\{
\begin{aligned}
&l_n^d(w)\propto w^{-n-1}\\
&m_{2,2}^d(w)\propto w^{-2}\ 
\label{comportement_asymptotique}
\end{aligned}
\right.
\end{equation}
for $w\gg 1$. The RG equation (\ref{Wetterichfinal}) makes thus
explicit the phenomenon of decoupling of massive modes: if the 
renormalized  square mass $M_k^2$ --- here $2 u_2 \rho_0$  --- of a massive mode increases when the
scale $k$ is lowered, the contribution of this mode to the flow
becomes negligible below a scale $k_c$ defined by $M_{k_c}\sim 1$.

3) The threshold functions are nonpolynomial functions of their
arguments. Thus the flow equations (\ref{flotu2rho}) and
(\ref{flot_eta_phi4}) are nonperturbative with respect to the coupling
constant $u_2$ as well as   to  $1/{\rho_0}$ which, as we shall see, is proportional to the coupling constant --- 
the temperature $T$ --- that parametrizes the perturbative expansion of  NL$\sigma$ model.

As we now show, the effective average action approach allows to
recover the perturbative results obtained at low-temperature around
$d=2$, at weak-coupling around $d=4$ and in a $1/N$ expansion.

\subsubsection{The weak-coupling expansion around $d=4$}
\label{chapitre_weak_coupling_ON}

Just below four dimensions, the nontrivial fixed point governing the
phase transition of the $O(N)$ model is very close to the Gaussian
fixed point. This justifies to expand the RG flow equations
(\ref{flotu2rho}) and (\ref{flot_eta_phi4}) both in the coupling
constant $u_2$ and in $\epsilon=4-d$. At lowest order, the function
$\eta$ is vanishing. Since $\rho_0$ remains finite, the quantity $2
u_2\rho_0$ is of order $\epsilon$ and one can perform a small mass
expansion.  The flows of the coupling $u_2$ and of $\rho_0$ are
obtained using $l_n^d(\omega)\simeq l_n^d(0)-n \omega l_{n+1}^d(0)$
for $\omega \ll 1$ and $l_2^{4}(0)=1$. This leads to:
\begin{equation} \left\{
\begin{aligned}
&\partial_t{\rho_0}=-(2-\epsilon)\rho_0+\frac{(N+2)}{16\pi^2}\ l_1^{4}(0)-
\frac3{8\pi^2} u_2\rho_0 \\ &\partial_t{u}_2=-\epsilon
u_2+\frac{N+8}{16\pi^2} u_2^2\ .
\end{aligned}
\right.
\label{eqaround4}
\end{equation}
At leading order, the roots of these equations are the gaussian fixed point ---
$u_2^\star=0$ and $\rho_0^\star=(N+2)l_1^{4}(0)/{32\pi^2}$ --- and a nontrivial
fixed point obtained for $u_2^\star={16 \pi^2 \epsilon}/{(N+8)}$ and
$\rho_0^\star=(N+2)l_1^{4}(0)/{32\pi^2}$. One easily deduces the critical
exponent $\nu$ from Eqs.~(\ref{eqaround4}), linearized around the
nontrivial fixed point:
\begin{equation}
\nu=\frac12+\frac\epsilon4\;\frac{N+2}{N+8}\ .
\end{equation}
It coincides with the one-loop expression obtained within a
perturbative weak-coupling approach of the corresponding GLW model
in $d=4-\epsilon$.

\subsubsection{The low-temperature expansion around $d=2$}
\label{chapitre_NLsigma_ON}

 Let us now show that Eq.~(\ref{Wetterichfinal}) also allows to
recover the one-loop results obtained around two dimensions in a
low-temperature expansion of the NL$\sigma$ model. We first make
contact between the parameters --- essentially the temperature --- of
the $O(N)$ NL$\sigma$ model and those of the effective average action
(\ref{developpement_general_action_on}).

The partition function of the $O(N)$ NL$\sigma$ model is given by:
\begin{equation}
\mathcal Z=\int \mathcal{D} \vec{\phi}\;
\delta\left({\vec{\phi}}^{\,2}-1\right)\,\exp\left( -\frac{ 1}{ 2T}{\int}
d^d\cg x\ (\partial\vec{\phi})^2\right)\,.
\end{equation} 

Let us replace the delta-constraint by a soft constraint. Using the
field redefinition $\vec{\phi}\to\vec{\phi}\sqrt{T}$ one obtains:
\begin{equation}
\mathcal Z=\int \mathcal{D} \vec{\phi}\;\exp\left[
-\frac{1}{2}{\int} 
d^d\cg x\left( (\partial\vec{\phi})^2 - \lambda(\vec{\phi}^{\,2}
T-1)^2\right)\right]
\end{equation} 
where the delta-constraint is recovered when $\lambda\to\infty$.
Comparing this expression with the {\it ansatz}
(\ref{developpement_general_action_on}) and using the relation
({\ref{kappa})  one obtains the relation:
\begin{equation}
T={1\over 2\rho_0}\ .
\label{relationTkappa}
\end{equation}
As a consequence, the low-temperature one-loop perturbative results
can be recovered from a $\displaystyle{1/ \rho_0}$ expansion. In fact,
since the dimensionless renormalized mass of the massive modes is
given by $2u_2\rho_0$, one has to perform a large-mass
expansion. Physically, this corresponds to the known fact that, around
two dimensions, the longitudinal modes of the $O(N)$ linear model are
frozen and only the --- transverse --- ``Goldstone'' fluctuations are
activated. This phenomenon corresponds to the decoupling of massive
modes. Technically, this is realized through the behavior of the
threshold functions. As already stated, the threshold functions
decrease as power-laws for large arguments,
Eq.~(\ref{comportement_asymptotique}). As a consequence, in the flow
equations (\ref{flotu2rho}), the contribution of the massive mode ---
which is proportional to $l_n^d(2\rho_0\lambda)$ --- is subdominant
compared with the contribution of the Goldstone modes --- proportional
to $l_n^d(0)$. Now, using the large-mass expansion
$m_{2,2}^{d=2}(\omega) = \omega^{-2}+O(\omega^{-3})$, one gets from
Eq.~(\ref{flot_eta_phi4}):
\begin{equation}
\eta\simeq \displaystyle{1\over 4\pi \rho_0}\ .
\label{anomalous}
\end{equation}
Using this expression of the anomalous dimension and the 
value $l_1^2(0)=1$, one gets:
\begin{equation}
\left\{
\begin{aligned}
&\partial_t{\rho_0}\simeq-\epsilon\rho_0+\frac{N-2}{4\pi}\\
&\partial_t{u}_2\simeq-2u_2+\frac{N-1}{4\pi} u_2^2\, l_2^2(0)\ .
\end{aligned}
\right.
\label{flotu2rhod-2}
\end{equation}

The flow equation for $\rho_0$ coincides exactly with the result
obtained in the one-loop low-temperature expansion of the NL$\sigma$
model for the temperature --- which is given by
Eq.~(\ref{relationTkappa}). The fixed point coordinates are given by:
$\rho_0^\star=(N-2)/{(4\pi \epsilon)}$ and
$u_2^\star={8\pi}/((N-1)l_2^2(0))$. This leads to the critical
exponents:
\begin{equation}
\label{expd=2}
\left\{
\begin{aligned}
&\eta=\frac{\epsilon}{N-2}\\ &\nu=\frac1\epsilon
\end{aligned}
\right.
\end{equation}
which identify with those given by the low-temperature perturbative
expansion of the NL$\sigma$ model at one-loop \cite{zinnjustin89}.

Note that the perturbative $\beta$ function for $u_2$ ---
Eq.~(\ref{eqaround4}) --- and for $\rho$ --- Eq.~(\ref{flotu2rhod-2})
--- are {\it universal}, \ie independent of the cut-off function
$R_k(\cg q)$. Indeed, these $\beta$ functions only depend on the values of the
threshold functions $l_n^{2n}(\omega)$ at $\omega=0$ and on $m_{2,2}^{d=2}(\omega)$
at large $\omega$ that, as shown in Appendix
\ref{annexe_universel}, do not depend on the cut-off function $R_k(\cg
q)$. The matching with the perturbative results obtained around $d=2$
and $d=4$ is a very important feature of the effective average action
method. First, it allows to interpolate smoothly between two and four
dimensions in a unified framework.  Second, it suggests that it is possible to reliably explore the behavior
of the system in any dimension $d$ and, in particular in $d=3$, see
below.

\subsubsection{The large-$N$ analysis}
\label{chapitre_Large_N_ON}

The flow equations (\ref{flotu2rho}) and (\ref{flot_eta_phi4}) can
easily be expanded in the large-$N$ limit. The leading contributions
come from the Goldstone modes, which appear with a multiplicative
factor $N$. The $\beta$ functions then read:
\begin{equation}
\label{flot_constantes_largeN}
\left\{
\begin{array}{l}
\partial_t{\rho_0}=(2-d)\rho_0+2Nv_d\;  l_1^d(0)\\
\\
\partial_t{u}_2=(d-4)u_2+2N\,v_d u_2^2\; l_2^d(0)
\end{array}
\right. 
\end{equation}
where we have anticipated that the anomalous dimension vanishes at
leading order, see below. The fixed point solutions are easily found
to be: $\rho_0^\star=2 N\ v_d l_1^d(0)/(d-2)$ and
$u_2^\star=(d-4)/(2v_dN l_2^d(0))$. From these results, we check that
the anomalous dimension behaves like $1/N$ and thus gives subdominant
corrections to the $\beta$ functions. We can, finally, compute the
critical exponents by diagonalizing the stability matrix. We then find
$\nu=1/(d-2)$, in agreement with the standard leading order result of
the $1/N$ expansion.

We now check that the effective average action method provides reliable
results in three dimensions.

\subsubsection{The critical exponents in three dimensions}

One of the main  interest of the effective average action method is its
ability to tackle with the physics in a nonperturbative  regime,
precisely when there is no small parameter, as it is the case in three
dimensions. We provide, in Table \ref{exposants_critiques_On}, the
critical exponents $\nu$ and $\eta$ obtained with this method, as
functions of the order of the derivative and field expansions of
$U_k(\rho)$, $Z_k(\rho)$ and $Y_k(\rho)$ (see
Eq.~(\ref{truncationmixte})). We have also included the results of
high-order perturbative expansion for comparison. The exponent $\nu$
is rather poorly determined with our simple truncation
(\ref{developpement_general_action_on}). However, the precision
improves rapidly when more terms are added to the {\it ansatz} for $\Gamma_k$. For
the best truncation, $\nu$ is determined at less than one percent
compared with the world best estimates.
\begin{table}[htbp]
\centering
\begin{tabular}{|l|l|l|l|l|}
\hline N&\multicolumn{2}{c|}{$\nu$}&\multicolumn{2}{c|}{$\eta$}\\
\hline \hline
1&0.520$^{a)}$&0.6290(25)$^{g)}$&0.057$^{a)}$&0.036(5)$^{g)}$\\
 &0.688$^{b)}$&0.6304(13)$^{h)}$&0.038$^{b)}$&0.0335(15)$^{h)}$\\
 &0.635$^{c)}$&&0.056$^{c)}$&\\ 
 &0.635$^{d)}$&&0.058$^{d)}$&\\ 
 &0.6307$^{e)}$&&0.0467$^{e)}$&\\
 &0.632$^{f)}$&&0.033$^{f)}$&\\
 \hline
2&0.613$^{a)}$&0.6680(35)$^{g)}$&0.058$^{a)}$&0.038(5)$^{g)}$\\
 &0.722$^{b)}$&0.6703(15)$^{h)}$&0.038$^{b)}$&0.0354(25)$^{h)}$\\
 &0.683$^{c)}$&&0.054$^{c)}$&\\ 
 &0.666$^{d)}$&&0.055$^{d)}$&\\ 
 &0.666$^{e)}$&&0.049$^{e)}$&\\ 
\hline
3&0.699$^{a)}$&0.7045(55)$^{g)}$&0.051$^{a)}$&0.0375(45)$^{g)}$\\
 &0.756$^{b)}$&0.7073(35)$^{h)}$&0.035$^{b)}$&0.0355(25)$^{h)}$\\
 &0.726$^{c)}$&&0.051$^{c)}$&\\ 
 &0.712$^{d)}$&&0.048$^{d)}$&\\ 
 &0.704$^{e)}$&&0.049$^{e)}$&\\ 
\hline
\end{tabular}
\caption{The critical exponents in three dimensions for the $O(N)$
model. ${a)}$ corresponds to the truncation where only the flow of
$\{Z,\rho_0,u_2\}$ is  considered. In ${b)}$ one adds $u_3$. In
${c)}$, one adds $\{u_3,u_4,Y(\rho_0)\}$ and in ${d)}$
$\{u_3,u_4,Y_0,Z'(\rho_0)\}$. $e)$ takes into account the full
dependence of $U_k$, $Z_k$ and $Y_k$ in the field \cite{gersdorff01}.
In  ${f)}$, the order $\partial^4$ terms of the derivative expansion
have been included  \cite{canet03b}. ${g)}$
corresponds to the five-loop resummed  perturbative results in
in $4-\epsilon$ \cite{guida98}. ${h)}$ are seven-loop perturbative
results in three dimensions \cite{guida98}.}
\label{exposants_critiques_On}
\end{table}
Although we shall not be concerned in the following in 
truncations using the {\it full} potential $U_k(\rho)$ and the {\it
full} kinetic terms $Z_k(\rho)$ and $Y_k(\rho)$ entering in
Eq.~(\ref{tronqderive}), we have indicated, in Table
\ref{exposants_critiques_On}, the critical exponents computed with
such {\it ans\"atze}. One notes that $\nu$ is in very good agreement
with seven-loop resummed series \cite{guida98} while the anomalous dimension is less
precisely determined until the order $\partial^4$ terms of the derivative
expansion have been included in the {\it ansatz}, see \cite{canet03b}.

\subsubsection{The  XY and multicritical Ising models in two dimensions}

Let us close this section devoted to the analysis of the $O(N)$ model
by a discussion of the results obtained in $d=2$ for the XY and Ising 
models. These are, in fact, two of its most spectacular successes
because they correspond to truly nonperturbative systems.

As well known, the physics of the XY model at finite temperature is
governed by topologically nontrivial configurations --- vortices ---
which are not taken into account in a low-temperature
treatment. According to Eq.~(\ref{flotu2rhod-2}), the flow for
$\rho_0$ --- or $T$ --- vanishes identically in $d=2$ and $N=2$ so
that the theory is free.  However, as well known, the model actually
exhibits a phase transition at a finite temperature $T_{\text{BKT}}$ --- the
Berezinskii-Kosterlitz-Thouless phase transition --- induced by the
deconfinement of the vortices, see
\cite{berezinskii70,kosterlitz73}. Remarkably, the simplest RG
equations (\ref{flotu2rho}) and (\ref{flot_eta_phi4}) already allow to
obtain the correct qualitative behavior of the XY model at finite
temperature: a very small $\beta$-function of $T$ is found between
$T=0$ and a finite $T_{\text{BKT}}$ \cite{grater95}. Recently, treating the full field-dependence of  $U_k$, $Z_k$ 
and $Y_k$, von~Gersdorff and Wetterich
\cite{gersdorff01} have recovered the correct behavior for the
correlation length of the XY model around $T_{\text{BKT}}$:
\begin{equation}
\xi\simeq \exp \left(\frac{\text{Cte}}{(T-T_{\text{BKT}})^\tau}\right)\ .
\label{longueur_correl_KT}
\end{equation}
The exact results are $\tau=1/2$ and $\eta=1/4$ for the anomalous
dimension at $T_{\text{BKT}}$ \cite{kosterlitz73,kosterlitz74} while von~Gersdorff and Wetterich have
found $\tau=0.502$ and $\eta=0.287$. This shows that the physics of
topological excitations is captured by the lowest orders of the
derivative expansion, without including explicitly these degrees of
freedom {\it \`a la} Villain~\cite{villain75}.  

As for the Ising model, it is known that, in two dimensions, it can
undergo infinitely many nontrivial kinds of phase transitions
associated with infinitely many multicritical fixed points
\cite{zamolodchikov86}. It is shown  in Ref.\cite{Itzykson89}
that they all correspond to strong coupling fixed points. They  are therefore very
 difficult to study by perturbative
means. By a systematic search of fixed points in the two-dimensional
scalar field theory, using an order $\partial ^2$ truncation of the
derivative expansion, Morris \cite{morris95b} has been able to find
explicitly the first ten fixed points of this series. He  has also shown
that no other fixed point exists but the multicritical fixed points.

\subsubsection{A difficulty related to the field expansion}
\label{difficulty}

Let us finally mention a difficulty linked to the field expansion of
the potential $U_k$ showing up, for instance, when the stable fixed
point of the $O(N)$ model is followed from $d=4$ down to $d=2$. When
the simplest $\phi^4$ truncation, Eq.~(\ref{developpement_general_action_on}), is used no problem occurs:
one can smoothly follow the stable --- critical --- fixed point from
$d=4$, where it identifies with that found in a weak coupling
expansion of the GLW model, down to $d=2$, where it coincides with
that obtained within a low-temperature expansion of the NL$\sigma$
model, Eq.~(\ref{flotu2rhod-2}). However, once the $\phi^6$ term is
added, a new nontrivial fixed point emerges from the Gaussian fixed
point in $d=3$. This is a  tricritical fixed point, {\it i.e.} a
fixed point with two directions of instabilities. As $d$ is lowered,
the critical and tricritical fixed points move closer together and
eventually coalesce in a dimension $2<d<3$. Actually, they both become
complex.  Note that when $d$ is further lowered, the two fixed points
become again real. In $d=2+\epsilon$, the stable fixed point can be
identified with that found within the low-temperature expansion of the
NL$\sigma$ model with the $\phi^4$ truncation. Thus, there exists a
small region between $d=2$ and $d=3$ where one fails to correctly
describe the fate of the stable fixed point of the model using the
$\phi^6$ truncation. However, this is just an artefact of the field
expansion, not of the method. To show this, let us describe what
happens when the order $p$ of the truncation is increased. First,  when
including a new monomial $\phi^p$ in the effective potential, a new --- multicritical --- fixed
point emerges from the Gaussian fixed point in the dimension $2 p/(p-2)$. Again the stable fixed
point coalesces with one of these multicritical points and reappears close
to $d=2$. Second, as $p$ increases,  the coalescence of the stable fixed point occurs at
smaller and smaller dimensions. Thus, one recovers a
better and better description when increasing the order of the
truncation. Also, it has been checked that when the full
field-dependence of the potential is kept, the problem fully
disappears and the stable fixed point can be followed smoothly between
$d=4$ and $d=2$ \cite{pruessner}. Finally, it is important to indicate
that, in the whole range of dimensions where the stable fixed point
exists within a field expansion, the critical exponents found within
this approach at sufficiently large order $p$ ($p\ge 10$) and those
found within a full potential computation are very close. The artefact
of the field expansion described here can be bypassed using either a
full potential computation or using a field expansion at sufficiently
high order. Actually, it is not surprising that difficulties occur
with the field expansion at low dimensions since the engineering
dimension of the field vanishes as $d\to 2$. This strongly suggests
that no power of the field can be safely discarded in $U_k$ when $d\to
2$. This is confirmed by the fact that the effective potential, which
is exactly known at $N=\infty$ for $d=3$ and $d=2$, is respectively a
polynomial of order six and an infinite series.

\subsection{Conclusion}

We have described, in this section, the main features of the effective
average action method. We now summarize them:

1) The effective average action method allows trivially to recover the
perturbative results around the upper --- $d=4$ --- and lower ---
$d=2$ --- critical dimensions and thus to make contact with these
approaches.

2) The results obtained via this method are nonperturbative in the
different parameters:  coupling constant and temperature.  In this
sense, it provides an alternative approach to the usual perturbative
methods. This is of great interest, especially when one suspects that
the perturbative series could be  not reliable as it is the case for
frustrated magnets.

3) Even with a very simple truncation of the effective average action,
it is possible to capture some genuine nonperturbative features ---
like nontrivial topological configurations --- that are unreachable
from a conventional low-temperature expansion. This aspect is
particularly important in the context of frustrated magnets since one
knows that the low-temperature expansion performed in $d=2+\epsilon$
does not provide the correct physics in $d=3$, a possible explanation
being the presence of vortex-like configurations in these systems.

\section{The $O(N)\times O(2)$ model}
\label{chapitre_on_o2}

We now come back to the study of frustrated magnets. We derive the
flow equations relevant to the study of frustrated magnets. The
derivation follows the same lines as in the $O(N)$ case (see Section
\ref{chap_deriv_eq_on} above).

\subsection{Truncation procedure}

As emphasized previously, since the NPRG
equation (\ref{Wetterichfinalln}) cannot be solved exactly, a
truncation for $\Gamma_k$ is needed. We consider here a truncation
involving only terms having at most two derivatives. At this order,
the most general form of the $O(N)\times O(2)$ effective average
action writes:
\begin{widetext}
\begin{equation}
\begin{split}
&\Gamma_k[\vec \phi_1,\vec \phi_2]=\int d^d \cg x \left( U_k(\rho,\tau)+\frac{1}{2}
Z_k(\rho,\tau)\left(\left(\partial \vec \phi_1\right)^2+ \left(\partial \vec
\phi_2\right)^2\right) + \frac{1}{4} Y^{(1)}_k(\rho,\tau) \left(
\vec\phi_1\cdot\partial \vec\phi_2- \vec\phi_2\cdot\partial
\vec\phi_1\right)^2+ \right.\\ &\left. +\frac 14 Y^{(2)}_k(\rho,\tau) \left(
\vec\phi_1\cdot\partial \vec\phi_1+ \vec\phi_2\cdot\partial
\vec\phi_2\right)^2 + \frac 14 Y^{(3)}_k(\rho,\tau) \left(
\left(\vec\phi_1\cdot\partial \vec\phi_1- \vec\phi_2\cdot\partial
\vec\phi_2\right)^2+ \left(\vec\phi_1\cdot\partial \vec\phi_2+
\vec\phi_2\cdot\partial \vec\phi_1\right)^2\right) \right) \ .
\label{action_generale}
\end{split}
\end{equation}
\end{widetext}
We recall that $\vec \phi_1$ and $\vec \phi_2$ are the two
$N$-component vectors that constitute the order parameter,
Eq.~(\ref{matriceparametreordre}) while  $\rho=\hbox{Tr}(^{t}
\Phi.\Phi)$ and $\tau = \frac{1}{2}\hbox{Tr} \left(^t\Phi.\Phi
-\openone\ \rho/2\right)^2$ --- with $\Phi=(\vec\phi_1,\vec\phi_2)$
--- are the two independent $O(N)\times O(2)$ invariants (see
Appendix \ref{annexe_invariants} for a more detailed discussion). The
truncation (\ref{action_generale}) is the analogue of
Eq.~(\ref{tronqderive}), in the case of matrix fields. Here
$U_k(\rho,\tau)$ is the potential part of the effective average action
while $Z_k(\rho,\tau)$ and $Y_k^{(i)}(\rho,\tau)$, $i=1,2,3$, are kinetic
functions.

At this level of approximation, the RG analysis requires to specify
the five functions $U_k$, $Z_k$ and $Y_k^{(i)}$, $i=1,2,3$. This is to
say an infinite number of coupling constants.  As in the case of the
vectorial $O(N)$ model, we further simplify the {\it ansatz} by
expanding these functions in powers of the fields. Again, we choose to
expand around a nonvanishing field configuration which minimizes
$\Gamma_k$. This constraint is satisfied when $\vec\phi_1$ and
$\vec\phi_2$ are orthogonal, with the same norm.  We choose:
\begin{equation}
\Phi^\Min(\cg x)=\sqrt{\tilde\kappa}
\begin{pmatrix}
1&0\\
0&1\\
0&0\\
\vdots&\vdots\\
0&0
\end{pmatrix}
\label{def_min}
\end{equation}
the physical results being  independent of this
particular choice. The quantity $\sqrt{\tilde\kappa}$ entering in
Eq.~(\ref{def_min}) is analogous to the quantity $\phi_0$ in the $O(N)$
case, see Eq.~(\ref{defmin}) and we refer to it in the
following as the magnetization.

While studying the critical properties of the system, we have
considered various truncations differing by the number of monomials in
$\rho$ and $\tau$ included in the field expansion. Our largest
truncations consist either in keeping all terms in $U_k$ up to the
eighth power of the fields and all terms in $Z_k$ and in the
$Y^{(i)}_k$'s including four powers of the fields or all terms in
$U_k$ up to the tenth power of the fields and the first term of the
expansions of $Z_k$ and of $Y^{(1)}_k$.  With these truncations, we
have verified that our results are stable with respect to addition of
higher powers of the fields.  However, in order to keep our
presentation concise, we have chosen here to consider a reduced
truncation that already enables to recover the different perturbative
results --- in $4-\epsilon$, $2+\epsilon$ and $1/N$ --- in their
respective domains of validity. Within this truncation, we expand
$U_k$ up to terms containing four powers of the fields and keep only
the leading terms of $Z_k$ and $Y^{(1)}_k$. We also completely discard
the two other functions $Y_k^{(2)}$ and $Y_k^{(3)}$.  This choice is
motivated by the fact that, as we will see in the next section, only
the function $Y^{(1)}_k$ contributes directly to the physics  of the
Goldstone modes and is thus important around two dimensions. Since one
of our aims is to recover the results obtained around two dimensions,
we keep this term in our {\it ansatz}.  We are then led to the simple
truncation:
\begin{equation}
\begin{split}
&\Gamma_k [\vec \phi_1,\vec \phi_2]= \int d^d \cg x\left( \frac Z 2
\left(\left(\partial\vec \phi_1\right)^2+\left(\partial \vec
\phi_2\right)^2\right)+\right. \\&\left. +\frac{\tilde\omega}{4}
\left(\vec \phi_1\cdot\partial \vec\phi_2-\vec \phi_2\cdot\partial
\vec\phi_1\right)^2+\frac{\tilde{\lambda}}{4}\left(\frac\rho 2 -
\tilde{\kappa}\right)^2 + \frac{\tilde{\mu}}{4} \tau \right)\ .
\label{troncation}
\end{split}
\end{equation}
 Let us now discuss the different terms appearing in this
expression. The coupling constants $\tilde \lambda$ and $\tilde \mu$
have been  already introduced in the GLW approach (see
Eq.~(\ref{hamglw})). The coupling constant $\tilde \kappa$ describes
the position of the minimum of the potential and appears in the
truncation because we expand $\gk$ around the nonvanishing field
configuration $\Phi^\Min$. As in the $O(N)$ case, $Z$ corresponds to
the field renormalization. Finally, the unusual kinetic term with
coupling $\tilde \omega$ corresponds to the current-term of
Eq.~(\ref{eq15}) introduced in the discussion of the NL$\sigma$ model
approach. This term is irrelevant by
power counting around four dimensions since it is quartic in the
fields and quadratic in derivatives. However its presence is {\it
necessary} around two dimensions to recover the results of the
low-temperature approach of the NL$\sigma$  model
since it contributes to the field renormalization of the Goldstone
modes.  Being not constrained by the usual power counting one  includes
this term in the  ansatz.  

The above effective action has all the
ingredients to describe accurately the physics at low-temperature
around two dimensions as well as at weak-coupling regime around four
dimensions.  We can therefore anticipate that this simple truncation
is actually rich enough to recover the perturbative results around
$d=2$ and $d=4$. Of course, since our main goal is to go beyond the
usual perturbative expansion, we have studied larger truncations and
have controlled the convergence of our results.

\subsubsection{The spectrum}

We now discuss the spectrum of excitations around the minimum
(\ref{def_min}). The spectrum is given by the eigenvalues and
eigenvectors of the matrix $\delta^2\Gamma_k/ \delta\phi_i^j
\delta\phi_k^l$ --- where $i,k\in\{1,2\}$ and  $j,l\in \{1,\dots,N\}$ ---
evaluated in the configuration (\ref{def_min}). We find that the $2N$
degrees of freedom  of the  order parameter $\Phi$ divide in four
types that are described in Fig.~\ref{fig_spectre}:

1)  a family of $2N-4$ massless --- Goldstone --- modes
which correspond to rotating rigidly the dihedral
$(\vec\phi_1,\vec\phi_2)$ by keeping either $\vec\phi_1$ or
$\vec\phi_2$ unchanged, see Fig.~\ref{fig_mode_goldstone1}.

The four remaining modes correspond to the
situations where the two vectors $(\vec\phi_1,\vec\phi_2)$ remain in
the same plane:

2) a massless --- Goldstone --- singlet mode corresponding to rotating
the dihedral within its plane, see Fig.~\ref{fig_mode_goldstone}. Together with the $2N-4$ other ones,
this gives the $2N-3$ Goldstone mode of the model.

3) a massive singlet of square mass $\tilde \lambda\tilde\kappa$
corresponding to a dilation of the two vectors, see Fig.~\ref{fig_mode_massif_1}.

4) a massive doublet of square mass $\tilde \mu\tilde\kappa$
corresponding to fluctuations of each vector of the dihedral, with the
constraint that the sum of the lengths of the vectors
$|\vec{\phi_1}|+|\vec{\phi_2}|$ remains unchanged, see Fig.~\ref{fig_mode_massif_2}.

\begin{figure}[htbp] 
\begin{center} 
\makebox[\linewidth]{ 
\subfigure[Massless (2N-4)-uplet]{
\label{fig_mode_goldstone1}
\includegraphics[height=.2\linewidth,origin=tl]{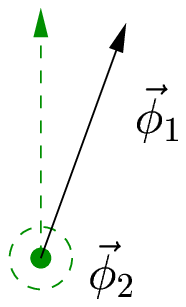}
\hspace{1cm}
\includegraphics[height=.2\linewidth,origin=tl]{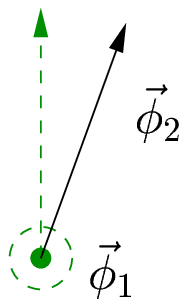}}}
\end{center}
\begin{center}
\makebox[\linewidth]{
\subfigure[Massless singlet]{
\label{fig_mode_goldstone}
\includegraphics[height=.3\linewidth,origin=tl]{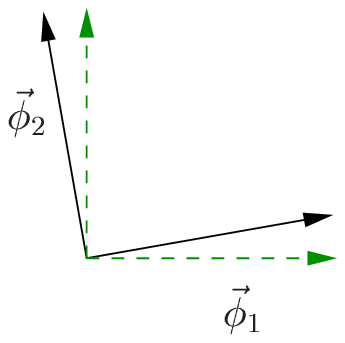}}
\hfill \subfigure[Massive singlet]{
\label{fig_mode_massif_1}
\includegraphics[height=.3\linewidth,origin=tl]{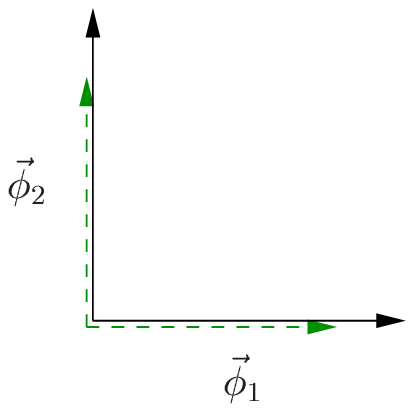}}}\newline
\end{center}
\begin{center}
\makebox[\linewidth]{ \subfigure[Massive doublet]{
\label{fig_mode_massif_2}
\includegraphics[height=.3\linewidth,origin=tl]{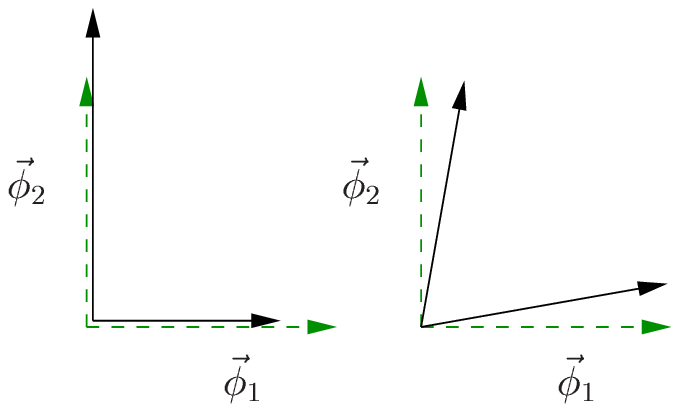}} }
\end{center}
\caption{ Schematic description of the deformations  of
$(\vec\phi_1,\vec\phi_2)$ associated with the four types of proper excitations
of the model. The dotted arrows display the ground
state configuration and the plain arrows display the relevant
deformations: a) massless $(2N-4)$-uplet, b) massless singlet, c) massive singlet, d) massive
doublet.}
\label{fig_spectre}
\end{figure}

In practice, it is very useful to work in the basis of the proper
excitations of the model since then, the propagator being diagonal,
the calculations are greatly simplified.  We therefore introduce $2N$
directions in the internal space, corresponding to the $2N$ proper
excitations of the model. They are given by:

\begin{equation}
\begin{aligned}
&\delta_{1,\cg p} = \frac{1}{\sqrt{2}}\left(\frac{\delta}{\delta
\phi_1^1(\cg p)} +
\frac{\delta}{\delta \phi_2^2(\cg p)}\right) \\ 
&\delta_{2, \cg p} =
\frac{1}{\sqrt{2}}\left(\frac{\delta}{\delta \phi_1^1(\cg p)} -
\frac{\delta}{\delta \phi_2^2(\cg p)}\right)  \\
&\delta_{3,\cg p} =
\frac{1}{\sqrt{2}}\left(\frac{\delta}{\delta \phi_2^1(\cg p)} +
\frac{\delta}{\delta \phi_1^2(\cg p)}\right)  \\ 
&\delta_{4,\cg p} =
\frac{1}{\sqrt{2}}\left(\frac{\delta}{\delta \phi_1^2(\cg p)} -
\frac{\delta}{\delta \phi_2^1(\cg p)}\right)  \\
&\delta_{5,\cg p} =
\frac{\delta}{\delta \phi_1^3(\cg p)}  \\ 
&\delta_{6,\cg p} =
\frac{\delta}{\delta \phi_2^3(\cg p)}  \\ 
&\quad\vdots \\
&\delta_{2N-1,\cg p} = \frac{\delta}{\delta \phi_1^N(\cg p)} \\
&\delta_{2N,\cg p} = \frac{\delta}{\delta \phi_2^N(\cg p)} \ .
\label{vecteurs_propres}
\end{aligned}
\end{equation}

In this basis, the two-point vertex function, \ie the inverse
propagator --- up to the $R_k$ term --- has the form:
\begin{align}
\Gamma&^{(2)}_{\{i,\cg q_1\},\{j,\cg q_2\}}\Big|_\Min=\frac{\delta(\cg{q_1}+\cg
{q_2})}{(2\pi)^d}\ . \nonumber\\&\ .\begin{pmatrix} Z\cg
{q_1}^2+\tilde{\lambda}\tilde{\kappa}& &&&&&&\\ &\hskip -8mm Z\cg
{q_1}^2+\tilde{\mu}\tilde{\kappa}&&&&0&\\ &&\hskip -8mm Z\cg
{q_1}^2+\tilde{\mu}\tilde{\kappa}&&&&\\ &&&\hskip -8mm
(Z+\tilde{\omega}\tilde{\kappa}) \cg {q_1}^2&&&\\ &&&&\hskip -8mm Z \cg
{q_1}^2&&\\ &\hskip -8mm 0&&&\ \hspace{0.cm} \ddots&\\ &&&&&\hskip -3mm Z
\cg {q_1}^2
\end{pmatrix}
\label{inverse_prop}
\end{align}

In the matrix (\ref{inverse_prop}), the first three lines correspond
to the massive modes and the last $2N-3$ to the massless modes.  Note
that a nonstandard kinetic term appears on the fourth line through an
additional field renormalization $\tilde \omega\tilde\kappa$ for the
Goldstone singlet. Let us add that if one keeps, in the truncation
(\ref{troncation}), contributions from the functions $Y^{(2)}$ and
$Y^{(3)}$ (see Eq.~(\ref{action_generale})), the field renormalizations
in the first three lines get  extra contributions similar to what is
obtained in the fourth line.  Note that $Y^{(2)}$ and $Y^{(3)}$ affect
only the field renormalization of massive modes. It is therefore not
necessary to take them into account in order to retrieve the leading
order behavior in a low-temperature expansion around $d=2$ which is
entirely governed by Goldstone modes. This is why we do not keep them
in our simplest truncation (\ref{troncation}).

\subsection{The flow equations}

We now display the flow equations for the coupling constants entering
in the truncation Eq.~(\ref{troncation}). We recall the four major
steps of this procedure (see Section \ref{chap_deriv_eq_on}):

{\it i)} The running coupling constants are defined as functional
derivatives of the {\it ansatz} of $\Gamma_k$, Eq.~(\ref{troncation}).

{\it ii)} The operator $\partial_t$ is then applied on these
definitions. By making use of the NPRG equation
(\ref{Wetterichfinalln}), flow equations for the coupling constants
are obtained as traces of vertex functions. These expressions are
evaluated by using the truncated form of $\Gamma_k$
Eq.~(\ref{troncation}).

{\it iii)} The flow equations are expressed in terms of threshold
functions.

{\it iv)} Dimensionless renormalized quantities are introduced.

\subsubsection{Definition of the coupling constants}

As in the vectorial model, the coupling constants are defined as
values of the vertex functions in the specific configuration $\Phi^\Min$
around which is made the field expansion, Eq.~(\ref{def_min}).

Let us start with the definition of $\tilde \kappa$. This coupling
constant parametrizes the ground state configuration $\Phi^\Min$. One
has, as in the $O(N)$ case, an implicit definition of $\tilde \kappa$:
\begin{equation}
\delta_{\alpha,\cg p=\cg 0}\, \Gamma_k \bigg |_{\hbox{\begin{tiny}
Min. \end{tiny}}}=0\; 
\label{kappa_frustre}
\end{equation}
with $\delta_{\alpha,\cg p}$ given by Eq.~(\ref{vecteurs_propres}). In the
following, as in the $O(N)$ case, we shall consider only the case
$\alpha=1$.

The other coupling constants are defined using the two-point vertex
function in different directions:
\begin{equation}
\left\{
\begin{aligned}
\tilde{\lambda}&=\frac {(2\pi)^d} {\tilde{\kappa}\delta(\cg 0)}\,
\delta_{1,\cg{0}} \delta_{1,\cg{0}} \Gamma_k \bigg|_\Min
\\ 
\tilde{\mu}&=\frac {(2\pi)^d}{\tilde{\kappa}\delta(\cg 0)}\,
\delta_{2,\cg{0}} \delta_{2,\cg{0}} \Gamma_k \bigg|_\Min
\label{couplinggg}\ .
\end{aligned}
\right.
\end{equation}
These two definitions come directly from the study of the spectrum
discussed previously (see Eq.~(\ref{inverse_prop})).

We finally define the coupling constants associated with the
momentum-dependent part of our truncation (\ref{troncation}), \ie the
field renormalization factor $Z$ and the current-term coupling
constant $\tilde \omega$:
\begin{equation}
\left\{
\begin{aligned}
 Z&=\frac{(2\pi)^d}{\delta(\cg 0)} \lim_{\cg p^2\to 0}\left.\frac{d}{d
\cg p^2}\bigg( \delta_{5,\cg p} \delta_{5,-\cg p} \;\Gamma_k 
\right|_\Min \bigg)\\ 
\tilde{\omega}&=\frac{(2\pi)^d}{\tilde\kappa\delta(\cg
0)}\left.\lim_{\cg p^2\to 0}\frac{d}{d \cg p^2} \bigg( \delta_{4,\cg p}
\delta_{4,-\cg p} \;\Gamma_k  \right|_\Min\bigg)-\frac Z
{\tilde\kappa}
\label{omega_frustre}
\end{aligned}
\right.
\end{equation}

\subsubsection{The $t$-derivation and the flow equations}

We now apply the operator $\partial_t$ to the definitions
(\ref{kappa_frustre}--\ref{omega_frustre}). In order to derive the
flow equations, we have to compute the functional derivatives of
$\partial_t \Gamma_k$ with respect to the fields. This is similar to
what has been done previously in the context of the $O(N)$ model (see
Section \ref{chapter_t_derivative} above), except that the tensorial
structure in the internal space is more involved so that the
computation of the traces is more cumbersome. We do not give the
details here.  We now  introduce the dimensionless renormalized
quantities defined as:
\begin{equation}
\left\{
\begin{aligned}
\kappa&=Z k^{2-d} \tilde{\kappa} \\
\lambda&= Z^{-2} k^{d-4}
\tilde{\lambda} \\ 
\mu&= Z^{-2} k^{d-4} \tilde{\mu} \\ 
\omega&=Z^{-2} k^{d-2} \tilde{\omega}
\end{aligned}
\right.
\end{equation}
as well as the threshold functions which are defined and discussed in
Appendix \ref{annexe_threshold}. We then get the following flow
equations \cite{tissier01}:
\vspace{3cm}
\begin{widetext}
\begin{subequations}
\begin{align}
\begin{split}
\frac{d\kappa}{ dt}=&-(d-2+\eta)\kappa+4 v_d\bigg[\frac 1 2
 l_{01}^{d}(0, 0,\kappa\, \omega) + (N-2) l_{10}^d(0,0,0) + \frac3 2
 l_{10}^d(\kappa\, \lambda, 0, 0) + \left(1 + 2\,\frac{\mu}{
 \lambda}\right)\, l_{10}^d(\kappa\, \mu, 0,0)+\\&+\frac{\omega}{\lambda}\,l_{0
 1}^{2 + d}(0, 0, \kappa \,\omega) \bigg]
\label{flot_kappa_frustre}
\end{split}\displaybreak[0]\\  
\begin{split}
\frac{d\lambda }{dt}=&(d-4+2\eta)\,\lambda+ v_d \bigg[2\lambda^2
\,(N-2) l_{20}^d( 0, 0, 0)+\lambda^2 \,l_{02}^d(0, 0, \kappa \,\omega)
+9\lambda^2 \,l_{20}^d(\kappa\,\lambda, 0, 0) +\\&+ 2 (\lambda +
2\mu)^2 \,l_{20}^d( \kappa\mu, 0, 0) +4 \lambda\omega l_{02}^{2 +
d}(0, 0,\kappa\,\omega)+ 4 \omega^2 l_{02}^{4 + d}(0, 0,\kappa
\,\omega)\bigg]
\label{flot_lambda_frustre}
\end{split}\displaybreak[0]\\
\begin{split}
\frac{d\mu }{dt}=&(d-4+2\eta)\,\mu -2 v_d\mu\bigg[-\frac{2}{\kappa}
\,l_{01}^d(0, 0, \kappa \,\omega) +\frac{3(2\lambda +
\mu)}{\kappa(\mu-\lambda)} l_{10}^d(\kappa\,\lambda,0,0) +\frac{8
\lambda + \mu}{ \kappa \,(\lambda - \mu)} l_{10}^d(\kappa \,\mu, 0,0)+\\& +
\mu l_{11}^d( \kappa \,\mu, 0,\kappa \,\omega) + \mu \,(N-2) l_{20}^d(0, 0,
0)\bigg]
\label{flot_mu_frustre}
\end{split}\displaybreak[0]\\
\begin{split}
\eta=&-\frac{d\;\ln Z}{dt}=2\frac{v_d}{d\kappa} \bigg[(4-d)\kappa
\,\omega l_{01}^d(0, 0,\kappa \,\omega) + 2\kappa^2 \,\omega^2 \,l_{02}^{2 +
d}(0, 0, \kappa \,\omega) + 2 m_{02}^d(0, 0, \kappa \,\omega)- 4 m_{11}^d(
0, 0, \kappa \,\omega)+\\&+2(-2 + d) \kappa \,\omega l_{10}^d(0, 0, 0) +2
m_{20}^d(0, 0, \kappa \,\omega)+ 2 \kappa^2 \,\lambda^2\, m_{2, 2}^d(\kappa
\,\lambda, 0, 0) +4 \kappa^2 \,\mu^2 \,m_{2, 2}^d(\kappa \,\mu, 0, 0) + \\& +
4 \kappa \,\omega \,n_{02}^d(0, 0, \kappa \,\omega) -8 \kappa \,\omega\,
n_{11}^d(0, 0, \kappa \,\omega) +4 \kappa \,\omega n_{20}^d(0, 0,\kappa\,
\omega)\bigg]
\label{flot_Z_frustre}
\end{split}\displaybreak[0]\\
\begin{split}
\frac{d\omega}{ dt}=&(d-2+ 2 \eta)\, \omega + \frac{4 v_d}{d \kappa^2}
\bigg[\kappa \omega\bigg\lbrace \frac{(4 -d)}{ 2} \,l_{01}^d(0, 0,
\kappa \,\omega) +\frac{(d-16 )}{2} \,l_{01}^d(\kappa \,\lambda, 0,
\kappa \,\omega) +\kappa \,\omega \,l_{02}^{2 + d}(0, 0, \kappa
\,\omega) -\\&- 3 \kappa \,\omega \,l_{02}^{2 + d}(\kappa \,\lambda,
0, \kappa \,\omega)+(d-2) \,l_{10}^d(0, 0, 0) - (d-8 )\,
l_{10}^d(\kappa \,\lambda, 0, 0)+8 \kappa \,\lambda \,l_{11}^d(\kappa
\,\lambda, 0, \kappa \,\omega) +2 \kappa \,\omega \,l_{20}^{2 +
d}(\kappa \,\mu, 0, 0) \\ &+2 \kappa \,\omega \,(N-2) \,l_{20}^{2 +
d}(0, 0, 0) \bigg\rbrace +m_{02}^d(0, 0, \kappa \,\omega) -
m_{02}^d(\kappa \,\lambda, 0, \kappa \,\omega) - 2 m_{11}^d(0, 0,
\kappa \,\omega) +2 m_{11}^d(\kappa \,\lambda, 0, \kappa \,\omega)+
\end{split}
\label{flot_omega_frustre}
\displaybreak[0]\\
\begin{split} \nonumber
& + m_{20}^d(0, 0,\kappa \,\omega)- m_{20}^d(\kappa \,\lambda, 0,
\kappa \,\omega)+\kappa^2 \,\lambda^2\, m_{22}^d(\kappa \,\lambda, 0,
0) +2\kappa^2 \,\mu^2 m_{22}^d(\kappa \,\mu, 0, 0) + 2 \kappa \,\omega
\,n_{02}^d(0, 0, \kappa \,\omega)-\\& -4 \kappa \,\omega
\,n_{02}^d(\kappa \,\lambda, 0, \kappa \,\omega) - 4 \kappa \,\omega
n_{11}^d(0, 0, \kappa \,\omega) +8 \kappa \,\omega n_{11}^d(\kappa
\,\lambda, 0, \kappa \,\omega) + 2 \kappa \,\omega \,n_{20}^d(0, 0,
\kappa \,\omega) - 4\kappa \,\omega n_{20}^d(\kappa \,\lambda, 0,
\kappa \,\omega)\bigg]
\end{split}
\end{align}
\label{recursion}
\end{subequations}
\end{widetext}

\section{Tests of the method and first results}
\label{chapitre_test}

This section is devoted to all possible tests of our method in the
$O(N)\times O(2)$ case. We show, in particular, how the various
perturbative results are recovered as it was already the case in the
$O(N)$ model. We also give our determination of $N_c(d)$ which is
compared with  the three-loop improved perturbative computation. Finally, we give our
determination of the exponents in the $N=$6 case and we compare them with those of the
Monte Carlo simulation.

\subsection{The weak-coupling expansion around $d=4$}

Around $d=4$, we expect a nontrivial fixed point close to the
gaussian. One can expand the flow equations at leading order in the
quartic coupling constants and in $\epsilon$, as we did in the $O(N)$
case (see Section \ref{chapitre_weak_coupling_ON}). As expected from
power counting, we find that the fixed point value of the coupling
constant $\omega$ associated with the current-term is vanishing at
leading order. This is also the case of $\eta$.  As in the $O(N)$
case, the square masses $\lambda \kappa$ and $\mu \kappa$ are of order
$\epsilon$ so that the threshold functions can be expanded in powers
of their arguments.  Once this expansion is performed one recovers the
standard one-loop $\beta$-functions for the coupling constants
$\lambda$ and $\mu$ given in Eq.~(\ref{recursionglw}) that we recall
here:
\begin{equation}
\left\{
\begin{aligned}
\beta_\lambda &= -\epsilon \lambda +\frac{1}{16\pi^2}\left(4 \lambda\mu +4 \mu^2 +\lambda^2(N+4) \right)\\
\beta_\mu &= -\epsilon \mu+\frac{1}{16\pi^2}\left(6 \lambda\mu + N \mu^2
\right)\ .
\end{aligned}
\right.
\end{equation}

One  can also expand the $\beta$ function for $\kappa$,
Eq.~(\ref{flot_kappa_frustre}):
\begin{equation}
\begin{split}
\beta_{\kappa}=-(2-\epsilon)\kappa+& \frac{l_1^4(0)}{8\pi^2}\left
 (N+1+\frac{2 \mu} \lambda \right)-\\ &-\frac{ 3 \kappa \lambda}{16
 \pi^2}-\frac{ \kappa \mu}{8 \pi^2}\left( 1+\frac{2\mu}\lambda\right)
\end{split}
\end{equation}
from which we can deduce the expression of $\nu$ at order $\epsilon$,
which coincide with the one-loop result of Eq.~(\ref{nu_on_o2}).

\subsection{The low-temperature expansion around  $d=2$}

As explained in the context of the $O(N)$ model (see
Section \ref{chapitre_NLsigma_ON}), in order to recover the NL$\sigma$
model results, we need to expand the flow equations at large masses.
Using the behavior of the threshold functions for large arguments 
(see Appendix \ref{annexe_threshold}), we
get \cite{tissier00}:
\begin{equation}
\left\{
\begin{aligned}
\frac{d\kappa}{dt}&=-(d-2+\eta)\kappa+\frac{N-2}{2\pi}+\frac1
{4\pi(1+\kappa \omega)}\\
\frac{d\omega}{dt}&=(-2+d+2\eta)\omega+\\&+\frac{1+\kappa\omega+(N-1)
\kappa^2\omega^2 +(N-2)\kappa^3\omega^3}{2\pi\kappa^2(
1+\kappa\omega)} \\
\eta&=\frac{3+4\kappa\omega+2\kappa^2\omega^2}{4\pi\kappa(1
+\kappa\omega)} \ .
\end{aligned}
\right.
\end{equation}
By making the change of variables:
\begin{equation}
\left\{
\begin{aligned}
\eta_1&=2\pi\kappa\\ \eta_2&=4\pi \kappa(1+\kappa\omega)\ 
\end{aligned}
\right.
\end{equation}
we recover the $\beta$-functions found in the framework of the
NL$\sigma$ model at one-loop order (see Eq.~(\ref{recursionnls})).

\subsection{The large-$N$ expansion in $d=3$}

As in the $O(N)$ case (see Section \ref{chapitre_Large_N_ON}), our
equations allow to recover the critical exponents at leading order in
${1/N}$.
\begin{figure}[htbp] 
\centering
\subfigure[$\eta$ as a function of $N$]{
\label{eta_de_n}
\includegraphics[width=.95\linewidth,origin=tl]{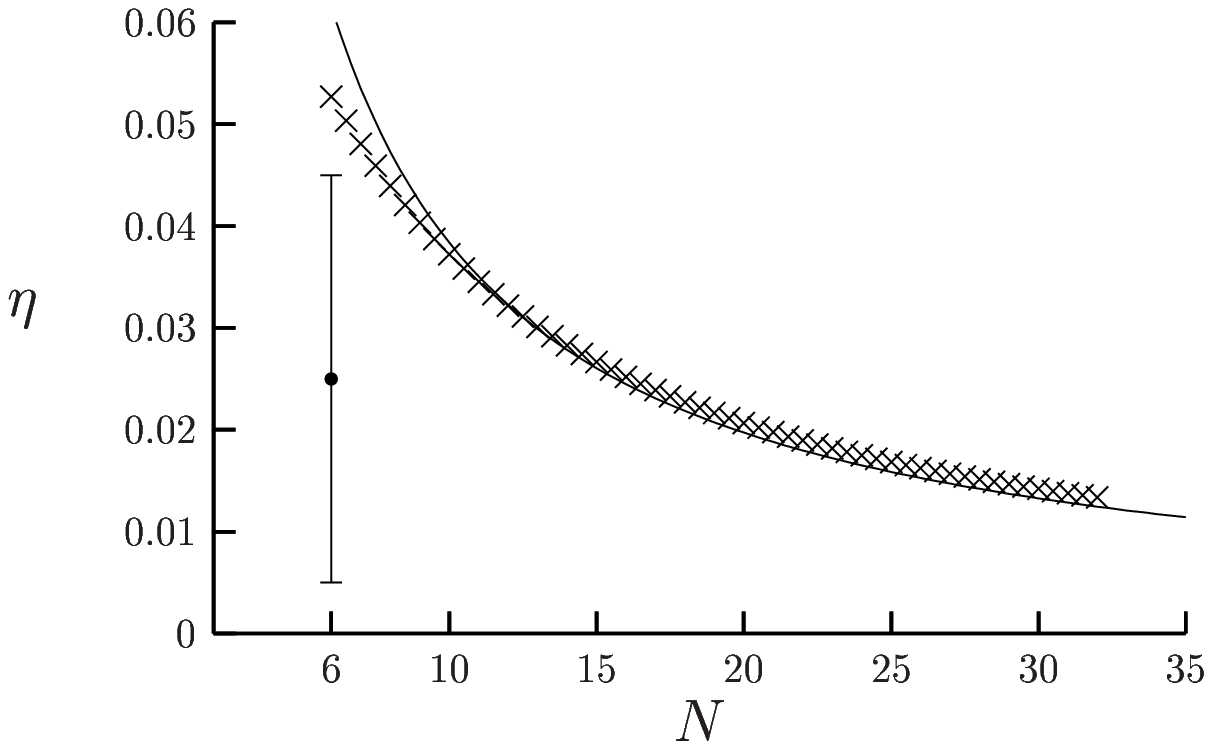}}\newline
\subfigure[$\nu$  as a function of $N$]{
\label{nu_de_n}
\includegraphics[width=.95\linewidth,origin=tl]{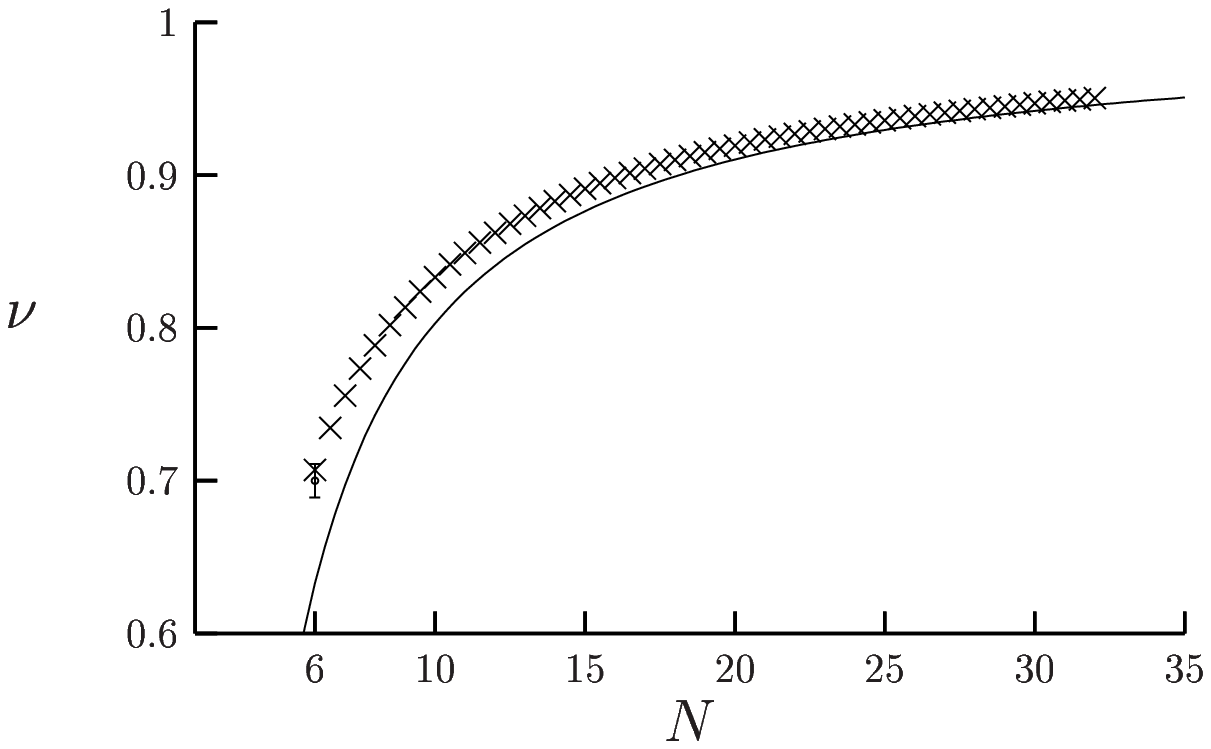}}
\caption{The exponents $\eta$ and $\nu$ as functions of $N$ in
$d=3$. The crosses represent our results and the full line the values
obtained from the $1/N$ expansion, Eqs. (\ref{exposent1surN}). The
circles and error bars are the Monte Carlo results obtained for $N=6$
\cite{loison00}.}
\label{etanu}
\end{figure}
We have computed $\eta$ and $\nu$ for a large range of values of $N$
and have compared our results with those caculated perturbatively at
order $1/N^2$,  Eq.~(\ref{exposent1surN}).  We find an excellent
agreement --- better than 1$\%$ --- for $\nu$, {\it for all} $N>10$,
see Fig.~\ref{etanu},   a domain of values of $N$  where one expects the  $1/N^2$ results 
to be very close to the exact values. We also quote in Table \ref{exposants16_32} our
results and those obtained by the six-loop calculation for $N=16$ and
$N=32$. 

\begin{table}[htbp]
\centering
\begin{tabular}{|c|c|c|c|}
\hline
$N$  &  Method        &     $\nu$      &  $\eta$   \\
\hline
\hline
 16   &1/N\cite{pelissetto01b}  &    0.885       &   0.0245  \\ 
\hline
      &  NPRG        &    0.898       &   0.0252  \\
\hline
      &six-loop &   0.858(4)\cite{pelissetto01a}, 0.863(4)\cite{calabrese03b}&    0.0246(2)\cite{calabrese03b}      \\
\hline
\hline
 32   &1/N\cite{pelissetto01b}  &    0.946       &   0.0125   \\ 
\hline
      &   NPRG             &    0.950       &   0.0134  \\
\hline
      &six-loop&   0.936(2)\cite{pelissetto01a}, 0.936(1)\cite{calabrese03b}      & 0.01357(1) \cite{calabrese03b}      \\
\hline
\end{tabular}
\caption{Exponents $\nu$ and $\eta$ computed from the $1/N$ expansion
\cite{pelissetto01b}, by our method (NPRG) and from the six-loop
 calculation \cite{pelissetto01a,calabrese03b}.}
\label{exposants16_32}
\end{table}

\subsection{The determination of  $N_c(d)$}

\label{chapitre_Nc}

Let us now interpolate between the results we have obtained around
$d=2$ and $d=4$ and discuss, in particular, the curve $N_c(d)$ that
separate the regions of first and second order.  

We have computed $N_c(d)$ with our best truncation and
with the cut-off function (\ref{cutoffstep}). 
\begin{figure}[htbp] 
\centering
\makebox[\linewidth]{
\includegraphics[width=0.85\linewidth,origin=tl]{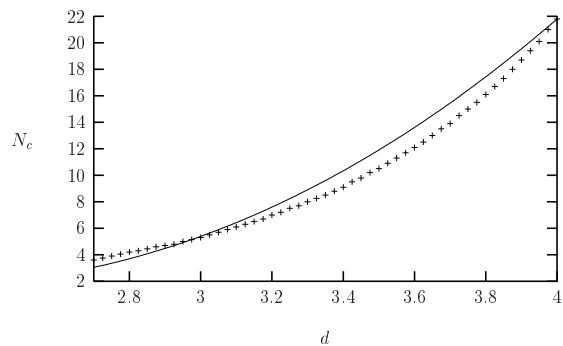}
}
\caption{The full line represents the curve $N_c(d)$ obtained by the
three-loop results improved by the constraint $N_c(d=2)=2$,
Eq.~(\ref{nc(d)troisboucles}). The crosses represent our calculation.}
\label{nc_de_d}
\end{figure}
 In Fig.~\ref{nc_de_d},  we give our result (crosses) from $d=4$ down to
$d=2.7$.  We also indicate the improved three-loop results given by
Eq.~(\ref{nc(d)troisboucles})  for comparison. The two curves agree
very well. Note that it is probably  a  coincidence that the curves  cross very
close to $d=3$. In this dimension, the NPRG method  leads to  $N_c(d=3)\simeq 5.1$ and the
improved three loop result: $N_c(d=3)\simeq 5.3(2)$.  Let us emphasize that, within the NPRG
 method,  the  quantity
 $N_c(d)$ is  very  sensitive to the order of the truncation~\cite{tissier01b}, 
 much more than the critical exponents. This means that one probably should not consider
the previous results as very reliable. In  this respect,  we recall the results obtained by means of
the six-loop calculation in $d=3$   \etal{}~\cite{calabrese03b}: $N_c(d=3)=6.4(4)$
 and by means of the $4-\epsilon$ expansion at five-loops \cite{calabrese03c}: $N_c(d=3)\simeq
6.1(6)$.

Let us finally mention that, for the reason already explained in
Section \ref{difficulty}, the field expansion we have performed at
order $\phi^{10}$ forbids us to follow the chiral  fixed point  $C_+$ in
dimensions typically between $d=2.5$ and $d=2.1$ and thus to determine
reliably the curve $N_c(d)$ in these dimensions.  As in the $O(N)$
case, this artefact could be overcome by keeping the full field
dependence of the effective potential $U_k(\rho,\tau)$.

\subsection{The critical exponents for  $N=6$}

 As already said, for $N=6$, the transition is either of second order
or  extremely weakly  of first order. In both cases  scaling  should exist on a large domain 
of temperature. The critical exponents obtained with our   best truncation are  given in Table
\ref{table_exp_crit_6_ERG}. Note that $\nu$ and $\eta$ are computed
directly while $\gamma$, $\beta$ and $\alpha$ are computed using the
scaling relations.
\begin{table}[htbp]
{{\begin{tabular}{|l|l|l|l|l|l|l|l}
\hline
Method  &$ \alpha$&$\beta$&$\gamma$&$\nu$&$\eta$\\
\hline
\hline
NPRG &-0.121&0.372&1.377&0.707&0.053\\
\hline
MC\cite{loison00}&-0.100(33)&0.359(14)&1.383(36)&0.700(11)&0.025(20)\\
\hline
\end{tabular}}
\caption{The exponents for $N=6$ obtained from  the NPRG  --- first line --- and from the Monte Carlo (MC)  simulation
--- second line.}
\label{table_exp_crit_6_ERG}}
\end{table}
Our   results agree very well with the numerical ones~\cite{loison00}. In particular,
the error on $\nu$, which is as usual the best determined exponent, is
only 1$\%$.  This constitutes a  success of the NPRG  approach  from  the
methodological point of view.}

\subsection{Conclusion}

Our method has successfully passed {\it all} possible tests.  This
gives us a great confidence in the reliability and the convergence of
our results. We are now in a position to explore the physics of
frustrated magnets in three dimensions.

\section{The physics in $d=3$}
\label{chapitre_d=3}

We now tackle with the physics in three dimensions. Before
embarking in this discussion, two points need to be clarified. The first concerns the
 existence of a fixed point for $N<N_c(d=3)$. The second one concerns the situation just below $N_c(d=3)$.

\subsection{The search of fixed points for $N<N_c(d)$}

 Let us first discuss the search,  within the NPRG method, of 
 fixed points in $d=3$ and  for $N<N_c(d=3)\simeq 5.1$. We recall that, for 
this critical value of $N$, the two fixed points $C_+$ and $C_-$ ---
see Fig.~\ref{ptfixe} --- coalesce. This means that these fixed points
--- that can be followed smoothly in the $(d,N)$ plane from the
gaussian in $d=4$ --- cease to be real below this  value. However, this
 does not imply  the absence  of other real fixed points.  One has to test the existence of fixed points  non
 trivially connected with  $C_+$ and $C_-$,  as 
advocated by Pelissetto \etal{}~\cite{pelissetto01a}. We  have thus searched  such fixed
 points both by directly looking for zeroes of the $\beta$-functions and 
by integrating numerically the RG  flow --- see below.  After an intensive search, we  have found {\it no} such
 fixed point.  This result and its relation with  that of Pelissetto \etal\   will be discussed in the following.

\subsection{The physics in $d=3$ just below $N_c(d)$: scaling with a pseudo-fixed point and  minimum
of the flow}

\label{chapitre_physics_d3}

In a fixed dimension $d$, the disappearance of the
nontrivial fixed points $C_+$ and $C_-$, when $N$ crosses $N_c(d)$, could
seem to be an abrupt process: the two fixed points collapse and
disappear.  Actually, when extended to the space of complex coupling
constants, this process is continuous since the only change is that,
when going from $N>N_c(d)$ to  $N<N_c(d)$, the fixed points acquire a small complex part. This
continuous character manifests itself as smooth changes of the RG flow
that can be explained thanks to continuity arguments.

\begin{figure}[htbp] 
\subfigure[$N>N_c(d)$ ]{
\label{regions_n-huge}
\includegraphics[height=.54\linewidth,origin=tl]{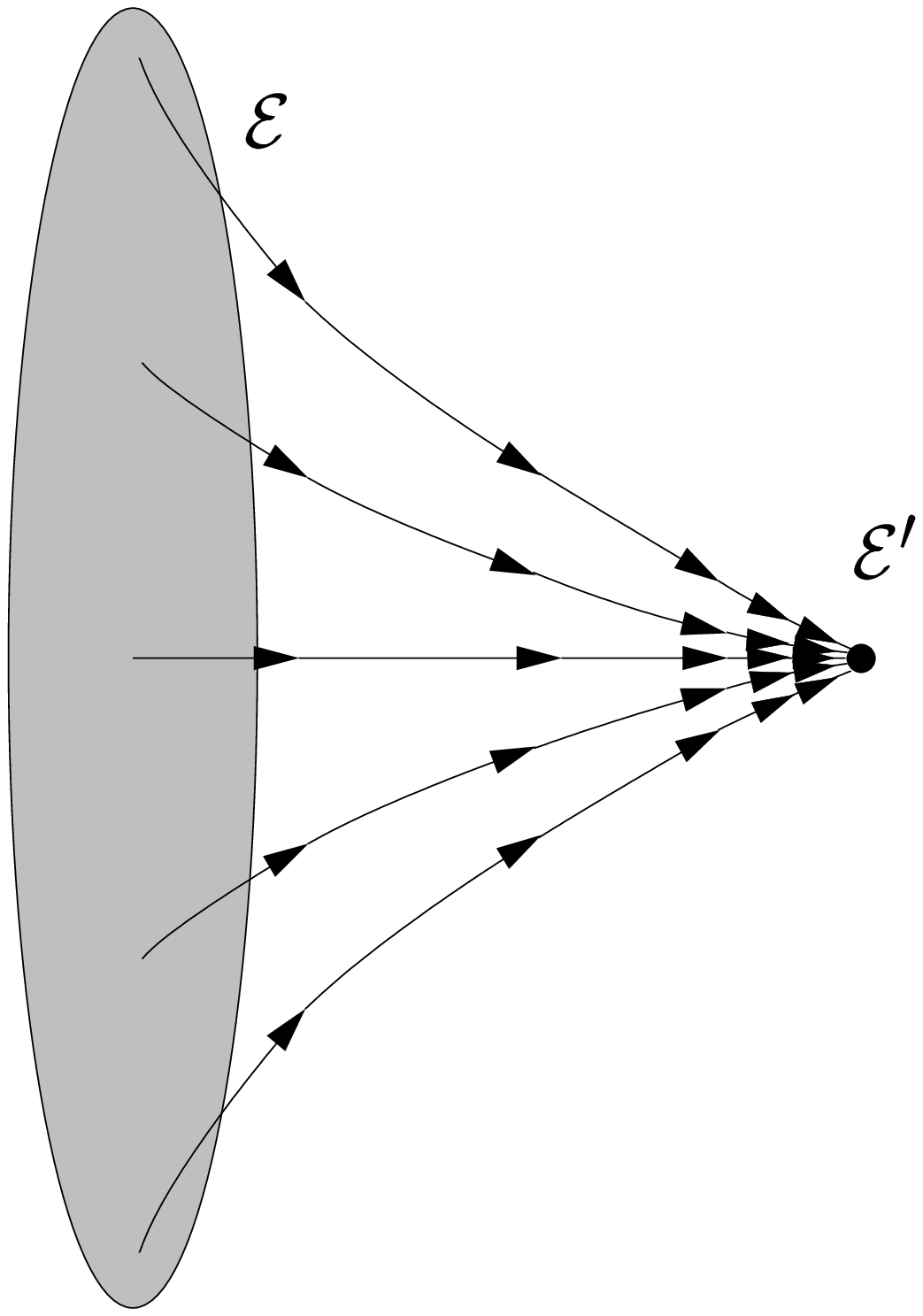}}%
\hfill
\subfigure[$N$ just below $N_c(d)$ ]{
\label{regions_n-grand}
\includegraphics[height=.54\linewidth,origin=tl]{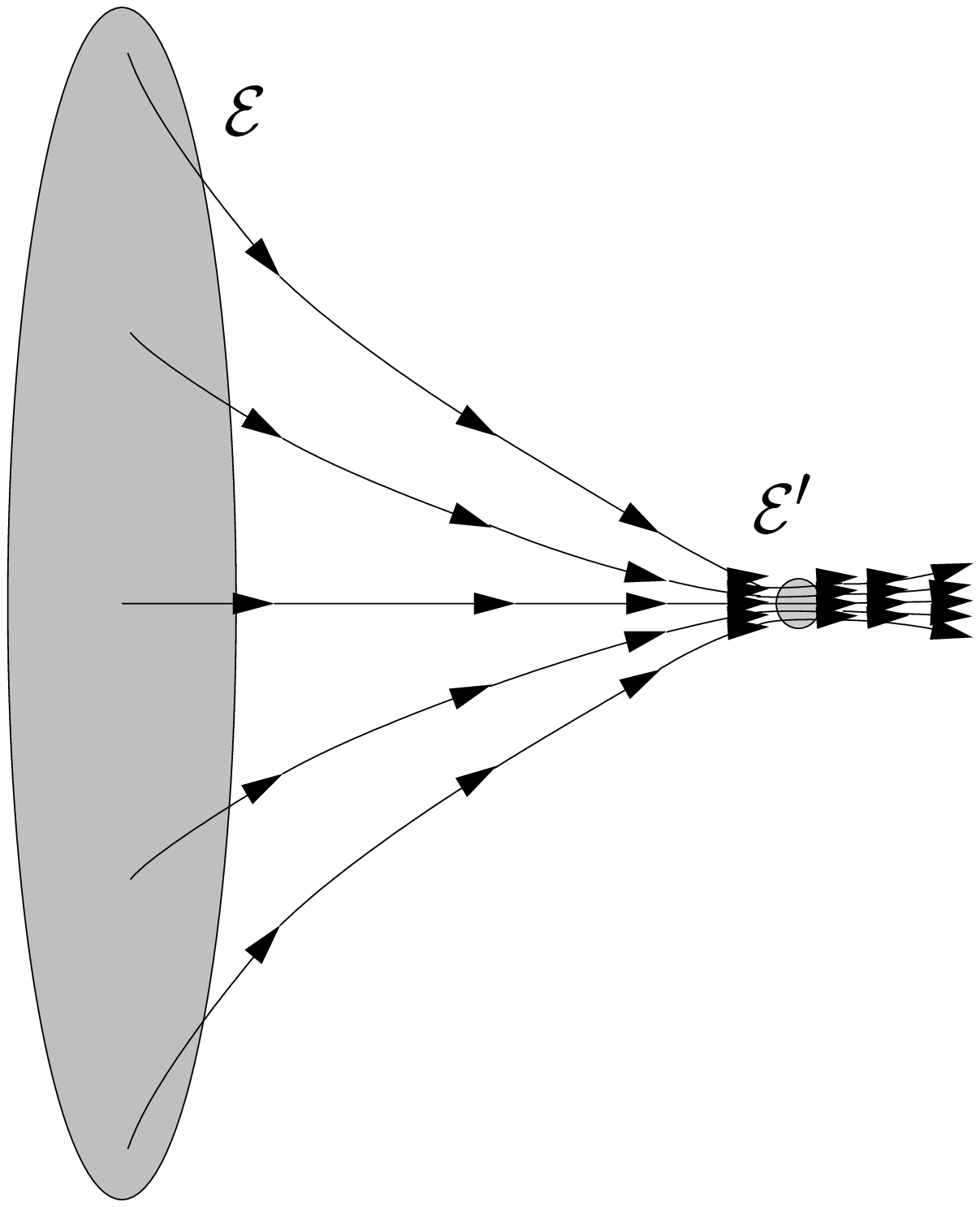}}
\caption{Schematic representation of the flow  a) for $N$ above  $N_c(d)$ and b) for $N$ just below $N_c(d)$. 
For the sake of clarity, we have represented  $\mathcal E'$
outside  $\mathcal E$ while it can  be included in it.}
\label{regions_pseudo_flot}
\end{figure}

To understand the evolution of the RG flow as $N$ is decreased, we
need to consider the space of {\it all} coupling constants, {\it i.e.}
the space such that to each point corresponds a microscopic Hamiltonian of 
a system. In this space, we
focus on the subspace $\mathcal E$ containing the representative
points, at $T=T_c$, of STA, STAR, V$_{N,2}$, BCT  and of all real materials
studied experimentally and, more generally, of all systems of physical
interest.  Let us now describe qualitatively the change of the RG
trajectories as $N$ crosses $N_c(d)$.

{\it i)} When $N$ is larger than $N_c(d)$, there exists a true stable fixed
point of the RG flow so that  all trajectories emerging from
$\mathcal E$ eventually end on this fixed point, see
Fig.~\ref{regions_n-huge}.  All systems exhibit scaling around the
transition and universality holds.

{\it ii)} As already stated, when $N$ is decreased slightly
below $N_c(d)$, the fixed point $C_+$ gets complex coordinates and
looses its direct physical meaning. In particular, the flow  no
longer stops  at a point, see Fig.~\ref{regions_n-grand}. Consequently,
the correlation lengths of systems in $\mathcal E$ do not diverge at
$T_c$. Strictly speaking, all systems  undergo  first
order phase transitions. However two facts  must be noted. Firstly, all the trajectories 
emerging from $\mathcal
E$ are attracted toward a small region in coupling constant space,
denoted by $\mathcal E'$ in Fig.~\ref{regions_n-grand}.  Secondly,
the flow in $\mathcal E'$ is very slow.

From the second observation, we deduce that for all systems in
$\mathcal E$ the correlation lengths at the transition are very large
--- although finite --- since they typically behave as the exponential
of the RG time spent around $\mathcal E'$, which is large.  Therefore, the transitions
are all extremely weakly of first order for systems in $\mathcal E$.
We thus expect scaling behaviors with  pseudo-critical exponents  for all physical quantities, with the
subtlety that this scaling aborts very close to $T_c$ where the true
first order nature of the transitions shows up.

As for the first observation ---  {\it i.e.} all trajectories are
attracted toward a small region $\mathcal E'$ ---, it allows to
conclude that all phase transitions are governed by a small region in
coupling constant space and that, therefore, universality almost
holds. In particular, the pseudo-critical exponents should be roughly
the same for all systems whose microscopic Hamiltonian corresponds to
a point in $\mathcal E$.

Let us study in greater detail the case where $N$ is just below
$N_c(d=3)$.  For such values of $N$, it is reasonable to approximate
$\mathcal E'$ by a point. The best approximation is clearly to choose
the point in $\mathcal E'$ that mimics best a fixed point, {\it i.e.}
the point where the flow is the slowest: the minimum of the flow
\cite{zumbach94}. To determine this so-called ``pseudo-fixed point'',
Zumbach \cite{zumbach94} has proposed to introduce a norm for the flow
and to determine the point where the norm is minimum.  He has
performed this approach in the context of a NPRG equation (LPA of the
Polchinski equation) where he has built the needed mathematical
structures. He has shown that, when a minimum exists, pseudo-critical exponents
characterizing pseudo-scaling can be associated with the pseudo-fixed
point, in the same way that true  exponents are associated with a true fixed
point (see Appendix~\ref{annexe_minimum} for more details).

 A natural assumption  to explain
the pseudo-scaling behaviors observed in real systems is that the
minimum of the RG flow mimics a true fixed point even for values of
$N$ not very close to $N_c(d=3)$.  For the Heisenberg systems, this 
position has been  advocated by Zumbach \cite{zumbach94} and by the present
authors \cite{tissier00}. 

Within our present approach we have
confirmed  the existence of a minimum of the flow, for values of $N$ just below
$N_c(d=3)$, leading to
pseudo-scaling and pseudo-universality \cite{tissier00}. By following
this  minimum we have confirmed that it persists down to
$N=3$ and have computed the associated pseudo-critical exponents, see
Table \ref{table_exp_critiques_3}. We also give in this table the
exponents found by Zumbach within the LPA of the Polchinski equation
for the same model \cite{zumbach94} and recall those found within the
six-loop approach of Pelissetto \etal{} \cite{pelissetto01a}.
\begin{table}[htbp]
\centering
\begin{tabular}{|l|l|l|l|l|l|l|l}
\hline
Method&Ref.&$\alpha$&$\beta$&$\gamma$&$\nu$&$\eta$\\
\hline
\hline
NPRG&\cite{tissier00}& 0.38&0.29&1.04&0.54&0.072\\
\hline
LPA&\cite{zumbach94}&  0.11&0.31&1.26&0.63&0.0\\
\hline
6-loop&\cite{pelissetto01a}&0.35(9)&0.30(2)&1.06(5)&0.55(3)&0.08\\
\hline
\end{tabular}
\caption{The critical and pseudo-critical exponents for
$N=3$. $\alpha,\beta$ and $\gamma$ have been computed assuming that
the scaling relations hold. The first line corresponds to our
nonperturbative approach, the second to Zumbach's work. In the third
line, we have recalled the six-loop results of Pelissetto \etal{} for
comparison.}
\label{table_exp_critiques_3}
\end{table}

The values that we have obtained within our calculation for the
critical exponents are not too far from --- some of --- those found
experimentally for group 2 of materials, see Eq.~(\ref{expoheisgroup2}),
as well as those found numerically for the STA, Table
\ref{table_exp_crit_heis_num}.  As usual, our truncation overestimates
$\eta$ and thus, at fixed $\beta$, underestimates $\nu$. It is
remarkable that the values of the pseudo-critical exponents we have
found at the minimum are in good agreement with those obtained within
the six-loop approach. This strongly suggests that there is a common
origin to these two sets of critical exponents. We shall come back on
this point later.

\subsection{Scaling with or without pseudo-fixed point: the Heisenberg
and XY cases}
\label{scaling_without}

Let us now argue that the preceding analysis, based solely on the
notion of minimum, is too naive  to give an explanation of the
pseudo-critical behaviors in the physically interesting cases.  Let us also  give a
qualitative picture that supplements  the concept of minimum. 

We have found that, when $N$ is  lowered below $N=3$, the
  minimum of the flow is less and less pronounced and that, for  some value of
  $N$ between 2 and 3, it completely disappears. Since several XY
systems exhibit pseudo-scaling in experiments or in numerical
  simulations, this means that the concept of minimum of the flow does
  not constitute the definitive explanation of scaling in absence of a
  fixed point. One encounters here the limit of the concept of minimum of the flow. 
First, it darkens the important fact  that the notion
  relevant to scaling is not the existence of a minimum  but
  that of a whole region in coupling constant space in which the
  flow is slow, {\ie}the $\beta$ functions are small. Put it
  differently, the existence of a minimum  does not
  guarantee that the flow is sufficiently slow to produce large
  correlation lengths. Reciprocally, one can encounter situations where
  the RG flow is slow, the correlation length being large so that
  scaling occurs even in absence of a minimum. The existence of a
  minimum is thus neither necessary nor sufficient to explain
  pseudo-scaling. Second, even when the minimum exists, reducing the region $\mathcal
E'$ to a point rules out the possibility of testing the violation of
universality. For instance, one  knows  that for $N=3$ universality
 is, in fact, violated, see Table \ref{table_exp_crit_heis_num},
 while a minimum of the RG flow is found. 
 This feature cannot be reproduced by the unique set of exponents computed
at the minimum.  The  opposite assumption,  done first 
 by Zumbach~\cite{zumbach94} and by the present authors~\cite{tissier00}, 
 was thus unjustified.

\begin{figure}[htbp] 
\subfigure[$N$ below $N_c(d)$ ]{
\label{regions_n-moyen}
\includegraphics[height=.4\linewidth,origin=tl]{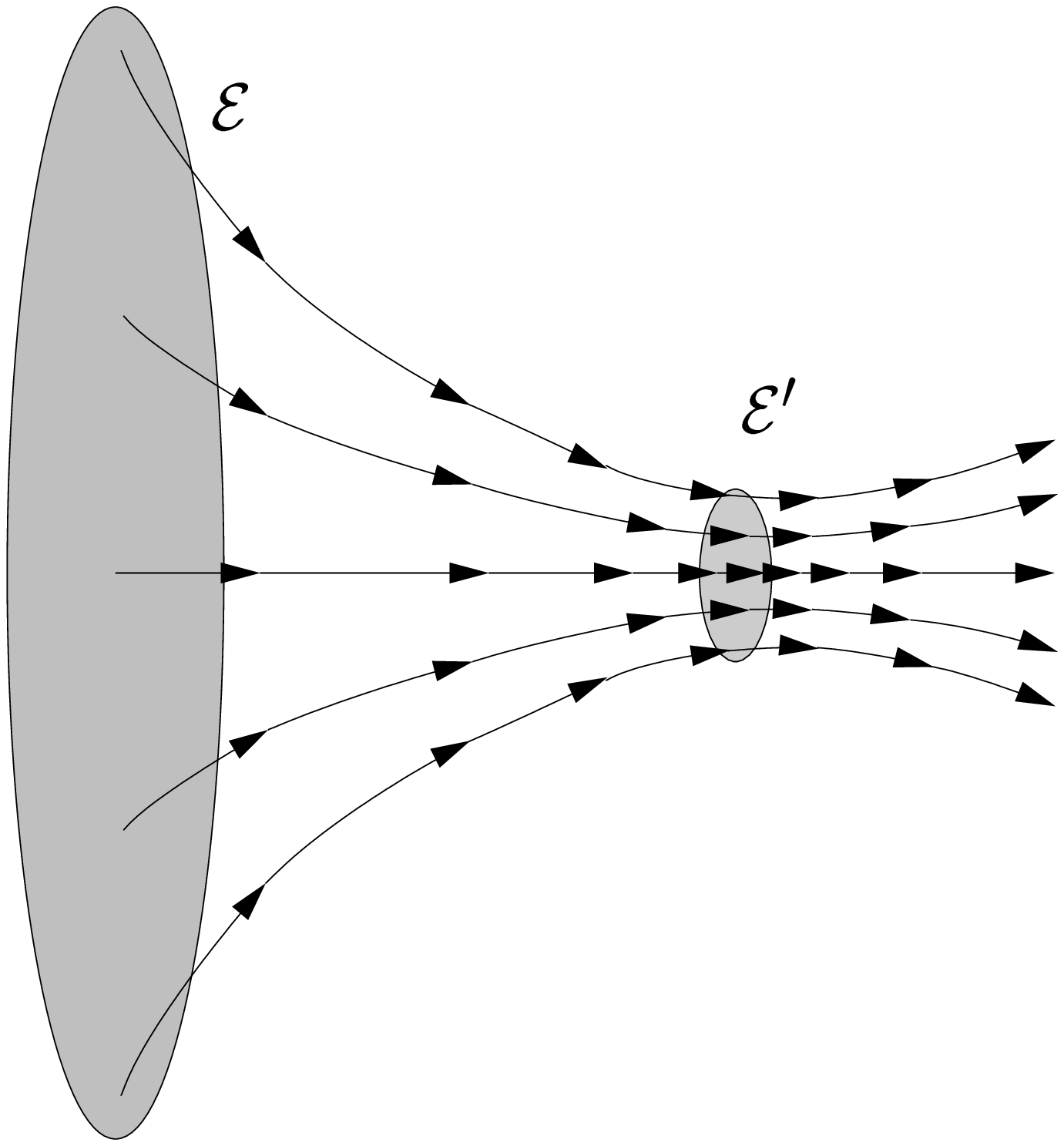}}%
\subfigure[$N$ well below $N_c(d)$ ]{
\hspace{1cm}
\label{regions_n-petit}
\includegraphics[height=.4\linewidth,origin=tl]{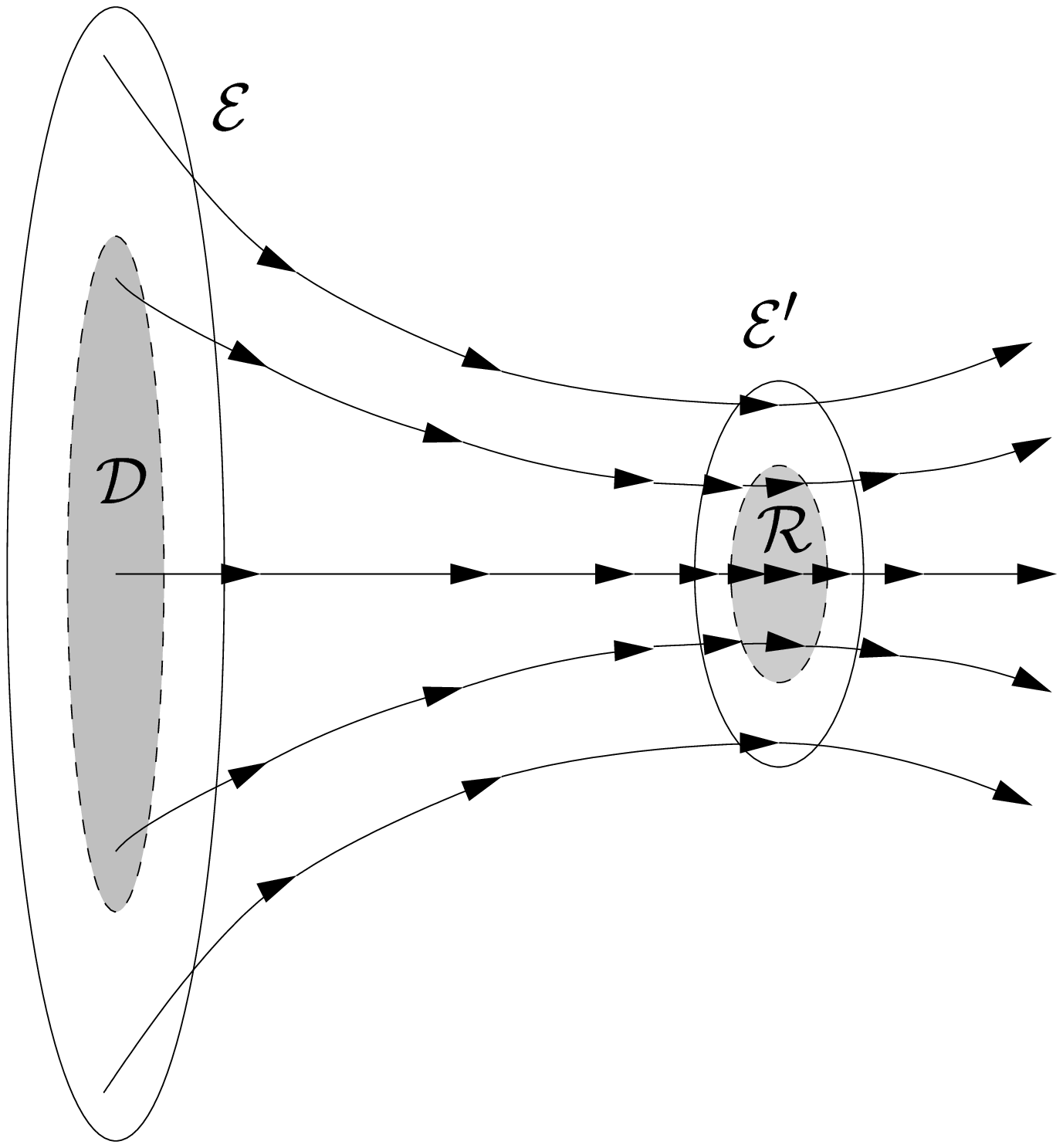}}%
\caption{Schematic representation of the flow a) for $N$ below $N_c(d)$ --- $N\simeq 3$ --- and
 b) for $N$ well below $N_c(d)$ --- $N\simeq 2$.  For the sake of clarity, we have represented
 $\mathcal E'$ outside $\mathcal E$ while it can be included in it.
 $\mathcal D$ and $\mathcal R$ are represented in grey. In a),
 $\mathcal D =\mathcal E$ and $\mathcal R =\mathcal E'$.}
\label{regions_pseudo_flot_2}
\end{figure}

Thus, even for very weak first order transitions, the beautiful
simplicity of second order transitions is lost and the finite extend
of the attractive region $\mathcal E'$ has to be taken into
account. To be precise, one  needs to define two subsets of $\mathcal E$
and $\mathcal E'$: $\mathcal D$ which is the region in $\mathcal E$  leading to  pseudo-scaling and $\mathcal R$,
 the subset of
$\mathcal E'$ which is the image of $\mathcal D$ in the RG flow, see
Figs.~\ref{regions_n-moyen} and \ref{regions_n-petit}.  Let us now consider
the characteristics of the flow when $N$ is varied.

Since for $N>N_c(d=3)$ all the systems in $\mathcal E$
undergo a second order phase transition, one  expects ---  thanks to continuity
arguments --- that  for $N$ slightly below  $N_c(d=3)$,  all systems in $\mathcal E$
exhibit pseudo-scaling and thus that  $\mathcal
D=\mathcal E$. At  the same time, $\mathcal E'$, the image  of $\mathcal
E$ is almost point-like --- see Fig.\ref{regions_n-grand} --- and universality holds.

As $N$ is decreased below $N_c(d)$, two phenomena occur.

{\it i)} While  $\mathcal D$ remains
equal to $\mathcal E$, the domain  $\mathcal E'$, which is initially point-like,
grows, see Fig.~\ref{regions_n-moyen}. This means  that  while
pseudo-scaling should be generically observed,  universality  starts to
be significantly violated: a whole spectrum of exponents should be
observed, the  size of $\mathcal E'$ providing  a measure of this  violation of
universality.

{\it ii)} For low values of $N$, the region  $\mathcal D$ leading to
pseudo-scaling  become smaller than $\mathcal E$, see
Fig.~\ref{regions_n-petit}. For systems defined by initial conditions in $\mathcal D$, the correlation lengths are
still relatively large but the pseudo-critical exponents can  vary  from system
to system according to the size of ${\cal R}$. For   systems  defined by initial conditions  in   $\mathcal E$ but 
not in $\mathcal D$,  the RG  flow is always fast, producing  small
correlation lengths at $T_c$. The corresponding systems
 undergo strong first order phase
transitions. Moreover, as $N$ decreases,  the flow in $\mathcal E'$ should
become more and more rapid so that, for systems in
$\mathcal E$, the correlation lengths at the phase transitions should
decrease. The transitions are thus expected to become more strongly of
first order for lower $N$.

The precise values of $N$ for which these changes of behaviors
occur as well as the shapes and extents of $\mathcal D$, $\mathcal R$
and $\mathcal E'$ can only be obtained from a detailed analysis of  both the  
microscopic Hamiltonian  and  of the  RG flow.  However, 
 independently  of
the details of the model under study, of the precise value of $N_c(d)$,
etc, one  expects  the following behavior:  as $N$ is decreased, a system that
 undergoes 
at large $N$ a second order
 transition undergoes, for $N$ just smaller than $N_c(d)$, a very weak
first order transition governed by the minimum. Then,  it should undergo 
 a weak first
order transition where the notion of minimum is no longer relevant and for
which universality does not hold anymore. Finally,  it should undergo
  a strong first
order phase transition.  In the spectrum of models
studied numerically, it is easy to see that the STAR, $V_{N,2}$ and  BCT
models with XY and Heisenberg spins nicely obey this prediction. For
$N=3$, they all show scaling and the phase transitions should be very
weakly of first order.  However, their exponents are clearly incompatible
with those of STA and with those associated with the minimum, see
Tables \ref{table_exp_crit_heis_num} and
\ref{table_exp_critiques_3}. The RG trajectories associated
with these systems are thus expected to pass through $\mathcal R$, but
far from the minimum.  One thus naturally expects that, when $N$ is
decreased down to $N=2$, no scaling behavior is observed for these
systems. This is indeed what is found in numerical simulations, see
Table~\ref{table_exp_crit_XY_num}. This strongly suggests that 
 $\mathcal D$ has shrinked between $N=3$ and $N=2$ and that
$N=3$ corresponds to Fig.~\ref{regions_n-moyen} and $N=2$ to
Fig.~\ref{regions_n-petit}.

\subsection{The integration of the RG flow for Heisenberg and XY systems}

\label{chapitre_integration}

In the previous section we have shown that the notion of minimum ---
or pseudo-fixed point --- in the RG flow is neither sufficient nor
necessary to explain the existence of scaling without a fixed
point. For this reason, one has to resort to another method to study
the physics of XY and Heisenberg frustrated magnets.  In practice, we
integrate numerically the RG flow around the transition temperature
$T_c$ and determine the behavior of the physical quantities such as
the correlation length, the susceptibility and the ``magnetization''
--- defined as $\sqrt{\tilde\kappa}$, see Eq.~(\ref{def_min}) --- as
functions of the reduced temperature $t_r=(T - T_c)/T_c$. 

\subsubsection{Three  difficulties}

Let us mention three  difficulties  encountered during the
integration of the flow. 

First, in principle, in the absence of universality, we should study each system
independently of the others. Thus, to correctly specify the initial
conditions of the RG flow, we should also keep all the microscopic
information relevant to the description of a given material. This
program remains, in the most general case, a difficult challenge
since this  would consist in keeping track of the lattice structure as
well as of the infinite number of coupling constants involved in the
microscopic Hamiltonian. However this is, in principle, possible. Actually, this
 has been done  with much
success  for certain classes of magnetic systems and fluids
described by $O(N)$ models \cite{seide99}  mostly within the
LPA \cite{parola85,parola95}.  Our truncations --- even the best one
--- are too restricted approximations to reach this goal since this  would at
least require to keep the {\it full} field dependence of the potential
$U_k(\rho,\tau)$.  We have thus used our flow equations to explain the
generic occurrence of pseudo-scaling in frustrated systems without
trying to describe the behavior of a specific system.  In practice, we
have computed the correlation length, magnetization and susceptibility
using a simplified version of our truncation keeping only the
potential part expanded up to order eight in the fields, a
field-independent field renormalization and discarding all the
current-terms involving four fields and two derivatives.  We have
checked that this {\it ansatz} leads to stable results with respect to
the addition of higher powers of the fields and inclusion of current
terms.

 Second, the truncations we have considered do
not allow to determine accurately the critical temperature. Indeed, 
in our approach we   perform  a local description of the
potential around the nontrivial minimum Eq.~(\ref{def_min}).   For a second order phase transition this does
 not matter since the nontrivial minimum, when it exists, is always the true one. However, for a  first order 
transition, the zero-field configuration, \ie with  $\vec\phi_1=\vec\phi_2=\vec 0$, plays a crucial role.  In 
effect, in this case, the transition temperature precisely
 coincides with the temperature  at  which the energy at the nontrivial minimum and 
 at  the zero-field configuration   are equal.  Since  we cannot  expect that our truncation  describes accurately
 the potential around the zero-field configuration, we  are
not able to compare the energy of this configuration with the
energy of  the  nontrivial minimum  and to determine the transition temperature accurately. 
 We discuss 
in more details this point  in Appendix \ref{chap_premier_ordre} and show that,  for a {\it weak} 
first  order transition,  this fact should not bias significantly our analysis.

The third  difficulty encountered in the integration of the flow is
that, in the absence of universality, the temperature dependence of
the physical quantities relies on the precise temperature dependence
of the microscopic coupling constants. We have used several {\it
ans\"atze} for the temperature dependence of the coupling constants
and have observed that, although it could be important for the details
of the results, it does not affect much the general conclusions. Thus,
we illustrate our results with the simplest {\it ansatz} consisting in
fixing all the couplings to temperature-independent values and by
taking a linear temperature dependence for $\kappa$ at the lattice
scale:
\begin{equation} 
\kappa_{k=\Lambda}=a + b T \ .
\label{dependance_T}
\end{equation}
For each temperature, we have integrated the flow equations and have
deduced the $t_r$-dependence of the physical quantities, such as the
``magnetization'', the correlation length, etc, around $T_c$. The
different coupling constants parametrizing the initial condition of
the flow have been varied to test the robustness of our conclusions.
This has allowed us to establish the following facts.

\subsubsection{The Heisenberg case}

\begin{figure}[tbp] 
\centering
\makebox[\linewidth]{
\label{log_aimantation}
\includegraphics[width=.95\linewidth,origin=tl]{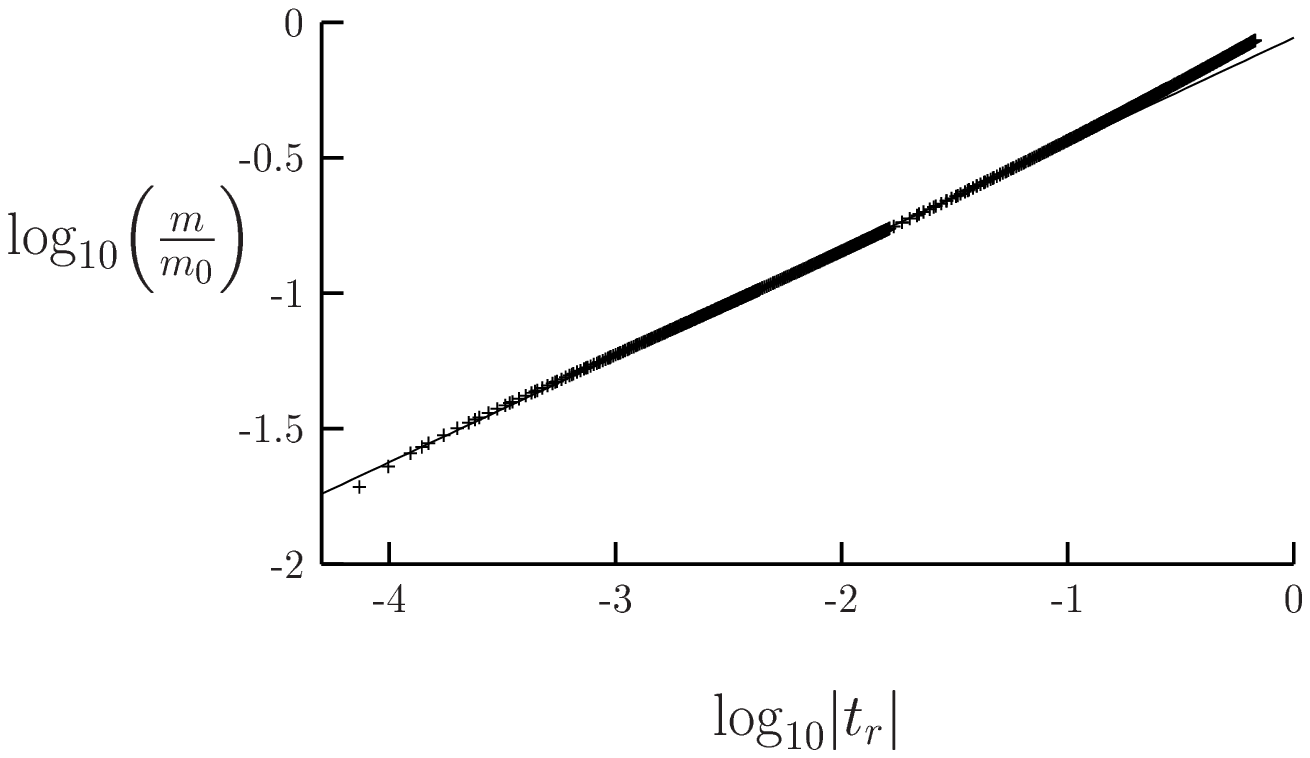}}
\makebox[\linewidth]{
\label{log_correlation}
\includegraphics[width=.95\linewidth,origin=tl]{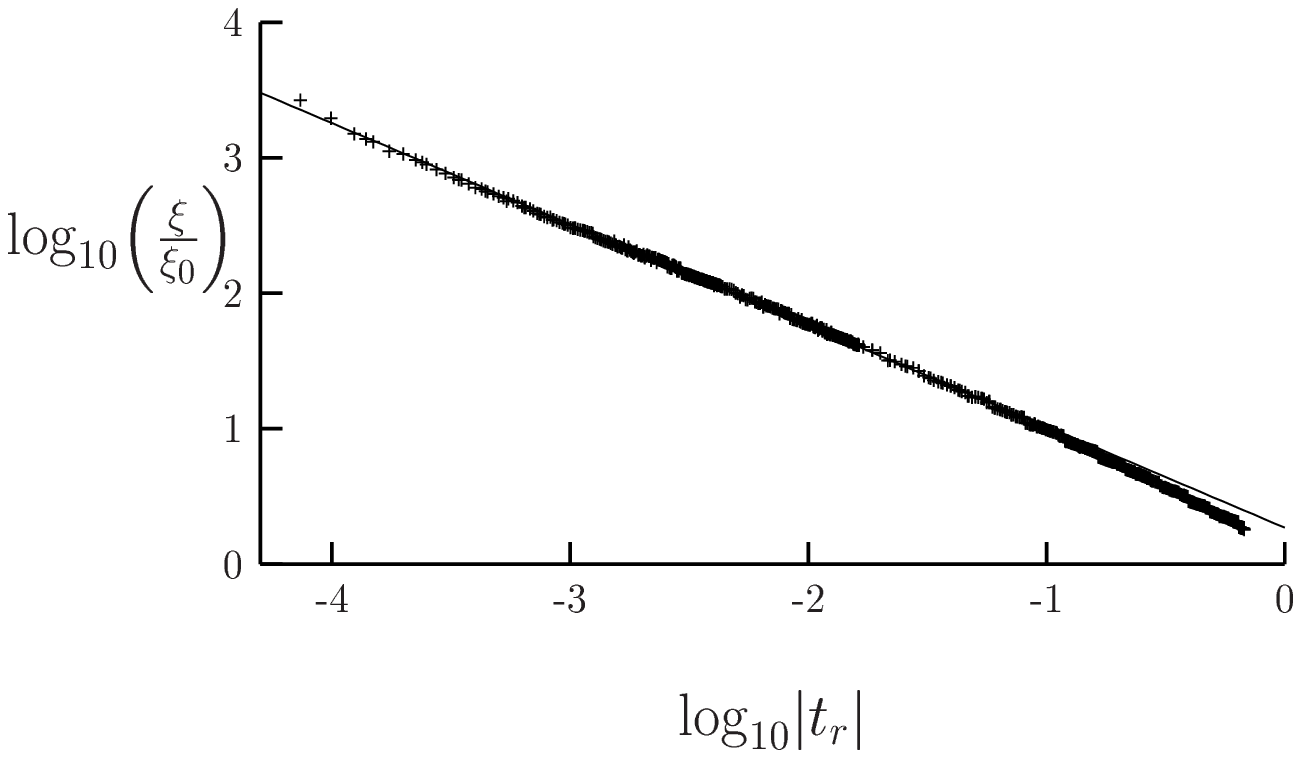}%
}
\caption{Log-log plot of the magnetization $m$ and of the correlation
length $\xi$ for $N=3$ as functions of the reduced temperature $t_r$.
The straight lines correspond to the best power law fit of the data.}
\label{log_m_xi}
\end{figure}

$\bullet$ For $N=3$, we can find initial conditions of the flow such
that for a wide range of reduced temperatures --- up to four decades
--- the physical quantities behave as power laws. From an experimental
viewpoint, this is all what is needed since scaling has been found on
temperature ranges that are even smaller. The kind of pseudo-critical
behaviors we find is illustrated on Fig.~\ref{log_m_xi}. 

$\bullet$ Varying the initial conditions of the flow, we observe that
this phenomenon happens in a wide domain of the coupling constant
space. This corresponds to the domain ${\cal D}$ previously defined,
see Fig.~\ref{regions_n-moyen}.

$\bullet$ Within ${\cal D}$, the pseudo-critical exponents vary smoothly:
$\beta$ varies typically between 0.27 and 0.42 and $\nu$ between 0.56
and 0.71.  These are only typical values since it has been impossible
to explore the whole space of coupling constants. Since for
$\beta\simeq 0.27$ one can find  $\nu\simeq 0.56$, the exponents of group 2 are
satisfactorily reproduced, see Tables~\ref{table_exp_crit_Heis_exp}
and \ref{table_exp_crit_heis_num}.  This shows in particular that
there exists, in ${\cal D}$, a set of ``microscopic'' coupling
constants that lead to the behavior observed in group 2.

\begin{figure}[tbp] 
\centering
\makebox[\linewidth]{
\label{log_aimantation_neq2}
\includegraphics[width=.95\linewidth,origin=tl]{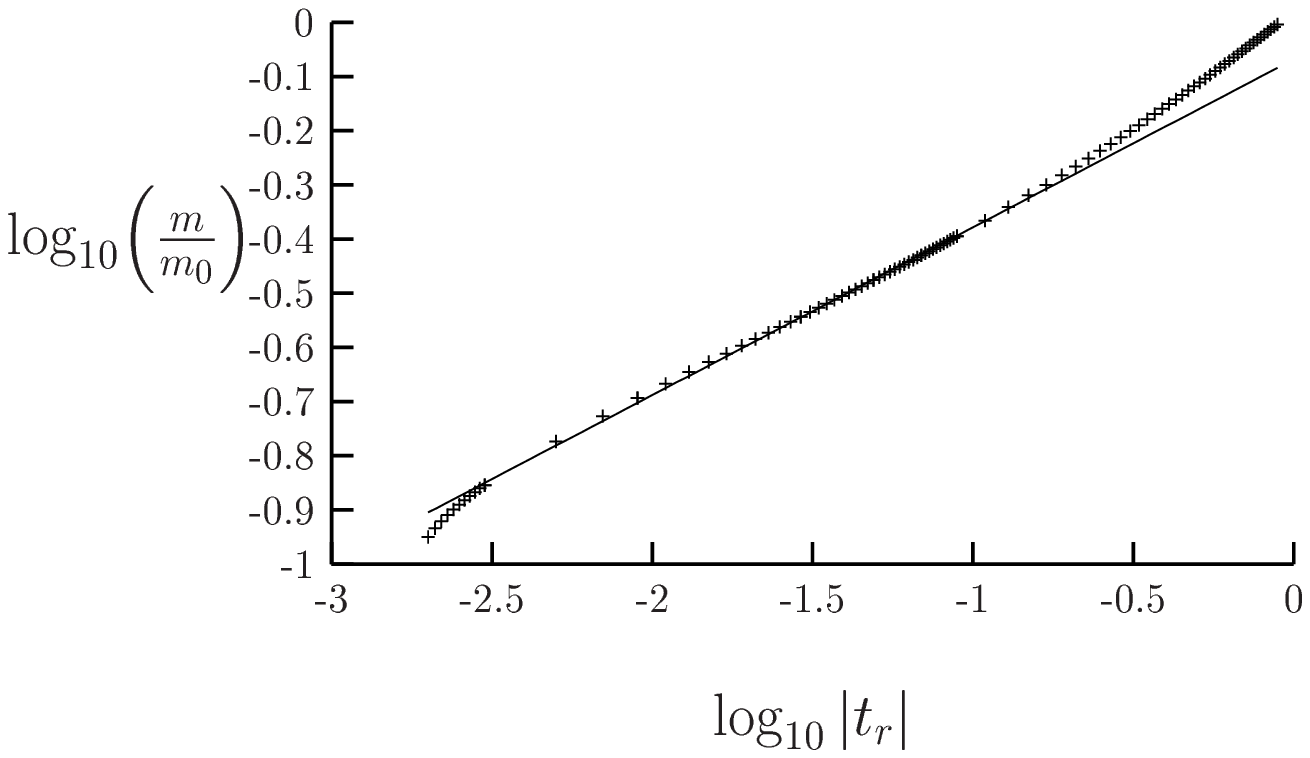}}
\makebox[\linewidth]{
\label{log_correlation_neq2}
\includegraphics[width=.95\linewidth,origin=tl]{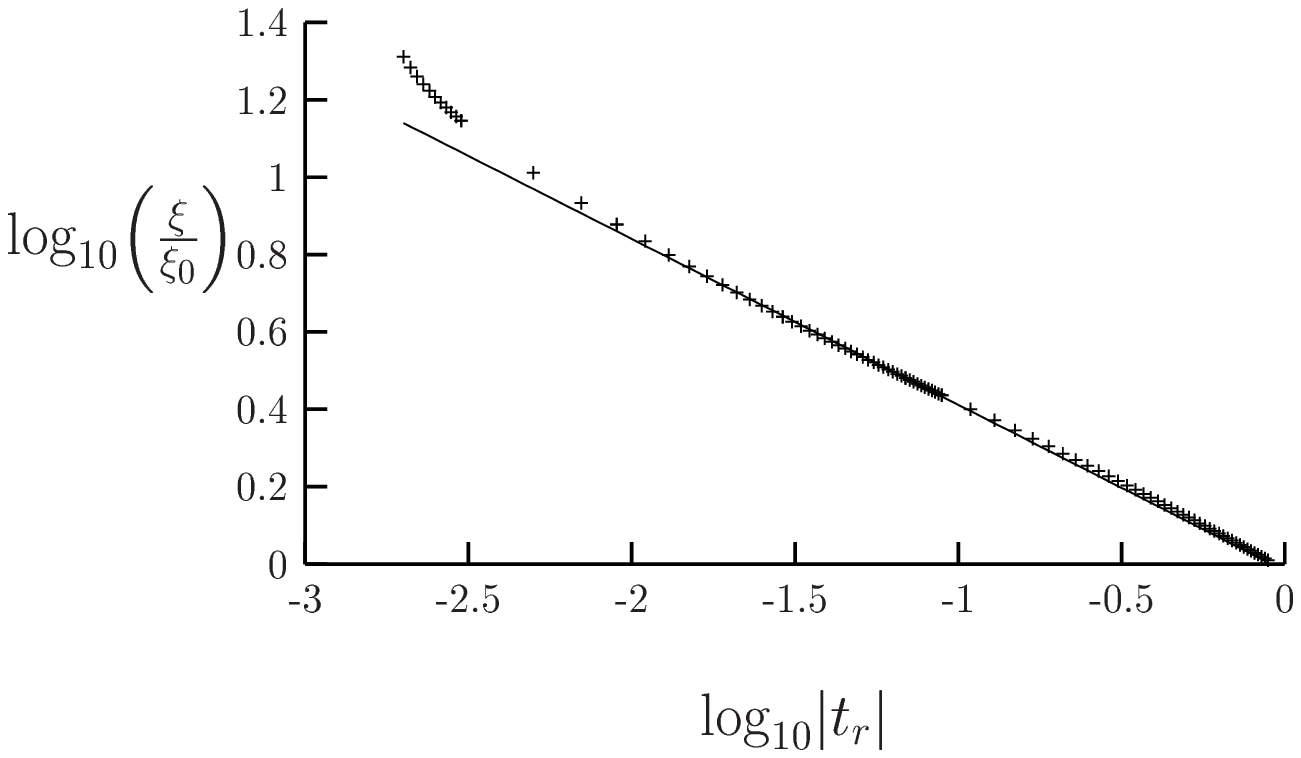}%
}
\caption{Log-log plot of the of the magnetization $m$ and of the
correlation length $\xi$ for $N=2$ as functions of the reduced
temperature $t_r$. The straight lines correspond to the best power law
fit of the data. The power-law behavior observed far from the critical
temperature breaks down for small $t_r$. The behavior of the
correlation length at small $t_r$ is an artefact of our truncation,
see Appendix \ref{chap_premier_ordre}}
\label{log_m_xi_neq2}
\end{figure}

$\bullet$ It is easy to find initial conditions leading to
pseudo-critical exponents in good agreement with those obtained in the
six-loop calculation, see Table \ref{table_expsixloop}. Actually, a
whole set of initial conditions lead to exactly the same (pseudo-) critical 
exponents as those found at six-loop $\beta=0.30(2)$,
$\nu=0.55(3)$. This corresponds to the region of the minimum of the
flow, see Table {\ref{table_exp_critiques_3}.

$\bullet$ In contrast, we have not found initial conditions of the RG
flow reproducing correctly the critical exponents of group 1, of STAR,
V$_{3,2}$ and BCT as well as negative values for $\eta$.  This can
originate {\it i)} in the overestimation of $\eta$ produced by our
truncation of $\Gamma_k$ in powers of the derivatives at order
$\partial^2$, Eq.~(\ref{action_generale}), {\it ii)} in the
impossibility to sample the whole coupling constant space, {\it iii)}
in the too simple temperature dependence of $\kappa_{\Lambda}$ that we
have considered, see Eq.~(\ref{dependance_T}).

$\bullet$ For a given value of one exponent, it is possible to find
several values for the other exponents. Thus we expect to find systems
sharing for instance almost the same $\beta$ but having quite
different values for $\nu$ and $\gamma$.

$\bullet$ At the border of ${\cal D}$, the temperature ranges over
which power laws hold become smaller and smaller.  In a log-log plot,
the $t_r$-dependence becomes less and less linear and the
pseudo-critical exponents more and more sensitive to the choice of $T_c$ made
for the fit. Finally, outside $\cal D$, no more power-law behavior is
observed.

$\bullet$ When we go from $N=3$ to $N=4$, we have observed, as expected,
that ${\cal D}$ becomes far wider and that the power laws hold generically
on larger temperature ranges. This is consistent with our discussion of
Section \ref{scaling_without}. Reciprocally, and as also expected, when  going
from $N=3$ to $N=2$, ${\cal D}$ becomes much smaller and the power
laws hold generically on smaller temperature ranges. Let us discuss this point in 
greater detail now.

\subsubsection{The XY  case}

$\bullet$ For $N=2$, one observes qualitatively the same type of
behaviors as for $N=3$.  However, as predicted above, ${\cal D}$ is
smaller and the power laws hold at best only on two decades of reduced
temperature, which is consistent with what is observed
experimentally. This is illustrated in Fig.~\ref{log_m_xi_neq2} where
we have represented log-log plots of the magnetization and correlation
length as functions of the reduced temperature. Note also the surprising behavior of the correlation  length that 
increases at small reduced temperature} (see Appendix \ref{chap_premier_ordre} for an explanation of this
 phenomenon.). 
 
$\bullet$ Within ${\cal D}$ the exponents vary on the intervals:
$0.25<\beta<0.38$ and $0.47<\nu<0.58$.

$\bullet$ We find initial conditions leading to exponents close
to those of group 2 (for Ho and Dy, see Table
\ref{table_exp_crit_heli_XY_exp}): $\beta=0.38$, $\nu=0.58$,
$\gamma=1.13$. These results are quite stable with respect to changes
of microscopic parameters.  This is in agreement with the stability of
$\beta$ in group 2. Interestingly, these initial conditions correspond
to small $\tilde{\mu}$ in our truncation Eq.~(\ref{troncation}), \ie
to initial conditions close the $O(4)$-invariant line: $\tilde\mu=0$,
see Fig.~\ref{ptfixe} where the $O(4)$ fixed point is denoted by V
\footnote{Let us emphasize that this $O(4)$ fixed point has nothing to
do with that of the Heisenberg system around $d=2$, see Section
\ref{chapitre_NLS_onXo2}.}. Thus, during a large part of the flow, the
trajectory remains close to the $O(4)$ fixed point before bifurcating
away from this point.  This is perhaps the reason why the value of
$\beta$ for  materials of group 2 is close to that associated with an
$O(4)$ behavior --- $\beta_{O(4)}=0.382$ --- a fact that has been
repeatedly noticed by experimentalists.  Note however that the other
exponents are not close to the $O(4)$ values: $\nu_{O(4)}=0.738,
\gamma_{O(4)}=1.449$.

$\bullet$ We also easily find initial conditions leading to
$\beta=0.25$, corresponding to group 1, essentially composed of STA
systems. The power laws then hold on smaller ranges of temperatures
and the critical exponent $\beta$ is more sensitive to the
determination of $T_c$ and to the initial conditions. For such values
of $\beta$, we find that $\nu$ varies between 0.47 and 0.49, which is
somewhat below the value found for CsMnBr$_3$, see
Table \ref{table_exp_crit_STA_XY_exp}.

$\bullet$ The two previous points suggest that both helimagnets ---
such as Ho or Dy --- and STA --- such as CsMnBr$_3$ --- can be
described by the same field theory but with exponents at the two ends
of the spectrum. It is actually also possible that helimagnets display
a different kind of physics because of the presence of long range
interactions or because of the presence of surface effects
\cite{thurston93}.

$\bullet$ As in the $N=3$ case, we can easily find initial conditions
leading to pseudo-critical exponents close to those found in the
six-loop calculation, Table \ref{table_expsixloop}.  For instance, for
initial conditions leading to $\beta=0.33$, we find typically:
$\nu=0.56$ and $\gamma=1.07$.

$\bullet$ As in the Heisenberg case, we have not been able to find
initial conditions of the RG flow leading to negative values of
$\eta$.

Let us now comment our results.

\subsubsection{Comments}

The main feature of the physics of Heisenberg and XY frustrated
magnets --- scaling behaviors {\it without} universality --- is
reproduced, at least qualitatively and, to some extent,
quantitatively.  This behavior finds a {\it natural} explanation:
there exists a whole domain ${\cal D}$ in the space of coupling
constants such that the RG trajectories starting in ${\cal D}$ are
``attracted'' toward a region ${\cal R}$ where the RG flow is slow so
that there is pseudo-scaling. Since ${\cal R}$ is {\it not} reduced to
a point, there exists a whole spectrum of exponents and not a {\it
unique} set.   The occurence of strong  first order phase transitions, that are observed 
in some materials and simulated systems, is explained by the  RG
trajectories starting out of  ${\cal D}$.

  Let us now stress that since universality is  lost, 
 the determination of the precise pseudo-critical exponents associated 
with a given material or system is obviously more difficult than the determination 
of the usual --- universal --- critical exponents characterizing a second order phase
 transition.
As already said, computing them would indeed require to know precisely the
microscopic structure of the materials or systems studied --- providing the initial conditions of the RG flow
--- and to take into account the full field-dependence of the
potential $U_k(\rho,\tau)$.

\section{Possible tests  of our scenario}

\label{chapitre_checkscenario}

There are several tests  that can be performed both experimentally and
numerically to confirm our proposals. Let us start by the Heisenberg
case.

$\bullet$ It is not clear, up to now, whether the materials of group 1
--- VCl$_2$ and VBr$_2$ --- are really three-dimensional  Heisenberg STA, at least for
a temperature range wide enough to measure exponents.  It would be
very interesting to re-study these materials and to measure all
exponents for each of them. This could allow to confirm experimentally
the absence of universality.

$\bullet$ Since we predict that they can be violated, there is clearly
a need to check the scaling relations as well as the negativity of
$\eta$.  The experimental determination of the exponents $\gamma$ and
$\nu$ for the two groups of Heisenberg materials is still much too
poor. It is also necessary to have an estimation of both the
systematic and statistical errors to strengthen our conclusion on the
negativity of $\eta$.  Let us however recall that the first order
nature of the transitions in Heisenberg systems is likely to be much
weaker than in XY systems. Thus the violations of both the scaling
relations and the positivity of $\eta$ should be much more difficult
to prove experimentally in this case.

$\bullet$  It remains mysterious why, in
CsMn(Br$_{0.19}$I$_{0.81}$)$_3$, such strange values of the exponents
$\gamma$ and $\nu$ have been found, see Table
\ref{table_exp_crit_Heis_exp}. As we have already discussed in point
{\it i)} of Section \ref{chap_exp_heis}, we find unconvincing the
arguments proposed in \cite{kakurai98} to explain them.  Remeasuring
these exponents could provide accurate results for $\gamma$ and $\nu$
from which universality and the negativity of $\eta$ could be tested.

$\bullet$ Most probably STAR and the V$_{3,2}$ model undergo both
first order transitions since $\eta$ is found negative for these
models. It would be extremely interesting to study a sequence of
models that interpolate between STA and STAR to see how the effective
exponents change with the deformation of the model.

$\bullet$ We have already noticed that the exponents found in the
six-loop calculation are very close to the pseudo-critical exponents found at
the minimum of the RG flow in our approach.  It is important to know
if this is just an accidental coincidence or if they correspond to the
same fixed point, real in one approach complex in the other.

Let us now discuss the XY case. Most of the points discussed in the
Heisenberg case can be transposed here: necessity to check the scaling
relations and the positivity of $\eta$, possibility to interpolate
between the STA and STAR.  Here, however, we are in a better position
to obtain conclusive results since the transition is expected to be
more strongly of first order.

$\bullet$ A better determination of $\nu$ in CsMnBr$_3$ would help to
confirm that $\eta$ is indeed negative.  We also expect to have a
weaker universality and thus a faster change of the exponents as the
microscopic details of the model are varied.  In particular, a precise
determination of $\alpha$ in the different materials of group 1 could
lead to incompatible exponents --- they are up to now only marginally
compatible --- and would give a direct proof of the lack of
universality.

$\bullet$  On the numerical side, the sequence of models interpolating
between STA and STAR should lead to rapidly varying exponents. Thus
the lack of universality in this case should be much simpler to prove
numerically than in the Heisenberg case.  For STA, it would also be extremely
interesting to determine $\eta$ independently by the two scaling
relations $\eta=2\beta /\nu-1$ and $\eta=2-\gamma/\nu$.  As far as we
know $\eta$ has mainly been determined using $\gamma/\nu$.   According to
our scenario the two determinations should not 
coincide. However, they are probably both negative.

\section{Consequences for perturbation theories}

\label{chapitre_consequences}

Frustrated magnets represent a unique controversial example of systems
for which almost all the possible perturbative and nonperturbative
approaches have been used, sometimes with a  very high  precision. This
allows to draw several conclusions about the relative predictive
power of these different methods. Firstly, it appears that the
{\it low-order}  results obtained within the NL$\sigma$ or GLW models fail to
correctly describe the physics in three dimensions. Indeed, we recall
that the one-loop result of the NL$\sigma$ model predicts a second
order phase transition with a $O(4)$ behavior while the GLW approach
leads to first order phase transitions for all values of $N$ smaller
than 21.8. Secondly, frustrated magnets probably provides the first
example where high-order perturbative results is questionable.  We shall now discuss the status
of the various perturbative approaches at the light of our results.

\subsection{The NL$\sigma$ model approach}

\begin{figure}[htbp] 
\begin{center}
\includegraphics[width=.9\linewidth,origin=tl]{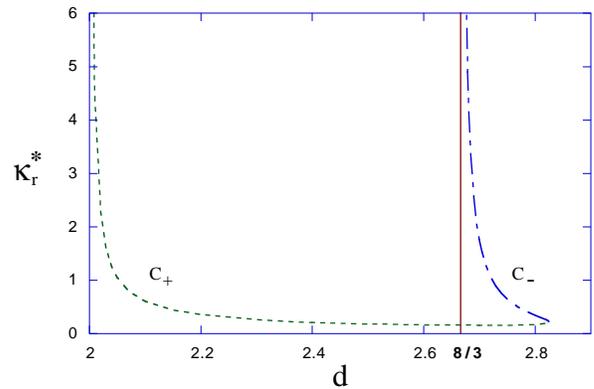}\hfill%
\end{center} 
\caption{The stable --- $C_+$ --- and unstable --- $C_-$ --- fixed points  as functions of the dimension $d$.
 The fixed points are parametrized  by the quantity  $\kappa_r^*$ which is proportionnal to the inverse temperature
 of the NL$\sigma$ model. The fixed point $C_-$ appears in a dimension $8/3$ and collapses  with the stable fixed
 point $C_+$ in $d\simeq 2.83$.  }
\label{collapse} 
\end{figure}

Let us first consider the NL$\sigma$ model approach, focusing on the
Heisenberg case since it is notorious that this approach does not work
for XY spins. The very likely existence of a line $N_c(d)$ going from
$d=2$ to $d=4$ confirms what has been already anticipated in Section
{\ref{chapitre_perturbatif}}: the predictions based on this approach
are incorrect as for the physics in $d=3$. Indeed, the shape of this
line implies  that the $O(4)$ fixed point predicted in the Heisenberg
case --- that likely exists at all orders of perturbation theory ---
very probably disappears between two and three dimensions. Actually,
following this fixed point,  that we call $C_+$ for an obvious reason,  from $d=2$  with the
simplest
$\phi^4$-like truncation, we have found several interesting
features. First, infinitesimally close to $d=2$, we find that  $C_+$ is characterized
by an exponent $\nu$ of the $O(4)$ universality class.  Second, as $d$ is increased, the exponent
$\nu$ associated with $C_+$  becomes more and more
different from that characterizing an $O(4)$ transition. Third, we
find that an  unstable fixed point, $C_-$,  shows up in a
dimension $d>2$. As the dimension is further increased, the fixed points
  $C_+$ and $C_-$  get
 closer together and eventually coalesce
in a dimension less than three.  This  phenomenon  is illustrated in Fig~(\ref{collapse}) in 
the case of the $O(3)\times O(3)$ model  at the lowest order in the
 field expansion
 \cite{tissier00b}.

The collapse of the fixed points  for different values of $N$ generates the curve
 $N_c(d)$. This curves is well known from the perturbative expansion performed around
 four dimensions. Within our approach,  this curve can be followed  when the dimension
 is lowered down to $d=2$.   There, for a given --- low ---  value of $N$, the curve  $N_c(d)$ 
provides  the value of $d_c(N)$ for which  the stable  fixed point obtained  within the NL$\sigma$
model  approach  collapses with another ---  unstable --- fixed point. Since this unstable fixed
point   is {\it not}  found in the low-temperature perturbative expansion we  therefore obtain  here
a nonperturbative  solution to the breakdown of the NL$\sigma$ model approach. For $N=3$,
one gets $d_c\simeq 2.8$. Note that obtaining an accurate determination of the  dimension $d_c$
where the fixed points collapse would require to consider better 
truncations in fields of
$\Gamma_k$ than those we have considered. However, as already explained in the $O(N)$ case, see Section \ref{difficulty}, the
stable fixed point coalesces in this case with one of the multicritical points. Thus  it is 
impossible, within our truncation, to follow it smoothly for $2.1\lesssim d\lesssim 2.5$. With our
best truncation, we are anyway able to give an estimate of this dimension: $d_c\simeq 2.6-2.7$
which fits well with the results  of Pelissetto \etal~\cite{pelissetto01b},  see Fig.~\ref{nc_de_d}.

Frustrated magnets thus provide a situation where there is a manifest
breakdown of the low-temperature expansion of the NL$\sigma$ model.
This is not the first occurence of such a breakdown. The case of the
two-component nonfrustrated $O(2)$ system has already exemplified the
inadequacy of the low-temperature expansion to explain the existence
of a phase transition for these systems in two dimensions, the
Berezinskii-Kosterlitz-Thouless phase transition
\cite{berezinskii70,kosterlitz73}. There is however an important
difference between the case of XY nonfrustrated spins and that of
Heisenberg frustrated spins. Indeed, in the former case, the
low-temperature expansion performed on the corresponding $O(2)$
NL$\sigma$ model leads to a free theory to all orders in the
temperature $T$ in $d\ge 2$. This result is however known to be
incorrect for XY spins themselves or for the systems that belong to
the same universality class --- like $^4$He --- that both undergo a
phase transition in $d\ge 2$. In this case, the unability of the
low-temperature expansion to correctly describe the physics makes no
doubt and one is invited to turn to other methods: Coulomb-gas
\cite{villain75} or spin-vortices \cite{berezinskii70,kosterlitz73}
formulations in two dimensions or GLW model approach in three
dimensions. On the contrary, in the case of Heisenberg frustrated
spins, the low-temperature expansion leads to a nontrivial behavior
--- due to the nonabelian character of the $SO(3)$ group --- so that
the inadequacy of the low-temperature perturbation theory is not so
obvious.

It remains to understand the very origin of this failure of the
low-temperature perturbation theory. In the case of XY nonfrustrated
spins, it clearly lies in the existence of nontrivial topological
configurations, called vortices, that are not taken into account in a
low-temperature expansion. In the case of Heisenberg frustrated
magnets, the influence of nontrivial topological configurations on the
phase transition in three dimensions has also been invoked (see
Section V-A). It remains however to confirm that these configurations
indeed play a  fundamental  role and to know, for instance, if they are responsible
for the first order character of the transitions in three
dimensions.

 This is a delicate question. Indeed, whereas  the perturbative approach to the 
NL$\sigma$ model misses  topologically nontrivial configurations, the GLW and effective
average action approaches are very likely sensitive to such 
vortices. In effect, both approaches correctly reproduce
the physics of three-dimensional XY nonfrustrated spin systems at the transition which
is very likely driven by vortices \cite{halperin81,antunes01}.  However, within these approaches, it is still
  not clear how the vortices are  taken into account. Therefore, disentangling vortices and spin-waves and 
understanding the respective role of each kind of excitation within the phase transition  remains a theoretical
 challenge.

\subsection{The GLW  model approach vs the  NPRG approach}

 We now discuss  the  relationship between  the weak-coupling  results obtained 
within the GLW model approach ---  in particular, the six-loop computation --- and our results. 
A natural question  arises: how is it  possible to reconcile these results  together  and 
what does this imply for the different approaches?

We have noted an important fact:  the critical  exponents found for $N=3$ in the six-loop
 calculation and in our
approach --- at the minimum of the flow  --- are very close (see Sections
\ref{chapitre_physics_d3} and \ref{chapitre_integration}).  We  have also 
found  very close exponents for $N=2$ (see Section
\ref{chapitre_integration}) with the only difference that there is no
minimum in the flow in this case. This is a rather strong indication
that the two sets of exponents have a common origin. This leads us to formulate some
proposals  to reconcile the two approaches.

The first one  is that  the  fixed point that appears as  real in the
six-loop calculation and complex in our approach is, actually, a complex one. This would  mean that 
the computations  of Pelissetto \etal{} and Calabrese \etal{}  is, actually, not converged as for the nature ---
 real or complex --- of this fixed point whereas it is  almost converged as for the exponents. We shall not
 speculate too much  about the
origin of this ---  hypothethical --- failure of the
weak-coupling approach. Let us  mention again however  that  the perturbative series obtained in the case of
 frustrated magnets appear to be  rather particular since  the critical properties deduced from them strongly
 depend on the order of the series.  We recall  that there is no nontrivial fixed point
 for $N=2$ and $N=3$ up to three loops; they only appear at five loops.  Also, the  six-loop results has been 
obtained in a region where
 the perturbative expansions are {\it not} Borel summable \cite{pelissetto01a}. It is clear that this question 
 deserves further investigations.  Frustrated magnets could appear
as the first exemple of a breakdown of  a weak-coupling perturbative analysis.

The second proposal is that, reciprocally, within the NPRG method, the lack of fixed point in the XY and Heisenberg
 cases is due to artefacts of the truncation in fields  and/or derivatives. Only the recourse to other kinds of
 expansions of   the effective action $\Gamma_k$ --- involving  either the {\it full} function  $U_k(\rho,\tau)$ 
or the {\it full} momentum dependence ---  could  lead to unambiguous statements. In this respect, we however recall
 that the LPA approach of Zumbach, that  involves  the full field-dependence of the potential, has led to no fixed
 point for $N=2$ and $N=3$.

\section{Conclusion and prospects}

\label{chapitre_conclusion}

 On the basis of their specific symmetry breaking scheme, it has been
proposed~\cite{garel76,yosefin85,kawamura85,kawamura86,kawamura87,kawamura88} that
the critical physics of XY and Heisenberg  frustrated systems in three dimensions
could be characterized by critical exponents associated with a {\it new} universality class. From this point of view,
 the study of frustrated magnets has been rather disappointing,  the experimental and numerical  contexts  excluding
  such an 
hypothesis. At the same time, the phenomenology of frustrated magnets has displayed
a  novel kind of critical behavior ---  {\it generic} scaling {\it without} universality~\cite{tissier01} --- 
requiring  the use  of 
new theoretical approaches.

High-order perturbative calculations in $d=~3$~\cite{pelissetto01a,calabrese02,calabrese03b}
provide an  explanation to the lack of universality in frustrated magnets: the focus character of
 the fixed point  induces spiral-like RG trajectories from which, according to Calabrese \etal\ \cite{calabrese03b},
 follows  varying effective critical exponents~\cite{calabrese02,calabrese03b}. We  have, however, underlined several
 drawbacks of this explanation. The major one lies in its 
lack of naturalness: a fine-tuning of initial conditions seems to be necessary to match with the phenomenology. 
  Another drawback of the perturbative approach is that, being restricted to investigate the physics in three
 dimensions, it cannot provide a general picture of what happens between two and four dimensions. In particular
 it provides no explanation to the failure of the NL$\sigma$ model approach. 

Within the framework of  a  NPRG  approach, the  generic and  
nonuniversal scaling finds  a natural explanation in terms of the slowness and
 ``geometry'' of the flow. This method  also explains   the mismatch between the
 different
 perturbative approaches by means of a mechanism of annihilation of fixed points 
 in a dimension
 between two and three  that invalidates   the low-temperature perturbative 
 approach performed from  the NL$\sigma$ model. As said along this article, more
 work, in particular the recourse to  ansatz  involving the full field-dependence or full momentum dependence of the 
effective action,  is  probably necessary to completely  understand the situation. This includes the clarification
 of the relation between our approach and the six-loop results. However, the main features of frustrated magnets 
appear now to be well described.  

The main result of this article is the explanation of the {\it generic} character of weak first order phase
 transitions in frustrated magnets.  Being given the closeness  between
these systems  and others systems --- see the Introduction --- it is  natural to  speculate about the degree
 of generality of  this phenomenon. 

 Within our approach, the  generic character of the weak first order phase transition 
 appear to be strongly related to the   proximity of  the number of components  $N$ of 
the system under study with $N_c(d=3)$. For frustrated systems, it appears that 
this value  is of  the same order than  the physically  relevant values of $N$, $N=2$ and $N=3$. This could be  a
 very specific 
property of  the frustrated systems.  We now argue that, on the contrary, this property is
likely to be common to many other systems.

Let us recall that the line $N_c(d)$ corresponds to the collapse of
two fixed points, one of them governing the phase transition. This
phenomenon cannot happen in theories with only one $\phi^4$ coupling
constant (\ie in $O(N)$ models) since, in this case, there is only one
fixed point apart from the gaussian. However, for theories with $c$
coupling constants, we expect $2^c$ perturbative ---  real or
complex --- fixed points in $d=4-\epsilon$ since, at one-loop, the $\beta$-functions
are quadratic in the coupling constants. When the number of components
of the field is varied, these fixed points move in the coupling
constant space and it is generically observed that they meet and
collapse for some critical value $N_c(d)$. Many examples are now known in
the literature. Let us review some of them.

 Let us first consider the generalization of the
model studied in this article consisting in $P$ orthonormal
$N$-component vectors. It  has  a $P$-dependent critical value of $N$ given
at one-loop by \cite{zumbach94}: $N_c\sim 10 P$. For $P=3$, one finds
at two loop order \cite{pelissetto01b}: \be N_c(4-\epsilon)=32.49-
33.72 \epsilon \ .  \ee

In the Abelian Higgs model coupled to a $N$-component scalar field,
relevant to superconductors, $N_c(d)$ is found at two-loop order to be
\cite{lubensky74,lawrie82,arnold94}:
\begin{equation}
N_c(d=4-\epsilon)=182.9 - 242.7 \epsilon \ .
\end{equation}

In a $SU(2)$ gauge model
coupled to bosons, it is given at two loop order by \cite{arnold94}: 
\begin{equation}
 N_c(d=4-\epsilon)=718 - 990.8 \epsilon \ .
\end{equation}

In a $O(p)$ gauge theory coupled to $N$ scalar fields (in the vector
representation) it is given at one-loop by \cite{ginsparg80}: $N_c\sim
40 p$.

In all these examples, we observe that $N_c(d)$ decreases very steeply
when $d$ decreases. This is in line with our expectation that large
$N$ and small $d$ favour continuous phase transitions. In particular,
as far as we know, in all NL$\sigma$ models relevant to systems whose
order parameter is continuous, a stable fixed point is found in
$d=2+\epsilon$ for all $N>2$. This is in particular the case for the
NL$\sigma$ model supposed to describe the physics of the abelian Higgs
model in $d=2$ \cite{hikami79,lawrie83}. This suggests that $N_c(d=2)$
is always smaller or equal to 2. It is interesting to notice that
this bound is probably reached in frustrated systems
\cite{pelissetto01b}, see Section \ref{three-loopresults}. It is thus
extremely probable that in many systems the curve $N_c(d)$ has a
similar shape as the one found in frustrated systems, see
Fig.~\ref{nc_de_d}. This suggests that many systems could exhibit
weakly first order transitions in $d=3$ {\sl without any fine-tuning
of parameters} \footnote{We could also add the $q$-state Potts model
which is known to have $q_c=4$ in $d=2$ and probably $2<q_c<3$ in
$d=3$. The existence of a critical value of $q$ corresponds also to a
collapse of fixed points \cite{newman84b}. Note however that the Potts
model has a discrete symmetry and thus cannot be analyzed as the
systems with a continuous order parameter for which Mermin-Wagner
theorem applies in $d=2$.}.  The effective average action method 
 should be ideally suited to study these situations.

\acknowledgements

 We thank D. Loison for useful remarks and  J. Vidal for a careful reading of manuscript and helpful remarks.

\appendix
\section{The positivity of the anomalous dimension}
\label{annexe_eta}
In this appendix, we sketch the proof showing that the anomalous
dimension $\eta$ must be positive in a second order phase transition
if the underlying theory is given by a usual $\phi^4$-like GLW
theory. This excludes, for instance, theories involving gauge fields
or replica field theories of disordered systems using the formal $N\to
0$ limit.  The argument goes as follows.  On one hand, using the
K\"{a}llen-Lhemann decomposition, it is possible to prove that the
field renormalization $Z$ is positive and less or equal to one: $0\le
Z\le1$ \cite{zinn_eta_pos}. On the other hand, using the RG equations,
it is possible to show that, around the fixed point describing the second order
 phase transition, $Z$ behaves with the RG scale $\mu$ as:
\begin{equation}
 Z(\mu)\sim\mu^\eta 
\end{equation}
when $\mu\to 0$, which corresponds to the long distance --- \ie
critical --- physics. By combining these two results we find $\eta\ge
0$.

\section{The invariants of the symmetry group}
\label{annexe_invariants}
We show,  in this appendix,  that all field
combinations invariant under  $O(N)\times O(2)$ can
be rewritten in terms of the two invariants $\rho$ and $\tau$
introduced in  Eq.(\ref{hamilton}) and given by $\rho= \hbox{Tr}(\,^t\Phi \Phi)$ and $\tau=\frac12\hbox{Tr}\left( ^t\Phi \Phi -\nbOne\  \rho/2\right)^2$. This property is {\it a priori} nontrivial
since we can easily build an infinite number of invariants by
considering, for instance, $\tr \left(^t\Phi.\Phi\right)^n$ for any
value of $n$ or  $\det \left(^t\Phi.\Phi\right)$.  The result 
 is easily obtained by using the properties of
the characteristic polynomial of the square matrix $X$:
\begin{equation}
P_X(\lambda)=\det(X-\lambda\openone) \ .
\end{equation}
In the case of a two by two matrix, the characteristic polynomial
reads:
\begin{equation}
P_X(\lambda)=\lambda^2-\lambda \;\tr X+\det X \ .
\end{equation}
The Cayley-Hamilton theorem states that any matrix is a root of its
characteristic polynomial:
\begin{equation}
P_X(X)=0\ .
\end{equation}
Applying this last result  to the two by two matrix
$^t\Phi.\Phi$ we  get:
\begin{equation}
\left(^t\Phi.\Phi\right)^2- \,^t\Phi.\Phi \;\tr \left(^t\Phi.\Phi\right)+\det
\left(^t\Phi.\Phi\right)=0\ .
\label{eq_caracteristique}
\end{equation}
By taking the trace of this equation, we see that $\det
\left(^t\Phi.\Phi\right)=\rho^2/4-\tau$. Moreover, if we multiply
Eq.~(\ref{eq_caracteristique}) by $^t\Phi.\Phi$ and take the trace of this
equation, we observe that $\tr\left(^t\Phi.\Phi\right)^3$ can be
expressed in terms of $\tr\left(^t\Phi.\Phi\right)^2$,
$\tr\left(^t\Phi.\Phi\right)$ and $\det\left(^t\Phi.\Phi\right)$ which,
themselves, can be expressed in terms of $\rho$ and $\tau$. By
iteration, we can show that all $O(N)\times O(2)$ invariants can
be expressed in terms of $\rho$ and $\tau$.  This property can be
generalized to the $O(N)\times O(P)$ model  (with $N\geq P$), which
 admits  $P$ independent invariants.

\section{The threshold functions}
\label{annexe_threshold}

We discuss in this appendix the different threshold functions $l$, $m$
and $n$ appearing in the flow equations, which encode the
nonperturbative properties of the theory. We consider here a general
case, where the threshold functions depend on three arguments. For
particular truncations --- for instance that discussed in the $O(N)$
vectorial model --- it may happen that some of these arguments are
vanishing. In such case, we do not write the associated argument so
that, for instance, $l^d_{n,0}(w,0,0)$ is denoted by $l^d_{n}(w)$

\subsection{Definitions}

The threshold functions are defined as:

\begin{subequations}
\begin{align}
l_{n_1,n_2}^d(w_1,w_2,a)=\nonumber \\
-\frac12\int_0^\infty dy\;&  y^{d/2-1}
\tilde{\partial}_t\left\{\frac{1}{(P_1+w_1)^{n_1}
(P_2+w_2)^{n_2}}\right\} ,\label{annexe_def_l}\\ \displaybreak[0]
m_{n_1,n_2}^d(w_1,w_2,a)&=\nonumber\\
-\frac12\int_0^\infty  dy\;& y^{d/2-1}
\tilde{\partial}_t\left\{\frac{y(\partial_y P_1)^2}{(P_1+w_1)^{n_1}
(P_2+w_2)^{n_2}}\right\} \label{annexe_def_m},\\\displaybreak[0]
n_{n_1,n_2}^d(w_1,w_2,a)&=\nonumber\\
-\frac12\int_0^\infty  dy \;&  y^{d/2-1}
\tilde{\partial}_t\left\{\frac{y\partial_y P_1}{(P_1+w_1)^{n_1}
(P_2+w_2)^{n_2}}\right\} ,\label{annexe_def_n}
\end{align}
\label{annexe_def_thres}
\end{subequations}
where we have introduced:
\begin{equation}
\left\{
\begin{array}{ll}
&P_1=P_1(y,a)=y(1+r(y)+a)\\ 
\\
&P_2=P_2(y)=y(1+r(y))
\end{array}
\right.
\end{equation}
with $r(y)$ being the  dimensionless cut-off:
\begin{equation}
r(y)=\frac{R_k(y k^2)}{Z y k^2}\ .
\end{equation}

We recall that the tilde in $\tilde\partial_t$ means that only the $t$
dependence of the function $R_k$ is to be considered. As a
consequence, we should not consider the $t$-dependence of the coupling
constants to perform this derivative. Therefore, in the preceding
equations:
\begin{align}
\tilde{\partial}_t P_i&=\frac{\partial R_k}{\partial t}
\frac{\partial}{\partial R_k} P_i\\
&=-y(\eta r(y)+2y r'(y)).
\end{align}
Now, threshold functions can be expressed as explicit integrals if we
compute the action  of $\tilde \partial_t$. To this end, it is
interesting to notice the equality: $\partial_y \tilde{\partial}_t P_i
= \tilde{\partial}_t \partial_y P_i$, so that:
\begin{equation}
\tilde{\partial}_t\partial_y r(y)=-\eta(r(y)+yr'(y))-2y(2r'(y)+yr''(y))
\end{equation}
We then get:
\begin{widetext}
\begin{gather}
\begin{split}
\hspace{-3.7cm}l_{n_1,n_2}^d(w_1,w_2,a)=-\frac12\int_0^\infty dy\; y^{d/2}
\;&\frac{\eta r(y) +2yr'(y)}{(P_1+w_1)^{n_1}
(P_2+w_2)^{n_2}}\left(\frac{n_1}{P_1+w_1}+\frac{n_2}{P_2+w_2}
\right)\, \label{expression_pour_l}
\end{split}\\ \displaybreak[0]
\begin{split}
n_{n_1,n_2}^d(w_1,w_2,a)=-&\frac12\int_0^\infty dy\; y^{d/2}
\frac{1}
{(P_1+w_1)^{n_1}(P_2+w_2)^{n_2}} \Bigg\{y(1+a+r(y)+yr'(y))(\eta r(y)+2y
r'(y))\ . \\\ . &\left(\frac{n_1}{P_1+w_1} +\frac{n_2}{P_2+w_2} 
\right)-\eta(r(y)+yr'(y))-2y(2r'(y)+yr''(y))
   \Bigg\}\ ,
\end{split}\\ \displaybreak[0]
\begin{split}
m_{n_1,n_2}^d(w_1,w_2,a)=-&\frac12\int_0^\infty dy\; y^{d/2}
\frac{1+a+r(y)+yr'(y)}
{(P_1+w_1)^{n_1}(P_2+w_2)^{n_2}}\Bigg\{y(1+a+r(y)+yr'(y))(\eta r(y)+2y
r'(y))\ . \\\ .&\Bigg(\frac{n_1}{P_1+w_1}+\frac{n_2}{P_2+w_2}
\Bigg)-2\eta(r(y)+yr'(y))-4y(2r'(y)+yr''(y))\ \label{expression_pour_m}
   \Bigg\}\ .
\end{split}
\end{gather}
\end{widetext}

Once a regulator $r(y)$ has been chosen, the threshold functions can be
computed numerically and, in some cases, analytically.

\subsection{Substitution rules}

 We give here the rules which relate the different
integrals appearing in the calculation to the threshold functions.
When calculating the flow equation for the coupling constants related
to the potential part, a single function $l$ appears:
\begin{widetext}
\begin{equation}
\begin{aligned}
\tilde\partial_t\int \frac{d^d\cg q}{(2\pi)^d} \; {\left(R_k(\cg 
q^2)+(Z +A)\cg q^2+W_1\right)^{-n_1} \left(R_k(\cg q^2)+Z\cg
q^2+W_2\right)^{-n_2} }&=\\ =-4v_d\;Z^{-n_1-n_2}
k^{d-2(n_1+n_2)}\;l_{n_1,n_2}^d&\Big(\frac{W_1}{Z k^2},\frac{W_2}{Z
k^2},\frac AZ \Big)\ .
\end{aligned}
\end{equation}
\end{widetext}
For the coupling constants associated with the derivative terms, two
more functions appear:
\begin{widetext}
\begin{equation}
\begin{aligned}
\frac{d\;}{d\cg p^2}\ \tilde\partial_t&\int \frac{d^d\cg q}{(2\pi)^d}
\;\cg q^\alpha\left(R_k((\cg {p+q})^2)+(Z +A)(\cg
{p+q})^2+W_1\right)^{-n_1} \left(R_k(\cg q^2)+Z
\cg q^2+W_2\right)^{-n_2} =\\ &=\frac{4 v_d \;n_1}d
Z^{-n_1-n_2} k^{d+\alpha-2(n_1+n_2+1)}\;
\Bigg\{-\alpha\;n_{n_1+1,n_2}^{d+\alpha-2}\Big(\frac{W_1}{Z
k^2},\frac{W_2}{Z k^2},\frac AZ
\Big)+\\ &+2 n_2\bigg(m_{n_1+1,n_2+1}^{d+\alpha}\Big(\frac{W_1}{Z
k^2},\frac{W_2}{Z k^2},\frac AZ
\Big)  -\frac A Z
 n_{n_1+1,n_2+1}^{d+\alpha}\Big(\frac{W_1}{Z k^2},\frac{W_2}{Z
k^2},\frac AZ \Big) \bigg)\Bigg\}
\end{aligned}
\end{equation}
\end{widetext}
\begin{widetext}
\begin{equation}
\begin{aligned}
\frac{d\;}{d\cg p^2}\ \tilde\partial_t\int \frac{d^d\cg q}{(2\pi)^d}
\;{ {\cg p}}. { {\cg q}}\;\cg q^\alpha&\left(R_k((\cg
{p+q})^2)+(Z +A)(\cg {p+q})^2+W_1\right)^{-n_1}
\left(R_k(\cg q^2)+Z \cg q^2+W_2\right)^{-n_2} =\\ & =\frac{8
v_d \;n_1}d Z^{-n_1-n_2} k^{d+\alpha-2(n_1+n_2)}\;
n_{n_1+1,n_2}^{d+\alpha}\Big(\frac{W_1}{Z k^2},\frac{W_2}{Z
k^2},\frac AZ \Big)\ .
\end{aligned}
\end{equation}
\end{widetext}
Notice that the powers of $k$ and $Z$ appearing in the preceding
expressions are chosen so that when the flow equations are reexpressed
in terms of dimensionless renormalized quantities, there is  no explicit
dependence on these parameters.

\subsection{Universal values of the threshold functions}
\label{annexe_universel}

 For particular arguments, the threshold functions take values
independent of the choice of the regulating function $r(y)$. This is
particularly important when we extract the first coefficients of the
perturbative $\beta$ functions out of the nonperturbative ones, since
the former are universal. From Eq.~(\ref{expression_pour_l}) we can compute the value of  $l_{n,0}^{2n}(0,0,a)$
 which enters in the $\beta$ function for the coupling constant of  the GLW model around four dimensions:
\begin{align}
l_{n,0}^{2n}(0,0,a)&=-n \int_0^\infty dy\
{r'(y)\over (1+a+r(y))^{n+1}}\nonumber\\
&=\left[(1+a+r(y))^{-n}\right]_0^\infty\\ &=\frac1{(1+a)^n}\ . \nonumber
\end{align}
The last equality follows from the asymptotic behaviors of $r(y)$ that are given by Eqs.~(\ref{IR}) and (\ref{UlVi}): 
\begin{equation}
\left\{
\begin{aligned}
&\lim_{y\to \infty}\  r(y)=0\\
&\lim_{y\to 0}\  y\  r(y)=1
\end{aligned}
\right.
\label{asymptotics}
\end{equation}
and  are independant of the actual form chosen for $r(y)$.

Similarly, one finds: 
\begin{equation}
l_{0,n}^{2n}(0,0,a)=1\ .
\end{equation}

Also, the threshold function $m_{2,2}^d(w,0,0)$ takes a universal
form for large argument $\omega$,  which  enters in  the $\beta$ function 
of the temperature in the NL$\sigma$ model around two dimensions. Using Eq.~(\ref{expression_pour_l}), one gets:
\begin{align}
\lim_{w\to\infty} w^2  m_{2,2}^2(w,0,0)&=\int_0^\infty
dy\,\partial_y\left(\frac {1+r(y)+yr'(y)}{1+r(y)}\right)^2\nonumber\\
&=1
\end{align}
where, again, we have used  the asymptotic behaviors of
$r(y)$, Eq.~(\ref{asymptotics}).

\subsection{Threshold functions from the $theta$ cut-off}

For certain regulating functions $r(y)$, it is possible to compute
analytically the threshold functions. Using such regulating functions
is very helpful in practice and simplifies considerably the numerical
procedures. In this section, we give the threshold functions
associated with the theta cut-off, see  Eq.~(\ref{cutoffstep}). One has, for $a=0$:
\begin{align}
\begin{split}
l_{n_1,n_2}^d(&w_1,w_2,0)=\frac2d\left(1-\frac \eta
{d+2}\right)\\ &
\frac1{(1+w_1)^{n_1} (1+w_2)^{n_2}}\left(\frac {n_1}{1+w_1}+\frac
{n_2}{1+w_2}\right)
\end{split}
\\ 
m_{2,2}^d(&w_1,w_2,0)=\frac1{(1+w_1)^{n_1} (1+w_2)^{n_2}}\ .
\end{align}

\section{The minimum of the RG flow}
\label{annexe_minimum}

In this appendix, we describe in more details the notions of
pseudo-fixed point and of minimum of the flow. We then explain how
these ideas have been implemented in practice to determine effective
exponents for very weakly first order phase transitions.

As described previously, the RG flow equations for STA with a large
number of spin components ($N>N_c(d)$) admit two fixed points.  When $N$
is decreased slightly below $N_c(d)$, the two fixed points acquire a
small complex part and loose their direct physical
relevance. Strictly speaking, there is no more attractor in the real
coupling constant space but the flow remains sensitive to the
presence of complex fixed points.  Zumbach \cite{zumbach93,zumbach94,zumbach94c}
proposed that a particular point, the minimum of the flow, should
mimic to some extent the behavior of an attractor.  This point is
defined as the location, in coupling constant space, where the flow is
the slowest and  the quantity:
\begin{equation}
A(\{g_i\})=\frac12\sum_i\beta_i^2
\end{equation}
--- where $\beta_i$ are the $\beta$-functions for the different
coupling constants $g_i$ --- is minimum. Let us stress on few properties of the minimum of
the flow:
\begin{itemize}
\item when a true fixed point exists, $A(\{g_i^\star\})=0$ and, in
this case, the minimum {\em is} a fixed point.
\item When two fixed points annihilate, we are left with a single
minimum of the flow sitting right at the position where the fixed
points have collapsed.
\item For trajectories getting close to the minimum, the RG time spent
in its vicinity is large and so is the correlation length.
\end{itemize}
We therefore see that a minimum shares some features with a true fixed
point.

One easily obtains the equation characterizing a minimum:
\begin{equation}
\label{eq_min_annexe}
\frac{\partial A}{\partial g_i}=\sum_j M_{i,j}\beta_j=0
\end{equation}
with 
\begin{equation}
M_{i,j}=\frac{\partial \beta_j}{\partial g_i}\ .
\end{equation}

Under the assumption that  the minimum of the flow mimics
correctly the attractor of the trajectories, it is natural to compute
the critical pseudo-critical exponents in the standard way. The anomalous dimension is
obtained by evaluating $\eta(\{g_i\})$ at the minimum of the flow and
$\nu$ by diagonalizing the matrix $M_{i,j}$ at this point. It is
important to notice that the pseudo-critical exponents thus obtained are
invariant under reparametrization of coupling constants, as it should
be, since Eq.~(\ref{eq_min_annexe}) transforms as components of a
vector.

\section{The discontinuous character of the phase transition}

 In this appendix, we discuss in more details the problems that we encounter in our  description of the 
 first order  phase transition that occurs in frustrated magnets. We also explain the surprising increase of
 the correlation length observed at small reduced temperatures (see
Fig.~\ref{log_m_xi_neq2}b).

To this end, let us discuss the following toy model of first order phase transition. We consider a scalar
 $\nbZ_2$-invariant  model  characterized by one field $\phi$
and by a ``$\phi^4-\phi^6$'' potential:
\begin{equation}
U(\phi)=r\frac {\phi^2}2-\frac{ \phi^4}2+\frac {\phi^6}6 \ . 
\label{pot_annexe}
\end{equation}
As usual, we assume that $r$  varies linearly with the
temperature. 

\begin{figure}[htbp] 
\begin{center}
\includegraphics[width=.9\linewidth,origin=tl]{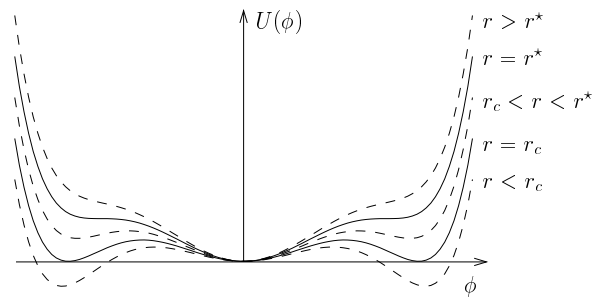}\hfill%
\end{center} 
\caption{Shape of the potential Eq.~(\ref{pot_annexe}) for different
temperatures, \ie different values of the parameter $r$. The plain
lines correspond to $r=r_c$ and $r=r^\star$, while the dotted lines
correspond to different generic values or $r$.}
\label{curve} 
\end{figure}

For low reduced  temperatures --- small $r$ --- the potential has a local minimum
for $\phi=0$ and a global minimum for: 
\begin{equation}
(\phi^{\Min}(r))^2=1+\sqrt{1- r }
\label{min_annex}
\end{equation}
so that the system exhibits a spontaneous magnetization, see
Fig.~\ref{curve}. 
\label{chap_premier_ordre}
When the temperature ---  $r$ ---  is increased, the
energy difference between the configurations $\phi=0$ and
$\phi=\phi^{\Min}(r)$ decreases and eventually vanishes for
$r=r_c=3/4$ which defines the critical temperature. For $r$ larger than $r_c$, the ground state of the system
is given by  the configuration $\phi=0$ so that the system has no more
spontaneous magnetization. Therefore, when one  crosses  the critical
temperature, ones  observes  a jump of the magnetization from
$\phi^{\Min}(r_c)$ to 0, which is the consequence of the competition
between two minima of the potential, see Fig.~\ref{curve}.

For $r>r_c$, the field configuration $\phi^{\Min}(r)$ which is no longer  the
ground state  becomes a metastable state. One sees  from Eq.~(\ref{min_annex}) that, for
$r>r^\star=1$, this  metastable state disappears and we are left with
$\phi=0$ as the only physically relevant state, see
Fig.~\ref{curve}.  Finally, it  must be noted  that, for $r=r^\star$, the
curvature of the potential at  the configuration $\phi^{\Min}(r^\star)$ vanishes:  $U''(\phi^{\Min}(r^\star))=0$.
 This means that the
susceptibility in the metastable state diverges at
$r=r^\star$. Similarly, one can show that the correlation length in
the metastable state also diverges.

Let us now come back to the NPRG method. In the truncation of the
effective average action that we use ---   an expansion in
powers of the fields  of the form Eq.~(\ref{troncation}) ---   we retain
only {\it local} informations on the potential around its nontrivial
minimum ---  which is  equivalent to the   configuration $\phi^{\Min}(r)$ discussed  above. In particular we do
 not accurately describe the physics  around the zero-field configuration $\vec\phi_1=\vec\phi_2=\vec 0$. We are
 thus unable to compare the energies of different local minima and to 
determine the temperature of transition  at which the energies of the two
minima are equal. Also, in a small domain of temperatures
--- equivalent here to $r_c<r<r^\star$ --- the configuration that we probe  
corresponds  actually to the metastable state and not to the true
equilibrium state. However, these  phenomena  should not induce a large bias in our analysis as long as the
 transition is {\it weakly} of first order since, in this case, the temperature range where metastable states 
exist is very small. This means that the error induced on the determination of the critical temperature is very 
small too.

Moreover in our study, as in the toy model above, we should observe  when $r$ 
reaches $r^\star$ ---  the temperature at which the 
metastable state must  disappear ---   the
associated divergence of the correlation length discussed
previously. This is precisely what we found in frustrated magnets for
small reduced temperature --- see Fig.~\ref{log_m_xi_neq2}b.

Note that this increase of the correlation length as well as the error
 associated with  our determination of the critical temperature  both rely on the 
 truncation that we have considered. These  problems  can be cured
by considering truncations of the form Eq.~(\ref{action_generale})
which retains the full field-dependence of the potential (see for
instance \cite{berges02} for a treatment of a first order phase
transition in a NPRG approach).

\end{document}